\documentclass[%
reprint,
amsmath,amssymb,
aps,
]{revtex4-2}

\usepackage{graphicx}
\usepackage{dcolumn}
\usepackage{bm}
\usepackage[colorlinks, linkcolor=blue, anchorcolor=blue, citecolor=blue]{hyperref}
\usepackage{url}
\usepackage{amsmath,amssymb}
\usepackage[mathlines]{lineno}
\usepackage{makecell}
\usepackage{multirow}
\usepackage{subfigure}
\usepackage{eurosym}
\usepackage{amsmath,amssymb,mathrsfs}
\usepackage{setspace}

\makeatletter

\newcommand{\Rmnum}[1]{\expandafter\@slowromancap\romannumeral #1@}
\makeatother

\begin{document}
	
	\preprint{APS/123-QED} 
	
	\title{Oscillating bound states in waveguide-QED system with two giant atoms}
	\author{F. J. L\"u}
	\affiliation{School of Physical Science and Technology, Southwest Jiaotong University, Chengdu 610031, China}
	\author{W. Z. Jia}
	\email{wenzjia@swjtu.edu.cn}
	\affiliation{School of Physical Science and Technology, Southwest Jiaotong University, Chengdu 610031, China}

	\date{\today}
	
	\begin{abstract}
	We study the bound states in the continuum (BIC) in a system of two identical two-level 
	giant atoms coupled to a one-dimensional waveguide. By deriving general dark-state conditions, 
	we clarify how coupling configurations and atomic parameters influence decay suppression. 	
	Through analysis of the long-time dynamical behaviors of atoms and bound photons, 
	we carry out a detailed classification of bound states and explore the connections between 
	these dynamical behaviors and the system’s intrinsic light-matter interactions.
	The system supports static bound states with persistent atomic excitations, and oscillating 
	bound states with periodic atom-photon or atom-atom excitation exchange. 	
	Under certain conditions, oscillating bound states can contain more harmonic components owing to the emergence of additional 
	quasi-dark modes, rendering them promising platforms for high-capacity quantum information processing.
        These findings advance the understanding of BIC in waveguide quantum electrodynamics with multiple giant atoms and 
        reveal their prospective applications in quantum technologies.
	\end{abstract}
	
	\maketitle
	
	
\section{\label{introduction}Introduction}
Giant atoms, defined by their nonlocal coupling to waveguides via multiple discrete connection points~\cite{Kockum-MI2021}, constitute a rapidly expanding subfield within waveguide quantum electrodynamics (wQED)~\cite{Roy-RMP2017,Gu-PhysReports2017,Sheremet-RMP2023}.
Giant atom architectures have been experimentally demonstrated across diverse quantum platforms, particularly in superconducting qubits~\cite{Andersson-NatPhys2019,Kannan-Nature2020} and macroscopic spin ensembles ~\cite{Wang-Natcom2022}. The non-dipole nature of such systems gives rise to remarkable phenomena, such as frequency-dependent decay rates and modified Lamb shifts ~\cite{Kockum-PRA2014}, decoherence-free interactions~\cite{Kannan-Nature2020,Kockum-PRL2018,Carollo-PRR2020,Du-PRA2023,Ingelsten-PRR2024}, and non-Markovian dynamics~\cite{Andersson-NatPhys2019,Guo-PRA2017,Guo-PRR2020,Xu-NJP2024,Roccati-PRL2024,Qiu-PRR2024,Yannopapas-Photonics2025,Sivakumar-PRA2025}.
Research on wQED systems, including those incorporating giant atoms, continues to uncover novel bound-state effects, including both bound states in the gap~\cite{John-PRL1990,Tong-JPhyB2010,Calajo-PRA2016,Shi-PRX2016,Hood-PNAS2016,Liu-NatPhy2017,Sundaresan-PRX2019,Zhao-PRA2020,Ferreira-PRX2021,Wang-PRL2021,Scigliuzzo-PRX2022,Wang-QST2022,Soro-PRA2023,Jia-OE2024} and bound states in the continuum (BIC)~\cite{Tufarelli-PRA2013,Hsu-NatRevMats2016,Fong-PRA2017,Facchi-PRA2019,Calajo-PRL2019,Sinha-PRL2020,Qiu-SciChina2023,Yu-PRA2025}. 
Their emergence is closely linked to 
strong light-matter interactions and the one-dimensional (1D) geometry inherent to such systems.
Particularly, if a wQED system supports multiple dark modes simultaneously, interference between them can form  
oscillating bound states~\cite{Lombardo-PRA2014,He-QT2025}. Owing to their
multi-point coupling structure, wQED systems incorporating giant atoms can readily form effective cavity-QED architectures with the 
atoms forming their own cavities, making them ideal platforms for generating such oscillating bound states in the 
continuum~\cite{Guo-PRR2020,Guo-PRA2020}. Subsequently, oscillating bound states of giant atoms under various conditions
have been further investigated, including structured waveguide environments~\cite{Lim-PRA2023,Zhang-PRA2023}, 
semi-infinite waveguides with boundary effects~\cite{Li-PRA2024}, and multi-level atoms coupled to external classical fields or cavity 
modes~\cite{Yang-PRA2025,Sun-PRA2025}. This phenomenon demonstrates that the non-Markovian dynamics arising from 
time-delay effects constitutes a valuable resource for enhanced quantum information storage and manipulation in higher-dimensional 
Hilbert spaces. Previous studies on oscillating bound states have focused mainly on single giant atoms, leaving the dynamics of 
multi-giant-atom systems largely unexplored.

In this work, we extend our analysis to a system comprising two identical two-level giant atoms coupled to a 1D waveguide, including both separate and braided configurations. First, we derive general dark-state conditions for two giant atoms, 
clarifying how coupling topology (separate versus braided) and the atomic parameters (including transition frequency $\Omega$, 
decay rate $\gamma$ at each coupling point, and number of coupling points $N$) influence decay suppression. 
The system hosts a diverse range of bound states with distinct dynamical behaviors, including static bound states with persistent 
atomic excitations, and oscillating bound states with periodic atom-photon or atom-atom excitation exchange. Due to interatomic 
interactions mediated by waveguide environment, the dynamical evolution of both atomic excitation probabilities and bound-photon 
distributions within oscillating bound states exhibits a richer diversity of patterns compared to that in single-giant-atom oscillating 
bound states. When all dark modes are confined to symmetric or antisymmetric subspaces, atomic collective modes and bound 
photons continuously exchange excitations, leading to synchronous oscillations between the two atoms. Conversely, if the system 
harbors both symmetric and antisymmetric dark modes simultaneously, interference between them causes the excitation probabilities 
of the two atoms to exhibit distinct oscillation patterns, forming asynchronous oscillating bound states. These include hybrid oscillating 
bound states where the two atoms possess different numbers of harmonic components, as well as exchange-type oscillating bound 
states characterized by periodic excitation exchange between the two atoms. 
Furthermore, such exchange-type oscillating bound states can emerge in the Markovian regime, where the total atomic excitation is 
nearly unity with negligible bound photons, and this is consistent with the predictions of decoherence-free 
interactions~\cite{Kockum-PRL2018}. Finally, we find that in some situations, the existence of quasi-dark modes significantly 
increases the harmonic components in oscillating bound states over longer time scales, establishing a promising pathway toward 
high-capacity platforms for quantum information storage and processing. These findings deepen our understanding of BIC in wQED systems with multiple giant atoms and reveal their potential applications in quantum technology.

The remainder of this paper is organized as follows. Section \ref{Model} introduces the theoretical framework, including the system Hamiltonian and equations of motion within the single-excitation subspace, from which we derive general expressions for atomic excitation amplitudes and photonic wave functions.
In Sec.~\ref{DSCondition}, we establish the dark-state conditions.
In Sec.~\ref{StaticBoundState}, we analyze the properties of static bound states, 
laying the foundation for understanding oscillating bound states.
Section \ref{OscillatingBS} is devoted to the properties of oscillating bound states. Specifically, the parameter points for such states are provided in Sec.~\ref{SpecialParameterOBS}; the general theoretical framework is presented in Sec. \ref{GeneralDescriptionOBS};  the properties of different types of oscillating bound states in separate and braided configurations are discussed in Secs. \ref{SepOBS} and \ref{BraOBS}, respectively; and
Sec.~\ref{PhenomenonQusiOBS} addresses the phenomenon of oscillating quasi-bound states.
Finally, further discussions and conclusions are given in Sec.~\ref{Conclusion}.
\section{\label{Model}Theoretical framework}
\begin{figure}[t]
	\centering
	\includegraphics[width=0.5\textwidth]{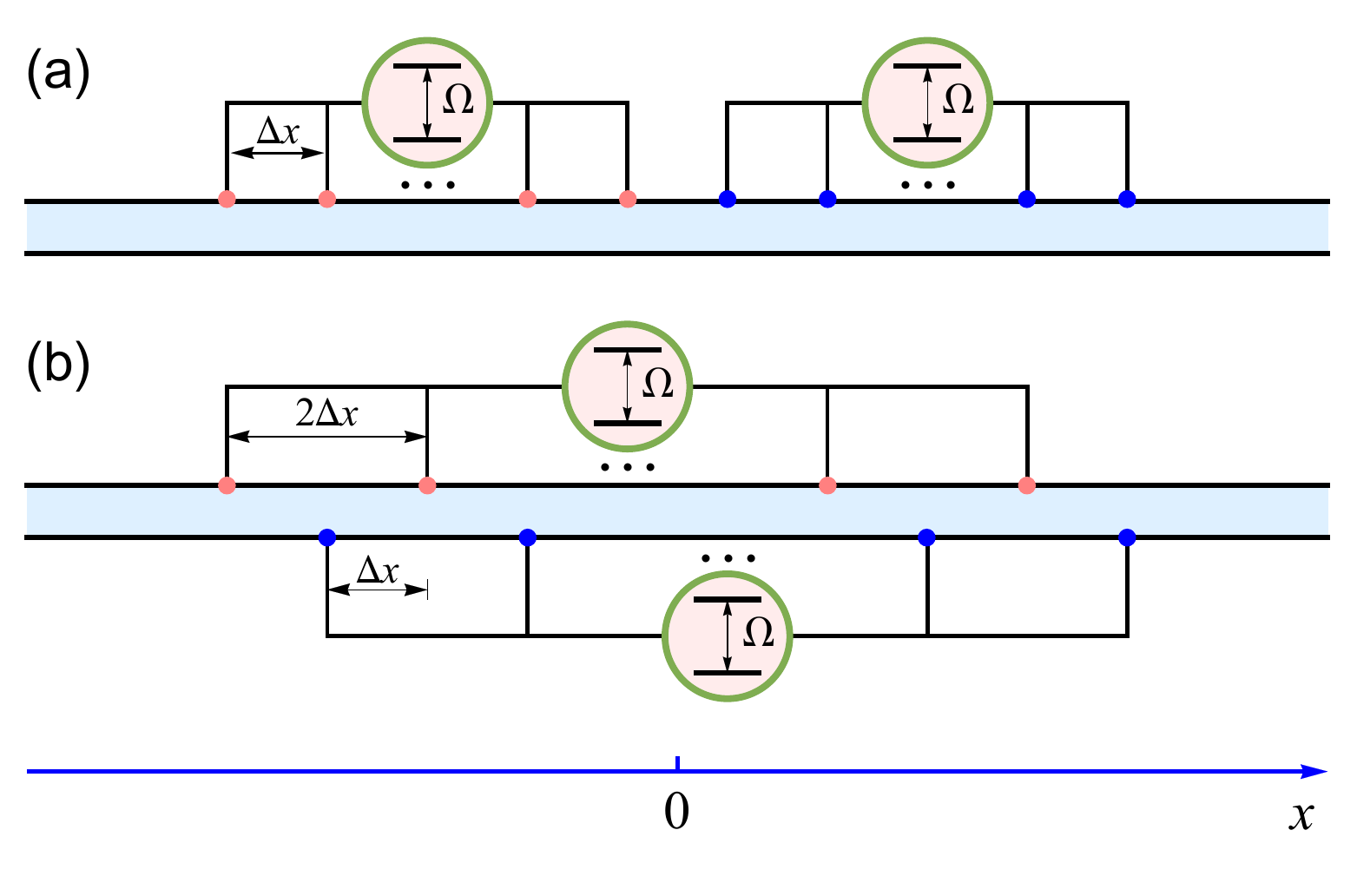}
	\caption{Sketch of two giant atoms coupled to an open waveguide for two distinct topologies:
	(a) two separate giant atoms; (b) two braided giant atoms.
	}
	\label{Schematics}
\end{figure}
The present study focuses on the wQED setup with two identical two-level giant atoms. Each atom couples to a 1D waveguide through $N$ connection points, forming a separate or braided configuration, as illustrated schematically in Figs.~\ref{Schematics}(a) and \ref{Schematics}(b).
The connection points, situated at locations $x_{ij}$ ($i=1,2$ and $j=1,2,\cdots,N$), are equidistant, with a separation of $\Delta x$.  
The midpoint of the system is designated as $x=0$.
Under the rotating-wave approximation (RWA), the Hamiltonian of the entire system can be expressed as follows ($\hbar=1$):
\begin{eqnarray}
	\hat{H}&=&\sum_{i}{\Omega\sigma_{i}^{+}\sigma_{i}^{-}}+\int_{-\infty}^{+\infty}\mathrm{d}k\nu_{k}\hat{a}_{k}^{\dagger}\hat{a}_{k}
	\nonumber
	\\
	&&+\sum_{ij}\int_{-\infty}^{+\infty}\mathrm{d}k{g(k)}\left(e^{\mathrm{i}k{x_{ij}}}\hat{a}_{k}\sigma_{i}^{+}+\mathrm{H.c.} \right),
\end{eqnarray}
where $\Omega$ is the transition frequency of the atoms. ${\sigma}_{i}^{+}$ and ${\sigma}_{i}^{-}$ are the raising and lowering operators of the $i$th atom. $\nu_{k}=v|k|$ are the frequencies of the bosonic fields (photons or phonons; henceforth, the term ``photon" will be used to denote both.) in the waveguide, where $k$ and $v$ represent wave vectors and group velocity, respectively. $\hat{a}_k^{\dagger}$ and $\hat{a}_{k}$ are the creation and annihilation operators of photons, satisfying $[\hat{a}_k,\hat{a}_{k'}^\dagger]=\delta(k-k^\prime)$. $g(k)$ represents the coupling strength at each coupling point.

We assume that the system is initialized in a single excited state 
$|\Psi(0)\rangle=(c_1|\text{eg}\rangle+c_2|\text{ge}\rangle)\otimes|\text{vac}\rangle$, with $|c_1|^2+|c_2|^2=1$.
$|\text{eg}\rangle\equiv\sigma_{1}^{+}|\text{gg}\rangle$ ($|\text{ge}\rangle\equiv\sigma_{2}^{+}|\text{gg}\rangle$) is the single-atom 
excitation state. $|\text{gg}\rangle$ is the ground state of the two-atom system, and $|\text{vac}\rangle$ is the vacuum state of 
photonic field in the waveguide. Thus, the problem is restricted to the single excitation subspace. The state vector that describes the 
system's dynamic evolution is expressed as follows:
\begin{equation}
|\Psi(t)\rangle=\sum_{i}\beta_i(t)\sigma_{i}^{+}|\emptyset\rangle+\int_{-\infty}^{+\infty}\mathrm{d}k\alpha_{k}(t)\hat{a}_{k}^{\dagger}|\emptyset\rangle,
\label{StateVector2A}
\end{equation}
where $|\emptyset\rangle\equiv|\text{gg}\rangle\otimes|\text{vac}\rangle$ is the ground state of the total system, $\beta_{i}(t)$ denotes excitation amplitude of the $i$th atom, and  $\alpha_{k}(t)$ is the probability amplitude of a photon in the $k$ mode. Note that the initial condition requires
$\beta_i(0)=c_{i}$ and $\alpha_k(0)=0$.
Tracing out the field modes, the evolution of atomic excitation amplitudes is given by the following equation (see Appendix \ref{DerivationEOM}): 
\begin{equation}
	\dot{\beta}_{i}(t)=-\mathrm{i}\Omega\beta_{i}(t)
	-\frac{\gamma}{2}\sum_{i'jj'}\beta_{i'}\left(t-\left|\tau_{ij,i'j^{\prime}}\right|\right)\Theta\left(t-\left|\tau_{ij,i'j^{\prime}}\right|\right).
	\label{EOMAtom2A}
\end{equation}
Here  $\gamma\equiv 4\pi\tilde{g}^{2}(\Omega)/v$ is the spontaneous emission rate at a single coupling point into the waveguide. 
Note that the Weisskopf-Wigner approximation has been performed such that  $g(k)\simeq g(\Omega/v)\equiv\tilde{g}(\Omega)$.
$\Theta(\bullet)$ denotes the Heaviside step function. $\tau_{ij,i'j^{\prime}}=(x_{ij}-x_{i'j^{\prime}})/v$ represents the time delay between the connection points $x_{ij}$ and $x_{i'j^{\prime}}$. 
By introducing $\hat{a}^{\dagger}(x)\equiv\int_{-\infty}^{\infty}\mathrm{d}k\hat{a}^{\dag}_{k}e^{-\mathrm{i}kx}/{\sqrt{2\pi}}$
and $|x\rangle\equiv\hat{a}^{\dagger}(x)|\emptyset\rangle$, 
the wave function of the light field in the waveguide $\varphi(x,t)=\langle x|\Psi(t)\rangle$ is given by (see Appendix \ref{DerivationEOM})
\begin{equation}
	\varphi(x,t)=
	-\mathrm{i}\sqrt{\frac{\gamma}{2v}}
	\sum_{ij}\beta_{i}\left(t-\frac{|x-x_{ij}|}v\right)
	\Theta\left(t-\frac{|x-x_{ij}|}v\right).
	\label{WFPhoton2A}
\end{equation}

With the help of the Laplace transform, the solution to Eq.~\eqref{EOMAtom2A} can be obtained as follows (see Appendix \ref{DerivationTM}): 
\begin{equation}
	\beta_{1,2}(t)=\frac{1}{\sqrt{2}}\left[c_{+}\beta_{+}\left(t\right)\pm c_{-}\beta_{-}\left(t\right)\right],
	\label{beta1and2}
\end{equation}
with $c_{\pm}=(c_1\pm c_2)/\sqrt{2}$.
\begin{equation}
	\beta_{\pm}\left(t\right)=\sum_{\lambda}A_{\pm,\lambda}{e^{s_{\pm,\lambda}t}}
	\label{BetaPMtGeneral} 
\end{equation}
is the time-dependent excitation amplitude of the symmetric (antisymmetric) atomic collective state 
$|\pm\rangle\equiv(|\text{eg}\rangle\pm|\text{ge}\rangle)/\sqrt{2}$ when the system is initialized in this state. 
The complex frequency $s_{\pm,\lambda}$ is determined by the following transcendental equation
\begin{equation}
	s+\mathrm{i}\Omega+\Sigma_{\pm}(s)=0.
	\label{TranscendentalEQ}
\end{equation}
And the amplitude corresponding to the mode $s_{\pm,\lambda}$ takes the form
\begin{equation}
	A_{\pm,\lambda}=\left[{1+\Sigma'_{\pm}(s_{\pm,\lambda})}\right]^{-1},
	\label{AmplitudeFun} 
\end{equation}
with $\Sigma'_{\pm}(s)\equiv\partial_{s}\Sigma_{\pm}(s)$.
\begin{equation}
	\Sigma_{\pm}(s)=\Sigma_{\mathrm{\Rmnum{1}}}(s)\pm\Sigma_{\mathrm{\Rmnum{2}}}(s)
	\label{SigmaPM}
\end{equation}
are the self-energy functions of the symmetric and anti-symmetric modes, respectively.  And the functions
\begin{equation}
\Sigma_{\mathrm{\Rmnum{1}}}(s)\equiv\Sigma_{11}(s)=\Sigma_{22}(s),~\Sigma_{\mathrm{\Rmnum{2}}}(s)\equiv\Sigma_{12}(s)=\Sigma_{21}(s)
\label{Sigma12}
\end{equation}
delineate the contribution of photon exchange between different connection points of the same atom and between different atoms, respectively.
The function $\Sigma_{ii'}(s)$ takes the form
\begin{equation}
	\Sigma_{ii'}(s)=\frac{1}{2} \gamma\sum_{jj'}e^{-s\left|\tau_{ij,i'j'}\right|}.
	\label{Sigmaiip}
\end{equation}
Specific analytical expressions for $\Sigma_{\mathrm{\Rmnum{1},\Rmnum{2}}}(s)$ and $\Sigma_{\pm}(s)$ are provided in Appendix~\ref{ExpressionSE}. 
\begin{figure}[t]
	\centering
	\includegraphics[width=0.45\textwidth]{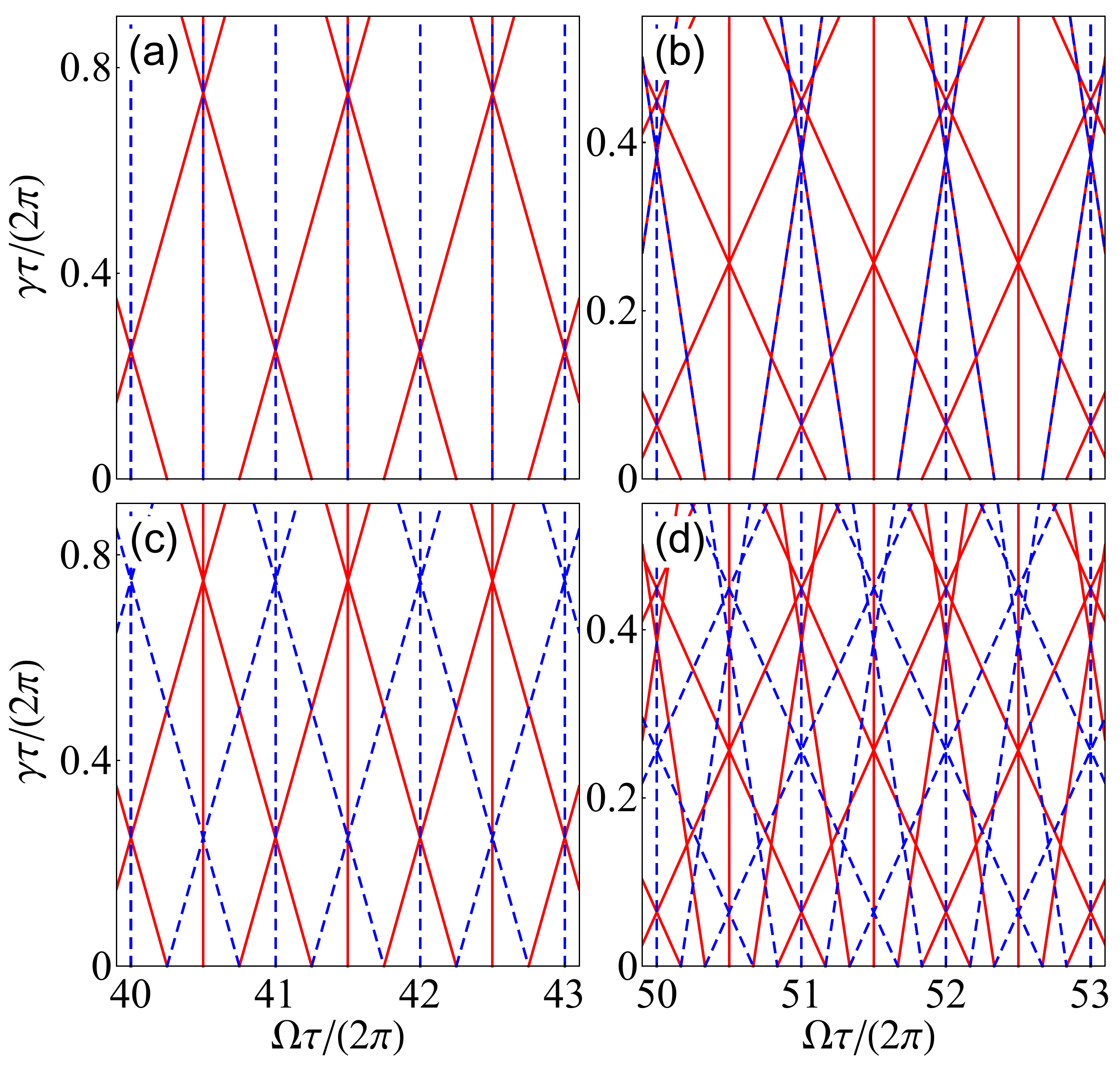}
	\caption{The $\Omega\text{-}\gamma$ parameter lines that support symmetric (red solid lines) and antisymmetric 
	(blue dashed lines ) dark modes. (a) and (b) Separate configuration with $N=2$ and $N=3$. (c) and (d) Braided 
	configuration with $N=2$ and $N=3$.}
	\label{DarkStateCondition}
\end{figure}
\section{\label{DSCondition}Dark-state conditions}
When the system is initialized in the state $|\pm\rangle$, parity conservation ensures that the 
atomic state vector retains the form $\beta_{\pm}(t)|\pm\rangle$ and the corresponding photonic wave function satisfies
$\varphi(-x,t)=\pm\varphi(x,t)$ throughout the dynamics. Here, $\beta_{\pm}(t)$ contains only symmetric or antisymmetric modes, whose complex frequencies and amplitudes are determined via Eqs.~\eqref{TranscendentalEQ} and \eqref{AmplitudeFun}.
Then, the results for an arbitrary initial state $c_{+}|+\rangle+c_{-}|-\rangle$ can be further obtained via superposition as 
 $c_{+}\beta_{+}(t)|+\rangle+c_{-}\beta_{-}(t)|-\rangle$.

In general, the solutions to Eq.~\eqref{TranscendentalEQ} are complex numbers with negative real parts. Therefore, the 
corresponding modes have nonzero relaxation rates and will completely disappear after a sufficiently long time. While those modes 
corresponding to the purely imaginary solutions will survive, because they decouple from the waveguide and form dark states. 
Therefore, the key to determining the dark-state conditions is to find the conditions for Eq.~\eqref{TranscendentalEQ} to have purely 
imaginary solutions, denoted as $s_{\pm,n}=-\mathrm{i}\omega_{\pm,n}$. After some calculations (see Appendix \ref{DerivationDSCondition}), it can be found that for both the 
separate and braided configurations, the possible frequencies of the symmetric (antisymmetric) dark modes take the form 
\begin{equation}
	\omega_{\pm,n}=\frac{n\pi}{N\tau}\equiv\omega_{n},
	\label{DarkFrequency}
\end{equation}
where $\tau\equiv\Delta x/v$ is the travel time between two neighboring connection points. Correspondingly, the dark-state condition satisfied by the parameters $\Omega$ and $\gamma$ can be expressed as  
\begin{equation}
	\Omega+\frac{1}{2}N\gamma\mathrm{cot}\left(\frac{1}{2}\omega_{q}\tau\right)=\omega_{n},~~\omega_{q}=\frac{q\pi}{N\tau}.
	\label{DarkConditionGeneral}
\end{equation}
The excitation amplitude for the atomic state $|+\rangle$ ($|-\rangle$) in a symmetric (antisymmetric) dark mode with frequency 
$\omega_n$ is given by
\begin{equation}
	A_{n}=\left[{1+\tfrac{1}{2}N\gamma\tau\csc^2\left(\tfrac{1}{2}\omega_{q}\tau\right)}\right]^{-1}.
	\label{AmplitudeDarkGeneral}
\end{equation}
In Eqs.~\eqref{DarkFrequency}-\eqref{AmplitudeDarkGeneral},
the values taken by the parameters $n$ and $q$ for different situations are summarized as follows: 
\begin{equation}
q=\left\{ 
\begin{array}{l}
	\mathrm{mod}[n,2N],~n\in\mathbb{Z}^{+},{n}/{N}\notin\mathbb{E},~\text{for ${\text{S}}_{+}$ and  ${\text{B}}_{+}$},
	\\
	\mathrm{mod}[n,2N],~n\in\mathbb{E}^{+},{n}/{N}\notin\mathbb{E},~\text{for ${\text{S}}_{-}$},
	\\
	\mathrm{mod}[n+N,2N],~n\in\mathbb{Z}^{+},{n}/{N}\notin\mathbb{O},~\text{for ${\text{B}}_{-}$}.
\end{array} \right.
\label{nANDq}
\end{equation}	
The symbols ``S'' and ``B'' refer to the separate and braided configurations, while the subscripts ``$+$'' and ``$-$''  
denote the symmetric and antisymmetric dark modes, respectively. 

Equation~\eqref{DarkConditionGeneral} reveals that the parameter points that enable the existence of bound states form a family of straight lines in the 
$\Omega\text{-}\gamma$ plane, as shown in Fig.~\ref{DarkStateCondition}. 
All points on a given line correspond to dark modes possessing the same frequency, whose value 
is specified by the $\Omega$-axis intercept of that line.
In addition, the slopes $K_q$ of the $\Omega\text{-}\gamma$ lines, given by
\begin{equation}
	K_q=\frac{2}{N}\tan\left(\frac{1}{2}\omega_q\tau\right),
	\label{Slope}
\end{equation}
show a $2N$-periodic dependence on the index $n$ of the dark modes [see Eqs.~\eqref{nANDq} 
and \eqref{Slope}]. Note that the parameter points $(\Omega, \gamma)$ should also satisfy the 
conditions $\Omega\gg N^2\gamma$ and $|\omega_n-\Omega|/\Omega\ll 1$ to ensure the validity of the 
RWA~\cite{Guo-PRR2020,Noachtar-PRA2022,Terradas-Brianso-PRA2022}. As can be verified 
from Eq.~\eqref{DarkConditionGeneral}, the first condition can guarantee the fulfillment of the second one.
Here we emphasize that the parameter ranges shown in Figs.~\ref{DarkStateCondition}(a)-\ref{DarkStateCondition}(d) strictly maintain the validity of the RWA.

In particular, when $q=N$ is satisfied, the conditions~\eqref{DarkFrequency} and \eqref{DarkConditionGeneral} reduce to
\begin{equation}
\omega_n\tau=\Omega\tau=m\pi~~(n=mN),
\label{DarkConditionSpecial}
\end{equation}
with $m\in\mathbb{O}^{+}$ for the $\text{S}_{\pm}$ and $\text{B}_{+}$ modes
($N\in\mathbb{E}^{+}$ is required for $\text{S}_{-}$) and $m\in\mathbb{E}^{+}$ for the $\text{B}_{-}$ modes. 
And the corresponding amplitude simplifies to
\begin{equation}
A_{n}=\left({1+\tfrac{1}{2}N\gamma\tau}\right)^{-1}.
\label{A_N}
\end{equation}
It should be emphasized that for the $\text{S}_{-}$ modes, the condition 
\eqref{DarkConditionSpecial} with $m\in\mathbb{E}^{+}$ is also applicable [corresponding to those vertical dashed lines that do not coincide with solid lines in Figs.~\ref{DarkStateCondition}(a) and \ref{DarkStateCondition}(b)]. 
The corresponding amplitude is
\begin{equation}
	A_{n}=\left[{1+\tfrac{1}{6}N\left(2N^2+1\right)\gamma\tau}\right]^{-1}.
	\label{ASminus}
\end{equation}
In this case, the dark-state frequency $\omega_n$ can still be given within the framework of Eq.~\eqref{DarkFrequency}, but the condition satisfied by $\Omega$ and the expression of $A_n$ are not special cases of those described by Eqs.~\eqref{DarkConditionGeneral}-\eqref{nANDq}.

Note that the above results summarize the general dark-state conditions for double giant atoms in scenarios where the coupling points are equally spaced. 
If each atom has two coupling points and only the modes matching the atomic transition frequency are considered, this condition reduces to the one presented in Ref.~\cite{Qiu-SciChina2023}, where the phase delay between neighboring coupling points must be an odd or even multiple of $\pi$.
\section{\label{StaticBoundState}Static bound states}
Here we first outline the properties of static bound states, which serve as a foundation for subsequent discussion on oscillating bound states.

When the system allows \textit{a single} symmetric (antisymmetric) dark-state mode with frequency $\omega_n$ and is initialized in the state
$|+\rangle$ ($|-\rangle$), in the long-time limit the atomic subsystem retains a residual steady excitation amplitude 
in this state [i.e., $\beta_{\pm}(t\to\infty)=A_{n}e^{-\mathrm{i}\omega_{n}t}$], accompanied by a symmetric (antisymmetric) photonic wave function $\varphi_{\pm}(x,t\to\infty)=\varphi_{\pm,n}(x)e^{-\mathrm{i}\omega_{n}t}$ confined between the left-most and right-most connection points. 
\begin{equation}
	\varphi_{\pm,n}(x)=\frac{1}{\sqrt{2}}\left[\varphi_{1,n}(x)\pm\varphi_{2,n}(x)\right]
	\label{DarkMode_PhotonW}
\end{equation}
is the stationary state wave function of the bound photons, with
\begin{equation}
	\varphi_{i,n}(x)=-{\mathrm{i}}\sqrt{\frac{\gamma}{2v}}A_n
        \sum_{j=1}^{N}e^{\mathrm{i}
	k_n{|x-x_{ij}|}},~~i=1,2.
	\label{DarkMode_PhotonW12}
\end{equation}
Here $k_n=\omega_n/v$ is the wave vector of the bound photons. Note that according to the relation $x_{1j}=-x_{2,N-j+1}$, we have $\varphi_{\pm,n}(x)=\pm\varphi_{\pm,n}(-x)$.
Thus, a static bound state of the form
$e^{-\mathrm{i}\omega_n t}|\mathcal{D}_{\pm,n}\rangle$ is produced, with
\begin{equation}
	\left|\mathcal{D}_{\pm,n}\right\rangle=A_n |\pm\rangle
	+\int\mathrm{d}x\varphi_{\pm,n}(x)|x\rangle.
	\label{DarkMode_n}
\end{equation}

The stationary state wave function $\varphi_{\pm,n}(x)$ can be explicitly
written as
\begin{equation}
\varphi_{\pm,n}(x)=\sum_{l=0}^{N-1}\varphi^{(l)}_{\pm,n}(x)\Theta\left(|x|-x_l\right)\Theta\left(x_{l+1}-|x|\right).
\end{equation}
Here we define $x_{l}=\left(l-1/2\right)v\tau$ ($l=1,2,\cdots,N$) and $x_{0}=0$. For parameters
satisfying condition \eqref{DarkConditionGeneral}, by choosing an appropriate global gauge, the wave function $\varphi^{(l)}_{\pm,n}(x)$ ($l=0,1,\cdots,N-1$) within the range $x_l<|x|<x_{l+1}$ can be expressed as the following real functions:
\begin{subequations}
\begin{equation}
	\varphi^{(l)}_{+,n}(x)=\mathcal{L}_{ln}\sin\left(k_{n}\left|x\right|-\theta_{ln}\right),
	~\text{for ${\text{S}}_{+}$,${\text{B}}_{+}$},
	\label{PhotonicWFexpliSPBP}
\end{equation}	
\begin{equation}
	\varphi^{(l)}_{-,n}(x)=-\text{sgn}(x)\mathcal{L}_{ln}\sin\left(k_{n}\left|x\right|-\theta_{ln}\right),
	~\text{for ${\text{S}}_{-}$},
	\label{PhotonicWFexpliSM}
\end{equation}		
\begin{equation}
	\varphi^{(l)}_{-,n}(x)=\text{sgn}(x)\mathcal{L}_{ln}'\cos
	\left(k_{n}\left|x\right|-\theta'_{ln}\right),
	~\text{for ${\text{B}}_{-}$}.
	\label{PhotonicWFexpliBM}
\end{equation}		
\end{subequations}
Here, the amplitudes are defined as  
\begin{subequations}
\begin{equation}
	\mathcal{L}_{ln}=(-1)^n\sqrt{\frac{\gamma }{v}}\frac{\sin\left(\theta_{ln}\right)}{\sin\left(\frac{1}{2}\omega_{n}\tau\right)}A_{n},
\end{equation}
\begin{equation}	
	\mathcal{L}'_{ln}=(-1)^n\sqrt{\frac{\gamma }{v}}\frac{\sin\left(\theta_{ln}'\right)}{\cos\left(\frac{1}{2}\omega_{n}\tau\right)}A_{n},
\end{equation}	
\end{subequations}
with $\theta_{ln}=\left(l+N\right)\omega_{n}\tau/2$ and $\theta_{ln}'=\left(l+N\right)(\omega_{n}\tau+\pi)/2$.
In addition, for an ${\text{S}}_{-}$ mode with $\omega_n\tau=\Omega\tau=m\pi$ (${n}/{N}=m\in\mathbb{E}$), we have
\begin{equation}
	\varphi^{(l)}_{-,n}(x)=\text{sgn}(x)\tilde{\mathcal{L}}_{ln}\sin\left(k_n\left|x\right|\right),
	\label{PhotonicWFexpliSMSpecial}
\end{equation}
with
\begin{equation}
	\tilde{\mathcal{L}}_{ln}=(-1)^{\frac{m}{2}}\sqrt{\frac{\gamma}{v}}\left(N-l\right)A_{n}.
\end{equation}

The field intensity in the long-time limit reads $|\varphi_{\pm,n}(x)|^2$. Furthermore, for wave functions of either form given in Eqs.~\eqref{PhotonicWFexpliSPBP}-\eqref{PhotonicWFexpliBM} and Eq.~\eqref{PhotonicWFexpliSMSpecial}, the 
excitation probability of bound photons admits the universal expression: 
\begin{equation}
	I_{n}=\int\mathrm{d}x\left|\varphi_{\pm,n}\left(x\right)\right|^2
	=\left[1+\frac{\sin(\omega_n\tau)}{2\omega_n\tau}\right]\tilde{I}_{n},
\end{equation}
with 
\begin{equation}
\tilde I_{n}=A_{n}\left(1-A_{n}\right).
\label{tildeI_n}
\end{equation}
When the condition $|\sin(\omega_n\tau)|\ll\omega_{n}\tau$ is satisfied,  $I_{n}\simeq\tilde{I}_{n}$ holds, and this 
approximation is valid for most parameters employed in this paper. It is easy to see that when $A_n=1/2$, the excitation of photons reaches its maximum value $I_n\simeq1/4$. The parameter points where the photon excitation rate reaches this value can be derived as follows: (i) For wave functions specified in Eqs.~\eqref{PhotonicWFexpliSPBP}-\eqref{PhotonicWFexpliBM}, 
the parameters are given by $\Omega\tau=\omega_n\tau-(1/2)\sin(\omega_q\tau)$ and $\gamma\tau=({2}/{N})\sin^2\left(\omega_q\tau/2\right)$ [$q$ is connected to $n$ through Eq.~\eqref{nANDq}];
(ii) For wave functions specified in Eq.~\eqref{PhotonicWFexpliSMSpecial}, we have $\Omega\tau=2m\pi$ and $\gamma\tau={6}/{[N(2 N^2 + 1)]}$.

\begin{figure*}[t]
	\centering
	\includegraphics[width=0.9\textwidth]{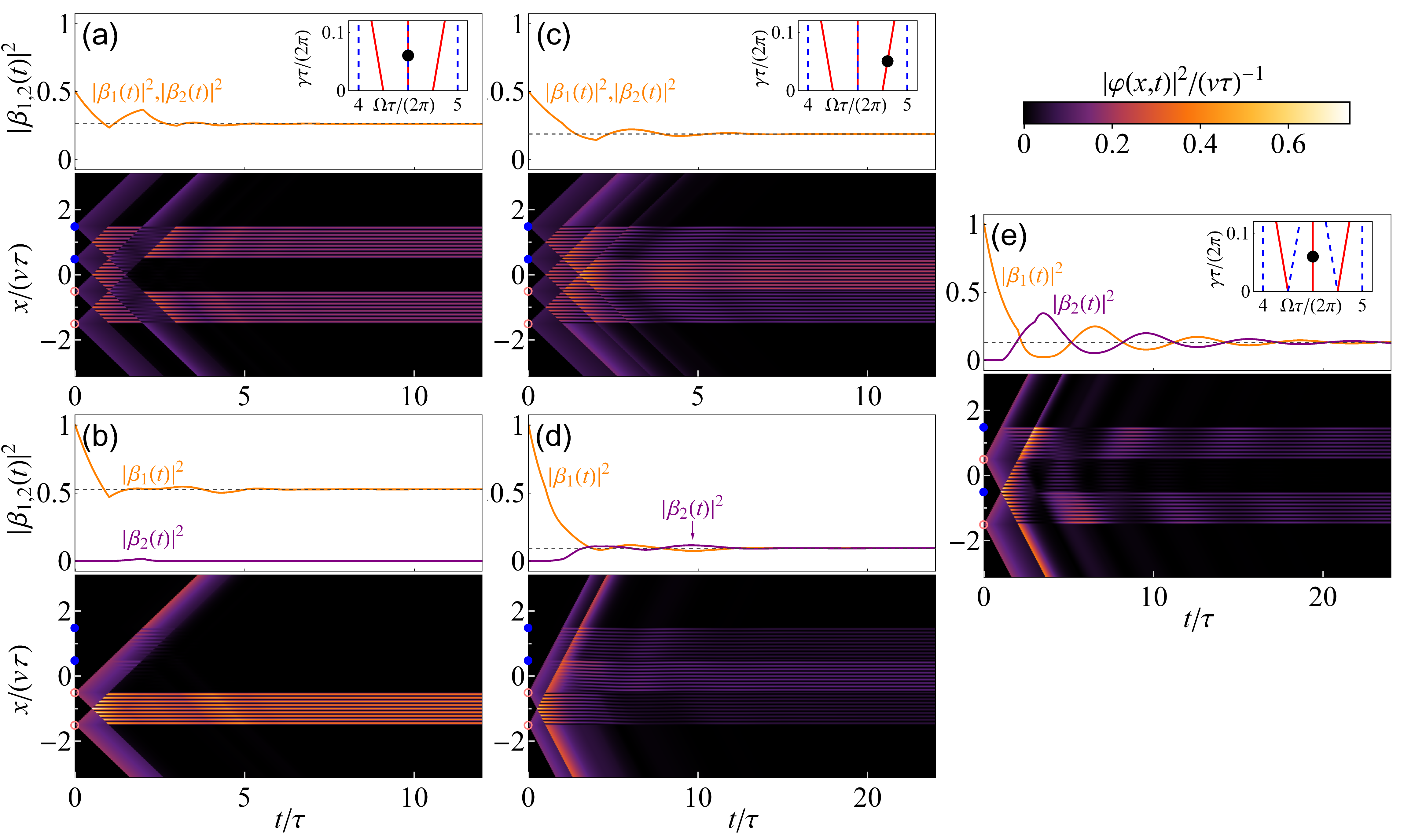}
	\caption{Static bound states for two-giant-atom systems with $N=2$. In each subfigure, the upper panel displays the time evolution of 
	the atomic excitation probability, while the lower panel shows the corresponding time evolution of the 
	field intensity in the waveguide.  
	(a) and (b) Separate configuration 
	with parameters $\Omega\tau/(2\pi)=4.5$ and $\gamma\tau/(2\pi)=0.06$ [see the inset of panel (a)],
	(c) and (d) Separate configuration with parameters $\Omega\tau/(2\pi)=4.8$ and $\gamma\tau/(2\pi)=0.05$ [see 
	the inset of panel (c)]. (e) Braided configuration with parameters $\Omega\tau/(2\pi)=4.5$ and 
	$\gamma\tau/(2\pi)=0.06$ [see the inset of  panel (e)]. 
	The initial states are $|+\rangle$ in panels (a) and (c), and $|\text{eg}\rangle$ in panels (b), (d), and (e).
	The gray dashed line in each panel
	indicates the steady-state value of the atomic excitation probability from analytical result. 
	The pink circles (blue disks) in each panel are used to label the positions of the connection points of the atom 1 (2).
	}
	\label{StaticBS}
\end{figure*}
Based on the above results, when the system allows multiple dark modes and its initial state is prepared
as $c_{1}|\text{eg}\rangle+c_{2}|\text{ge}\rangle=c_{+}|+\rangle+c_{-}|-\rangle$, the corresponding bound states
can be readily obtained via the superposition principle. The static bound states of interest in this section emerge 
at two types of parameter points: (i) those featuring a pair of degenerate symmetric and antisymmetric dark 
modes (only appear in separate configurations), which support two-atom bound states formed by the simple superposition of 
two independent single-atom bound states; and (ii) those with only a single symmetric or antisymmetric dark mode, 
which sustain \textit{genuine} two-atom bound states characterized by interatomic bound photons. 
We will elaborate on this in detail in the following discussion.

For the separate configurations with $n\in\mathbb{E}^{+}$ and
${n}/{N}\notin\mathbb{E}$, the corresponding parameter points are on the overlapping solid and dashed lines in Figs.~\ref{DarkStateCondition}(a) and \ref{DarkStateCondition}(b). 
According to Eqs.~\eqref{DarkFrequency}, \eqref{AmplitudeDarkGeneral}, and \eqref{nANDq}, a pair of symmetric and antisymmetric 
dark modes (i.e., ${\text{S}}_{\pm}$ modes) with the same frequency $\omega_{n}$ and amplitude $A_n$ are allowed. 
This holds true except at the intersection points with other $\Omega\text{-}\gamma$
lines, where additional dark modes with distinct frequencies emerge.
The degeneracy of these two dark modes implies 
the fulfillment of $\Sigma_{+}(-\mathrm{i}\omega_n)=\Sigma_{-}(-\mathrm{i}\omega_n)$ and 
thereby $\Sigma_{\mathrm{\Rmnum{2}}}(-\mathrm{i}\omega_n)=0$ [see Eqs.~\eqref{TranscendentalEQ} and \eqref{SigmaPM}],
signifying an absence of photon-mediated interactions between atoms. 
Unsurprisingly, the dark-state condition \eqref{DarkConditionGeneral} exactly reduces to that of a single giant atom, as 
given in Ref.~\cite{Guo-PRR2020}. When initialized in an arbitrary superposition state 
$c_{1}|\text{eg}\rangle+c_{2}|\text{ge}\rangle=c_{+}|+\rangle+c_{-}|-\rangle$, 
the system evolves into a static bound state of the form 
$e^{-\mathrm{i}\omega_n t}(c_{+}|\mathcal{D}_{+,n}\rangle+c_{-}|\mathcal{D}_{-,n}\rangle)$.
In this state, the excitation probability of the atom is given by $\left|c_{i}\right|^2 A^2_{n}$ ($i=1,2$). 
The photonic wave functions $\varphi_{1,n}(x)$ and $\varphi_{2,n}(x)$ are each localized within their distinct coupling regions and 
exhibit no spatial overlap [with $\varphi_{1,n}(x)=\sqrt{2}\varphi_{+,n}(x)\Theta(-x)$ and 
$\varphi_{2,n}(x)=\sqrt{2}\varphi_{+,n}(x)\Theta(x)$]. 
Therefore, the expression for the field intensity does not contain interference terms, with 
$|c_{1}|^2|\varphi_{1,n}(x)|^2+|c_{2}|^2|\varphi_{2,n}(x)|^2$. The corresponding excitation probability of the 
field confined between the coupling points is a constant $I_n$.
These results indicate that a two-atom bound state for this case is an incoherent superposition of two single-atom bound states. 
A specific example of these parameter points is shown in the inset of Fig.~\ref{StaticBS}(a).
When the system is initialized in the symmetric state $|\text{+}\rangle$, the two atoms 
undergo synchronous evolution due to parity conservation. In the long-time limit, the system evolves into two 
identical single-atom bound states, characterized by equal atomic 
excitation probability $A_n^2$/2 and localized photon distribution around each atom [see Fig.~\ref{StaticBS}(a)]. In contrast, when initialized in the state $|\text{eg}\rangle$, photon trapping occurs around the initially excited atom, 
resulting in a single-atom bound state with atomic 
excitation probability $A_n^2$, while the other atom remains persistently in its ground state
[see Fig.~\ref{StaticBS}(b)].

Now, we consider the scenario where the system of two separate giant atoms supports only a single 
symmetric (anti-symmetric) dark mode with frequency $\omega_n$. 
As shown in Figs.~\ref{DarkStateCondition}(a) and \ref{DarkStateCondition}(b), the corresponding parameter points lie on the solid (dashed) lines that do not overlap with the dashed (solid) lines, characterized by $n\in\mathbb{O}^{+}$ ($n/N\in\mathbb{E}^{+}$), while excluding any intersection points with other lines.
When the system is initialized in the state $c_{1}|\text{eg}\rangle+c_{2}|\text{ge}\rangle=c_{+}|+\rangle+c_{-}|-\rangle$
and satisfies the condition for the symmetric (antisymmetric) dark mode,
the system in the long-time limit will evolve into a static bound state 
$e^{-\mathrm{i}\omega_n t}c_{+}|\mathcal{D}_{+,n}\rangle$ ($e^{-\mathrm{i}\omega_n t}c_{-}|\mathcal{D}_{-,n}\rangle$). 
Under this state, both the atoms have the same excitation probability 
$\left|c_{+}\right|^2 A^2_{n}/2$ ($\left|c_{-}\right|^2 A^2_{n}/2$). The field intensity takes the form $|c_{+}|^2|\varphi_{+,n}(x)|^2$
[$|c_{-}|^2|\varphi_{-,n}(x)|^2$], which is symmetric about the origin. Correspondingly, the excitation 
probability of the bound photons is $|c_{+}|^2 I_{n}$ ($|c_{-}|^2 I_{n}$). 
As specific examples, at the parameter point indicated in the inset of Fig.~\ref{StaticBS}(c), a symmetric dark steady state is achieved 
in the long-time limit regardless of whether the system is initialized in state $|+\rangle$ or $|\text{eg}\rangle$ [see Figs.~\ref{StaticBS}(c) 
and \ref{StaticBS}(d)]. In these states, the trapped photons are mainly distributed in the interatomic region, thereby mediating 
effective interatomic interactions. Thus, the corresponding energy correction function $\Sigma_{\mathrm{\Rmnum{2}}}(-\mathrm{i}\omega_n)$ becomes nonzero, characterizing \textit{genuine} two-giant-atom bound state.

For the braided configuration, the parameters enabling dark modes are indicated by solid lines (symmetric modes $\text{B}_{+}$, with 
$q=\mathrm{mod}[n,2N]$) and dashed lines (antisymmetric modes $\text{B}_{-}$, with 
$q=\mathrm{mod}[n+N,2N]$) in Figs.~\ref{DarkStateCondition}(c) 
and \ref{DarkStateCondition}(d). At these parameter points, like the cases shown in Figs.~\ref{StaticBS}(c) 
and \ref{StaticBS}(d), only the symmetric or antisymmetric dark mode survives after sufficiently long evolution, giving rise to identical atomic excitations and symmetric photonic distributions [see Fig.~\ref{StaticBS}(e), where the case of an initial state $|\text{eg}\rangle$ is taken as an example]. Furthermore, in this configuration, the region between any two neighboring coupling points is interatomic. Thus, all bound states exhibit the characteristics of \textit{genuine} two-atom bound states, where trapped photons mediate interatomic interactions [with $\Sigma_{\mathrm{\Rmnum{2}}}(-\mathrm{i}\omega_n)\neq 0$].
\section{\label{OscillatingBS}Oscillating bound states}
\subsection{\label{SpecialParameterOBS}Parameter points for oscillating bound states}
For an intersection 
point of (at least) two $\Omega\text{-}\gamma$ lines corresponding to static bound states with frequencies
$\omega_{n_1}$ and $\omega_{n_2}$ (assuming $n_1<n_2$),
the system permits coexistence of non-degenerate dark states, leading to the formation of oscillating bound states in the long-time limit. Notably, for a single giant atom, oscillating bound states emerge only when the number of coupling points is three or more ($N\geq 3$)~\cite{Guo-PRR2020}. In contrast, for the case of two giant atoms, 
such oscillating bound states emerge even in the minimal configuration with $N=2$, as the nonlinear cotangent 
term in condition \eqref{DarkConditionGeneral} remains effective for $\text{S}_{+}$ and $\text{B}_{\pm}$ 
modes [as illustrated in Figs.~\ref{DarkStateCondition}(a) and 
\ref{DarkStateCondition}(c)]. The key reason is that two identical giant atoms, even with only
two coupling points each, can generate an effective cavity QED structure, enabling the observation of Rabi-oscillation 
like phenomena.
 
A general expression describing these parameter points can be found in Appendix~\ref{ParameterOBSGeneral}.
Here, we focus on the case where a pair of $\Omega\text{-}\gamma$ lines have opposite slopes, making them symmetric 
about the vertical line through their intersection point.
In this case, the conditions 
\begin{equation}
q_1=N+\tilde{q},~q_2=N-\tilde{q}
\label{tilde_q}
\end{equation}
[i.e., $\bar{q}=(q_1+q_2)/{2}=N$, $q_1$ and $q_2$ are connected to $n_1$ and $n_2$ through Eq.~\eqref{nANDq}] are satisfied, ensuring that $K_{q_1}=-K_{q_2}$ [see Eq.~\eqref{Slope}]. 
Correspondingly, the coordinates of the intersection point can be expressed as 
\begin{equation}
	\Omega=\bar{\omega},
	~~\gamma=\frac{2\tilde{\omega}}{N}\cot\left(\frac{1}{2}\omega_{\tilde{q}}\tau\right),
	\label{SymmetricParameterPoints}
\end{equation}
with 
\begin{equation}
\bar{\omega}=\frac{m\pi}{\tau},~~\tilde{\omega}=\frac{p\pi}{\tau}-\omega_{\tilde{q}},~~\omega_{\tilde{q}}=\frac{\tilde{q}\pi}{N\tau}.
\label{omegaparameter}
\end{equation}
At these parameter points, the system permits two dark modes, characterized by the indices 
\begin{equation}
   n_{1,2}=(m\mp p)N\pm\tilde{q} 
\end{equation}
and the corresponding frequencies 
\begin{equation}
\omega_{n_{1,2}}=\bar{\omega}\mp\tilde{\omega}=(m\mp p)\frac{\pi}{\tau}\pm\omega_{\tilde{q}}.
\label{Frequency_n12}
\end{equation}

After some analysis (see Appendix \ref{ParameterOBSGeneral}), one can determine both the admissible values of $m$ and, for each 
specified $m$, the corresponding values of $p$. And the values of $\tilde{q}$ can be
fixed by Eqs.~\eqref{nANDq} and \eqref{tilde_q}. The relevant results under various conditions are summarized as follows:

(i) If both  $n_1$ and $n_2$ ($n_1<n_2$) represent modes belonging to the same category among $\text{S}_{\pm}$ and $\text{B}_{+}$, we have   
\begin{eqnarray}
&&m=2,3,\cdots;~p=m-1,m-3,\cdots~(p>0); 
\nonumber
\\
&&\tilde{q}=\left\{ 
\begin{array}{l}
	1,2,\cdots,N-1~\text{for ${\text{S}}_{+}$ and ${\text{B}}_{+}$},
	\\
	1,3,\cdots,N-2~(N\in\mathbb{O}^{+})~\text{for ${\text{S}}_{-}$},
	\\
	2,4,\cdots,N-2~(N\in\mathbb{E}^{+})~\text{for ${\text{S}}_{-}$}.
\end{array}\right.
\label{mpq1}
\end{eqnarray}

(ii) For the braided configuration, if both $n_1$ and $n_2$ ($n_1<n_2$) represent antisymmetric dark modes  ($\text{B}_{-}$ modes), we have
\begin{eqnarray}
&&m=1,2,\cdots;~p=m,m-2,\cdots~(p>0); 
\nonumber
\\
&&\tilde{q}=1,2,\cdots,N-1.
\label{mpq2}
\end{eqnarray}

(iii) For the braided configuration, if $n_1$ and $n_2$ ($n_1<n_2$) correspond to a $\text{B}_{+}~\left(\text{B}_{-}\right)$ and a $\text{B}_{-}~\left(\text{B}_{+}\right)$ modes, respectively, we have
\begin{eqnarray}
&&m=\tfrac{1}{2},\tfrac{3}{2},\cdots;~p=\tfrac{1}{2},\tfrac{3}{2},\cdots,m;
\nonumber
\\
&&\tilde{q}=\left\{ 
\begin{array}{l}
	\tilde{q}=1,2,\cdots,N-1~\text{for $p\ge 1$},
	\\
	\tilde{q}=1,2,\cdots,\lfloor N/2\rfloor~\text{for $p=1/2$}.
\end{array} \right.
\label{mpq3} 
\end{eqnarray}

The parameters summarized above can be used to 
fix all possible intersection points of \textit{at least} a 
pair of $\Omega\text{-}\gamma$ lines with opposite slopes (i.e., coexisting non-degenerate dark modes with equal amplitudes). Notably, in many situations, such intersection points 
may support more than two dark modes, as other lines could also pass through these intersection
points.

In our system, the Markovian condition is 
\begin{equation}
\gamma\tau\ll N^{-3},
\label{MarkvianCondition}
\end{equation}
which means that the time delay for photons traveling from the leftmost coupling point to the rightmost one [$(2N-1)\tau\sim N\tau$] is much shorter than the dissipation timescale of the atoms [$1/(N^2\gamma)$].
On the other hand, it can be found that parameter points for oscillating bound states, specified by Eq.~\eqref{SymmetricParameterPoints}, 
satisfy $\gamma\tau\gtrsim N^{-3}$. Thus we can conclude that these states always appear in the non-Markovian regime.
\subsection{\label{GeneralDescriptionOBS}General description of oscillating bound states}
\begin{table*}[t]
	\renewcommand{\arraystretch}{1.25}
	\centering\caption{Classification of oscillating bound state (OBS) for two separate giant atoms. 
	S1: synchronous single-frequency ($\tilde\Omega$) OBS; 
	S2: synchronous double-frequency ($\tilde\Omega$ and $\tilde\Omega/2$) OBS; 
	H2-0: hybrid-type OBS with double-frequency ($\tilde\Omega$ and $\tilde\Omega/2$) oscillation in atom 1 and static population in atom 2; 
	H2-1: hybrid-type OBS with double-frequency oscillation ($\tilde\Omega$ and $\tilde\Omega/2$) in atom 1 and single-frequency oscillation ($\tilde\Omega$) in atom 2; 
	E2: Exchang-type OBS with double-frequency ($\tilde\Omega$ and $\tilde\Omega/2$) oscillation in each atom.}
	\label{tablesep}\begin{ruledtabular}
		\begin{tabular}{cccccc}
			&Initial state&$m,p$&$N,\tilde{q}$&Type of OBS
			\\
			\hline
			&\multirow{2}{*}{$|+\rangle$}
			&$m\in\mathbb{E^+};~p=1,3,\cdots,m-1$
			&$N\in\mathbb{Z^+};~\tilde{q}=1,2,\cdots,N-1$
			&S1
			\\
			&
			&$m\in\mathbb{O^+};~p=2,4,\cdots,m-1$
			&$N\in\mathbb{Z^+};~\tilde{q}=1,2,\cdots,N-1$
			&S2
			\\
			\cline{2-5}
			&\multirow{4}{*}{$|-\rangle$}
			&\multirow{2}{*}{$m\in\mathbb{E^+};~p=1,3,\cdots,m-1$}
			&$N\in\mathbb{O^+};~\tilde{q}=1,3,\cdots,N-2$
			&S2
			\\
			&
			&
			&$N\in\mathbb{E^+};~\tilde{q}=2,4,\cdots,N-2$
			&S2
			\\
			\cline{4-5}
			&
			&\multirow{2}{*}{$m\in\mathbb{O^+};~p=2,4,\cdots,m-1$}
			&$N\in\mathbb{O^+};~\tilde{q}=1,3,\cdots,N-2$
			&S1
			\\
			&
			&
			&$N\in\mathbb{E^+};~\tilde{q}=2,4,\cdots,N-2$
			&S2
			\\
			\cline{2-5}
			&\multirow{8}{*}{$|\text{eg}\rangle$}
			&\multirow{4}{*}{$m\in\mathbb{E^+};~p=1,3,\cdots,m-1$}
			&$N\in\mathbb{O^+};~\tilde{q}=1,3,\cdots,N-2$
			&H2-0
			\\
			&
			&
			&$N\in\mathbb{O^+};~\tilde{q}=2,4,\cdots,N-1$
			&E2
			\\
			&
			&
			&$N\in\mathbb{E^+};~\tilde{q}=1,3,\cdots,N-1$
			&E2
			\\
			&
			&
			&$N\in\mathbb{E^+};~\tilde{q}=2,4,\cdots,N-2$	
			&H2-0
			\\
			\cline{4-5}
			&
			&\multirow{4}{*}{$m\in\mathbb{O^+};~p=2,4,\cdots,m-1$}
			&$N\in\mathbb{O^+};~\tilde{q}=1,3,\cdots,N-2$
			&H2-0
			\\
			&
			&
			&$N\in\mathbb{O^+};~\tilde{q}=2,4,\cdots,N-1$
			&S2
			\\
			&
			&
			&$N\in\mathbb{E^+};~\tilde{q}=1,3,\cdots,N-1$
			&H2-1
			\\
			&
			&
			&$N\in\mathbb{E^+};~\tilde{q}=2,4,\cdots,N-2$
			&Single-atom
		\end{tabular}
	\end{ruledtabular}
	\label{OBSforSep}
\end{table*}
\begin{table*}[t]
	\renewcommand{\arraystretch}{1.25}
	\centering\caption{Classification of oscillating bound states (OBS) for two braided giant atoms. 
	S1: synchronous single-frequency ($\tilde\Omega$) OBS; 
	S2: synchronous double-frequency ($\tilde\Omega$ and $\tilde\Omega/2$) OBS; 
	E1: Exchang-type OBS with single-frequency ($\tilde\Omega$) oscillation in each atom; 
	E2: Exchang-type OBS with double-frequency ($\tilde\Omega$ and $\tilde\Omega/2$) oscillation in each atom.}
	\label{tablebra}
	\begin{ruledtabular}
		\begin{tabular}{ccccc}
			&Initial state&$m,p$&Type of OBS
			\\
			\hline
			&\multirow{2}{*}{$|+\rangle$}
			&$m\in\mathbb{E^+};~p=1,3,\cdots,m-1$
			&S1
			\\
			&
			&$m\in\mathbb{O^+};~p=2,4,\cdots,m-1$
			&S2
			\\
			\cline{2-4}
			&\multirow{2}{*}{$|-\rangle$}
			&$m\in\mathbb{O^+};~p=1,3,\cdots,m$
			&S1
			\\
			&
			&$m\in\mathbb{E^+};~p=2,4,\cdots,m$
			&S2
			\\
			\cline{2-4}
			&\multirow{3}{*}{$|\text{eg}\rangle$}
			&$m\in\mathbb{Z^+};~p=1,3,\cdots~(p\le m)$
			&E2
			\\
			&
			&$m\in\mathbb{Z^+};~p=2,4,\cdots~(p\le m)$
			&S2
			\\
			&
			&$m=\mathbb{O^+}/2,\cdots;~p=1/2,3/2,\cdots,m$
			&E1
			\\
		\end{tabular}
	\end{ruledtabular}
	\label{OBSforBra}
\end{table*}
At a parameter point simultaneously support multiple dark modes, the
system with initial state $c_{1}|\text{eg}\rangle+c_{2}|\text{ge}\rangle=c_{+}|+\rangle+c_{-}|-\rangle$
can finally evolve into the following oscillating bound state:
\begin{eqnarray}
\left|\Psi_{\text{OBS}}(t)\right\rangle&=&\sum_{\lambda=\pm}c_{\lambda}\sum_{n\in\mathcal{N}_{\lambda}}e^{-\mathrm{i}\omega_n t}\left|\mathcal{D}_{\lambda,n}\right\rangle
\nonumber
\\
&\equiv&\left|\Psi^{(\text{a})}_{\text{OBS}}(t)\right\rangle+\left|\Psi^{(\text{p})}_{\text{OBS}}(t)\right\rangle,
\label{StateOSB}
\end{eqnarray} 
where $\mathcal{N}_{\pm}$ is the set of indices $n$ labelling the symmetric (antisymmetric) dark modes.
The states of the atomic and photonic subsystems take the form
\begin{subequations}
\begin{equation}
\left|\Psi^{(\text{a})}_{\text{OBS}}(t)\right\rangle=\sum_{\lambda=\pm}c_{\lambda}\sum_{n\in\mathcal{N}_{\lambda}}A_n 
e^{-\mathrm{i}\omega_n t}|\lambda\rangle,
\label{StateOBSatom}
\end{equation} 
\begin{equation}
\left|\Psi^{(\text{p})}_{\text{OBS}}(t)\right\rangle
=\int\mathrm{d}x\sum_{\lambda=\pm}c_{\lambda}\sum_{n\in\mathcal{N}_{\lambda}}\varphi_{\lambda,n}(x)e^{-\mathrm{i}\omega_n t}|x\rangle.
\label{StateOBSphoton}
\end{equation} 
\end{subequations}

According to Eq.~\eqref{StateOBSatom}, we can obtain the  
excitation probability of each atom in the long-time limit:
\begin{eqnarray}
	\left|\beta_{1,2}(t)\right|^2
	&\to&\frac{1}{2}\sum_{\lambda=\pm}\left|c_{\lambda}\right|^2
	\sum_{n,n'\in\mathcal{N_{\lambda}}}A_n A_{n'}\cos\left(\omega_{nn'}t\right)
	\nonumber
	\\
	&&\pm\eta\sum_{\substack{n\in\mathcal{N}_{+}}}\sum_{\substack{n'\in\mathcal{N}_{-}}}
	{A_n A_{n'}}\cos\left(\omega_{nn'}t+\phi\right),
	\label{Beta12OSBGeneral}
\end{eqnarray}
with $\omega_{nn'}=\omega_{n}-\omega_{n'}$, $\eta=|c^{*}_{+}c_{-}|$, and $\phi=\text{Arg}(c^{*}_{+}c_{-})$.
When $n\neq n'$, the quantity $|\omega_{nn'}|$ denotes the frequency of the component that maintains
persistent oscillation. The total excitation probability of the two atoms reads
\begin{eqnarray}
	I_{\mathrm{a}}(t)&=&\left|\beta_{1}(t)\right|^2+\left|\beta_{2}(t)\right|^2
	\nonumber
	\\
	&\to&\sum_{\lambda=\pm}\left|c_{\lambda}\right|^2
	\sum_{n,n'\in\mathcal{N_{\lambda}}}A_n A_{n'}\cos\left(\omega_{nn'}t\right).
	\label{AtomExcitationOSBGeneral}
\end{eqnarray}
According to Eq.~\eqref{StateOBSphoton}, the field intensity related to the bound photons takes the form 
\begin{eqnarray}
         |\varphi(x,t)|^2&\to&\left|\left\langle x \Big|\Psi^{(\text{p})}_{\text{OBS}}(t)\right\rangle\right|^2
         \nonumber
         \\
	&=&\sum_{\lambda=\pm}\left|c_{\lambda}\right|^2
	\sum_{n,n'\in\mathcal{N_{\lambda}}}\varphi_{\lambda,n}(x) \varphi_{\lambda,n'}(x) \cos\left(\omega_{nn'}t\right)
	\nonumber
	\\
	&&+\eta\sum_{\substack{n\in\mathcal{N}_{+}}}\sum_{\substack{n'\in\mathcal{N}_{-}}}
	{\varphi_{+,n}(x) \varphi_{-,n'}(x)}
	\nonumber
	\\
	&&\times\cos\left(\omega_{nn'}t+\phi\right).
	\label{FieldIntensityOSBGeneral}
\end{eqnarray}
Above, we have utilized the fact that $\varphi_{\pm,n}(x)$ are real functions.
The total excitation probability of bound photons can be further derived as
\begin{eqnarray}
	I_{\mathrm{p}}(t)&=&\int\left|\varphi(x,t)\right|^2\mathrm{d}x
	\nonumber
	\\
	&\to&\sum_{\lambda=\pm}\left|c_{\lambda}\right|^2\Bigg[\sum_{n\in\mathcal{N}_{\lambda}}I_{n}-
	\sum_{\substack{n,n'\in\mathcal{N}_{\lambda} \\ n\neq n'}}A_nA_{n'}\cos\left(\omega_{nn'}t\right)\Bigg].
	\nonumber
	\\
	\label{PhotonExcitationOSBGeneral}
\end{eqnarray}
In deriving the above equation, we have used the following relation
\begin{equation}
	\int\varphi_{\lambda,n}\left(x\right)\varphi_{\lambda',n'}\left(x\right)\mathrm{d}x=
	[I_n\delta_{nn'}-(1-\delta_{nn'})A_nA_{n'}]\delta_{\lambda\lambda'},
\end{equation}
which holds under the RWA.
Thus, the total excitation probability of the atoms and the photons remains time-independent:
\begin{equation}
	I_{\mathrm{t}}(t)=I_{\mathrm{a}}(t)+I_{\mathrm{p}}(t)\to\sum_{\lambda=\pm}\left|c_{\lambda}\right|^2
	\sum_{n\in\mathcal{N_{\lambda}}}\left(A^2_{n}+I_{n}\right),
	\label{TotalExcitationGeneral}
\end{equation}
owing to the fact that the oscillating bound state does not decay.

The two-atom oscillating bound states can be  
classified into two categories: \textit{synchronous} and \textit{asynchronous} oscillating bound states.
A synchronous oscillating bound state contains only 
symmetric components $|\mathcal{D}_{+,n}\rangle$ or antisymmetric components $|\mathcal{D}_{-,n}\rangle$.
According to Eq.~\eqref{StateOSB}, this type of states emerge under the following two conditions: 
(i) The initial state is a symmetric state $|+\rangle$ (antisymmetric state $|-\rangle$) and the parameter point supports 
at least two non-degenerate symmetric (antisymmetric) dark modes. The corresponding long-time 
state evolves as $\sum_{n\in\mathcal{N}_{+}}e^{-\mathrm{i}\omega_n t}|\mathcal{D}_{+,n}\rangle$ 
($\sum_{n\in\mathcal{N}_{-}}e^{-\mathrm{i}\omega_n t}|\mathcal{D}_{-,n}\rangle$).
(ii) The initial state is neither symmetric nor antisymmetric (i.e., $|\Psi(0)\rangle=c_{+}|+\rangle+c_{-}|-\rangle$, with $c_{+}\neq 0$ 
and $c_{-}\neq 0$), and the parameter point supports at least two 
non-degenerate symmetric (antisymmetric) dark modes while excluding those of the opposite symmetry.
The oscillating bound state in the long-time limit then becomes $c_{+}\sum_{n\in\mathcal{N}_{+}}A_n e^{-\mathrm{i}\omega_n t}|\mathcal{D}_{+,n}\rangle$ 
($c_{-}\sum_{n\in\mathcal{N}_{-}}A_n e^{-\mathrm{i}\omega_n t}|\mathcal{D}_{-,n}\rangle$).  
In both cases, the long-time excitation probability dynamics of the two atoms remains identical, and the  
wave functions of the confined photons preserve parity symmetry.
 
In contrast, an asynchronous oscillating bound state contains both 
symmetric components $|\mathcal{D}_{+,n}\rangle$ and antisymmetric components $|\mathcal{D}_{-,n}\rangle$.
The generation of this type of bound state requires that the system is initialized in a superposition state 
$c_{+}|+\rangle+c_{-}|-\rangle$ (with $c_{+}\neq 0$ and $c_{-}\neq 0$), and the parameter point supports at least two non-degenerate dark modes that include both symmetric and antisymmetric types. 
The corresponding oscillating bound state in the long-time limit then takes the form $\sum_{\lambda=\pm}c_{\lambda}\sum_{n\in\mathcal{N}_{\lambda}}e^{-\mathrm{i}\omega_n t}\left|\mathcal{D}_{\lambda,n}\right\rangle$,
giving rise to distinct dynamical behaviors between the two atoms [see Eq.~\eqref{Beta12OSBGeneral}].
Further analysis (see Secs.~\ref{SepOBS} and \ref{BraOBS}) will show 
that this kind of oscillating bound states can be categorized into \textit{hybrid-type} (only appear in the separate configuration) 
and \textit{exchange-type}.

Tables \ref{OBSforSep} and \ref{OBSforBra} summarize the types of oscillating bound states at the parameter points given in Sec.~\ref{SpecialParameterOBS}. The initial state is chosen as a symmetric (antisymmetric) state $|+\rangle$ ($|-\rangle$) or a single-atom excitation state $|\text{eg}\rangle$.
In addition, Secs.~\ref{SepOBS} and \ref{BraOBS} offer specific examples along with in-depth analyses 
of these states.
\subsection{\label{SepOBS}Oscillating bound state in  the separate configuration}
\begin{figure*}[t]
	\centering
	\includegraphics[width=\textwidth]{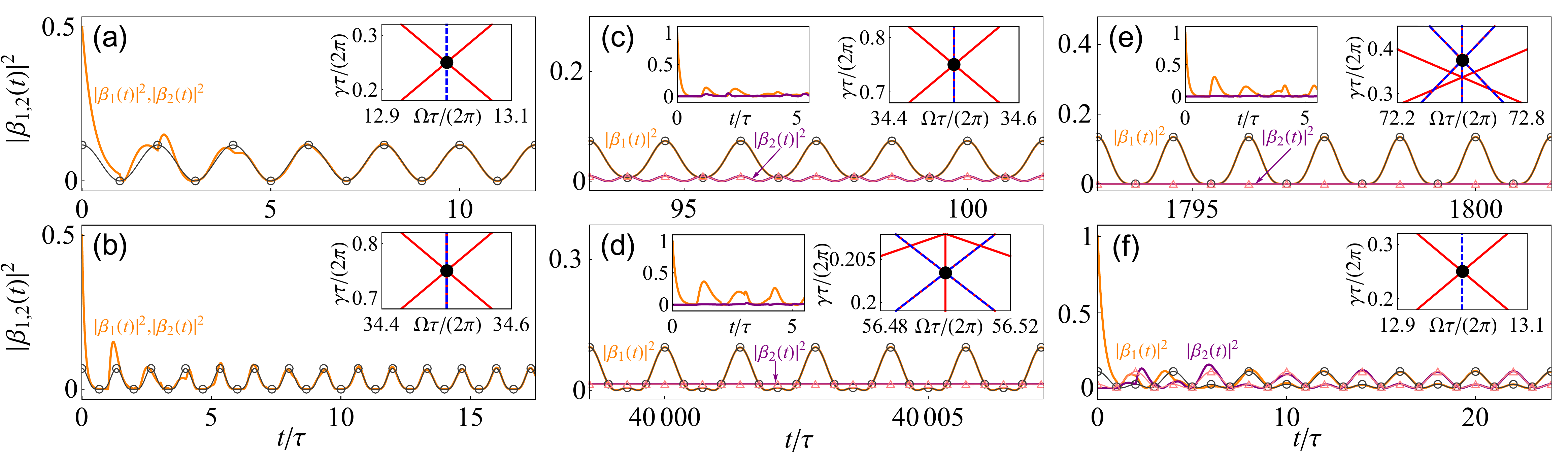}
	\caption{Time evolution of the atomic excitation probabilities of two separate giant atoms
	under parameters for oscillating bound states. 
	The initial states are $|+\rangle$ in panels (a) and (b), and $|\text{eg}\rangle$ in
	panels (c)-(f). Parameters: (a) $N=2$, $m=26$, $p=1$, and $\tilde{q}=1$ (i.e., $n_1=51, n_2=53$), corresponding to 
	the parameter point $\Omega\tau/(2\pi)=13$, $\gamma\tau/(2\pi)=0.25$ [see the inset of panel (a)]. 
	(b) $N=2$, $m=69$, $p=2$, and $\tilde{q}=1$ (i.e., $n_1=135, n_2=141$), corresponding to 
	$\Omega\tau/(2\pi)=34.5$, $\gamma\tau/(2\pi)=0.75$ [see the inset of panel (b)]. 
	(c) The parameters are the same as those used in panel (b).  
	(d) $N=5$, $m=113$, $p=2$, and $\tilde{q}=3$ (i.e., $n_1=558, n_2=572$, corresponding to  
	$\Omega\tau/(2\pi)=56.5$, $\gamma\tau/(2\pi)=0.203$ [see the right inset of panel (d)]. 
	(e) $N=4$, $m=145$, $p=2$, and $\tilde{q}=2$ (i.e., $n_1=574, n_2=586$), corresponding to 
	$\Omega\tau/(2\pi)=72.5$, $\gamma\tau/(2\pi)=0.375$ [see the right inset of 
	panel (e)]. (f) The parameters are the same as those used in panel (a).  In each panel, the gray thin line marked with 
	circles and the pink thin line marked with triangles represent the analytical results under the long-time limit.
	}
	\label{OscillatingBS_SepAmp}
\end{figure*}
\subsubsection{\label{SepSOBS}Atomic dynamics for synchronous oscillating bound state} 
In Figs.~\ref{OscillatingBS_SepAmp}(a) and \ref{OscillatingBS_SepAmp}(b), we present two distinct types of synchronous oscillating bound states for two separate giant atoms.  

The parameter point shown in the inset of Fig.~\ref{OscillatingBS_SepAmp}(a) satisfies $m\in\mathbb{E}^{+}$ and 
$p=1,3,\cdots,m-1$. Thus, the system with the initial state $|+\rangle$ supports two symmetric dark modes with frequencies 
$\omega_{n_1}=\bar\omega-\tilde\omega$ and $\omega_{n_2}=\bar\omega+\tilde\omega$ [see analysis in Sec.~\ref{SpecialParameterOBS}, corresponding to the solid lines in the inset of Fig.~\ref{OscillatingBS_SepAmp}(a)]. Both modes exhibit
equal excitation amplitudes for the state $|+\rangle$ [cf. Eqs.~\eqref{AmplitudeDarkGeneral}, \eqref{tilde_q}, and \eqref{SymmetricParameterPoints}]:
\begin{equation}
	A_{n_1}=A_{n_2}=\left[{1+2\tilde{\omega}\tau\csc\left(\omega_{\tilde{q}}\tau\right)}\right]^{-1}\equiv 
	{A}.
	\label{AmplitudeOSB1}
\end{equation}
Thus, according to Eq.~\eqref{Beta12OSBGeneral}, we obtain identical atomic excitation probabilities
$\left|\beta_1\left(t\right)\right|^2=\left|\beta_2\left(t\right)\right|^2$ in the long-time limit, as given by 
\begin{equation}
	{A^2}\left[1+\cos\left(\tilde{\Omega} t\right)\right],
	\label{SepS2}
\end{equation}
with 
\begin{equation}
\tilde\Omega=\omega_{n_2}-\omega_{n_1}=2\tilde\omega.
\end{equation}
The atomic excitation probabilities of this kind of synchronous single-frequency oscillating bound state (S1-type, as 
summarized in Table.~\ref{tablesep}) are displayed in Fig.~\ref{OscillatingBS_SepAmp}(a). 

At the parameter point with $m\in\mathbb{O}^{+}$ and $p=2,4,\cdots,m-1$
[see the inset of Fig.~\ref{OscillatingBS_SepAmp}(b)], for the initial state 
$|+\rangle$, the system supports two symmetric dark modes with frequencies 
$\omega_{n_1,n_2}=\bar\omega\mp\tilde\omega$ [see analysis in \ref{SpecialParameterOBS}, 
corresponding to the pair of solid lines with opposite slopes in the inset of 
Fig.~\ref{OscillatingBS_SepAmp}(b)], exhibiting identical excitation amplitudes 
$A_{n_1}=A_{n_2}= A$. Moreover, at this parameter with $\Omega\tau=m\pi$ ($m\in\mathbb{O}^{+}$), 
the system sustains an additional symmetric dark mode 
with frequency $\omega_{\bar{n}}=\bar\omega=m\pi/\tau$ 
[$\bar{n}=\left(n_1+n_2\right)/2=mN$, corresponding to the vertical solid line in the inset of Fig.~\ref{OscillatingBS_SepAmp}(b)]
and amplitude [cf. Eqs.~\eqref{A_N} and \eqref{SymmetricParameterPoints}]:
\begin{equation}
A_{\bar{n}}=\left[{1+\tilde{\omega}\tau\cot\left(\tfrac{1}{2}\omega_{\tilde{q}}\tau\right)}\right]^{-1}\equiv A'.
\end{equation}
Thus, in the long-time limit, the atomic excitation probabilities $\left|\beta_1(t)\right|^2$ and $\left|\beta_2(t)\right|^2$ exhibit identical temporal evolution, and they can be expressed as [cf. Eq.~\eqref{Beta12OSBGeneral}]:
\begin{equation}
	 B_{0}+B_1\cos\left(\tfrac{1}{2}\tilde{\Omega}t\right)+B_2\cos\left(\tilde\Omega t\right),
	 \label{Probability3f}
\end{equation}
with $B_0=A^2+{A'}^2/2$, $B_1=2AA'$, and $B_2=A^2$. 
The atomic excitation probabilities of such synchronous double-frequency (with $\tilde\Omega$ and $\tilde\Omega/2$) 
oscillating bound state (S2-type, as 
summarized in Table.~\ref{tablesep}) are displayed in Fig.~\ref{OscillatingBS_SepAmp}(b).  
\subsubsection{\label{SepNSOBS-H}Atomic dynamics for asynchronous oscillating bound state \Rmnum{1}: hybrid type} 
The formation of hybrid-type oscillating bound states necessitates the existence of 
degenerate symmetric and antisymmetric dark modes. Some of these parameter points are illustrated in the right
insets of Figs.~\ref{OscillatingBS_SepAmp}(c)-\ref{OscillatingBS_SepAmp}(e), which
are determined by the following parameters:

(i) $m\in\mathbb{O}^{+}; p=2,4,\cdots,m-1$ and $N\in\mathbb{E^+};~\tilde{q}=1,3,\cdots,N-1$.

(ii) $m\in\mathbb{O}^{+}; p=2,4,\cdots,m-1$ and $N\in\mathbb{O^+};~\tilde{q}=1,3,\cdots,N-2$.

(iii) $m\in\mathbb{O}^{+}; p=2,4,\cdots,m-1$ and $N\in\mathbb{E^+};~\tilde{q}=2,4,\cdots,N-2$.

For the initial state $|+\rangle$, each of the three parameter types supports three symmetric dark modes: 
two with frequencies $\omega_{n_1,n_2}=\bar\omega\mp\tilde\omega$ and amplitudes $A_{n_1} = A_{n_2} = A$,
and one with frequency $\omega_{\bar{n}}=\bar{\omega}$ and amplitude $A_{\bar{n}} = A'$ 
[corresponding to the solid lines in the right insets of Figs.~\ref{OscillatingBS_SepAmp}(c)-\ref{OscillatingBS_SepAmp}(e)].
In contrast, for the initial state $|-\rangle$, the allowed antisymmetric dark modes are as follows: 
a single mode at $\omega_{\bar{n}} = \bar{\omega}$ for type (i); two modes at 
$\omega_{n_1,n_2} = \bar{\omega} \mp \tilde{\omega}$ for type (ii); and both the mode at $\omega_{\bar{n}} = \bar{\omega}$ 
and the modes at $\omega_{n_1,n_2} = \bar{\omega} \mp \tilde{\omega}$ for type (iii) 
[corresponding to the dashed lines in the right insets of Figs.~\ref{OscillatingBS_SepAmp}(c)-\ref{OscillatingBS_SepAmp}(e)]. 
Each antisymmetric mode shares the same amplitude as its symmetric counterpart at the same frequency.

According to the above results, for initial state $|\text{eg}\rangle=(|+\rangle+|-\rangle)/\sqrt{2}$,
the excitation amplitude $\beta_{1}(t)=[\beta_{+}(t)+\beta_{-}(t)]/{2}$
still retains three frequency components (at $\bar\omega\pm\tilde\omega$ and $\bar\omega$) for cases (i)-(iii). While $\beta_{2}(t)=[\beta_{+}(t)-\beta_{-}(t)]/{2}$ comprises fewer frequency components due to complete destructive interference between each pair of degenerate symmetric and antisymmetric dark modes.
Consequently, according to Eq.~\eqref{Beta12OSBGeneral}, $|\beta_{1}(t)|^2$ exhibits double-frequency (with $\tilde\Omega$ and $\tilde\Omega/2$) oscillation
described by Eq.~\eqref{Probability3f}, with
\begin{equation}
	 \left\{ 
	 \renewcommand{\arraystretch}{1.5}
\begin{array}{l}
        B_0=\frac{1}{2}A^2+A'^2,~B_1=2A'A,~B_2=\frac{1}{2}A^2
	\\
	B_0=2A^2+\frac{1}{4}A'^2,~B_1=2AA',~B_2=2A^2
	\\
	B_0=2A^2+A'^2,~B_1=4AA',~B_2=2A^2
\end{array}\right.
\end{equation}
for the parameter points of types (i)-(iii). While $|\beta_{2}(t)|^2$ 
takes the form
\begin{equation}
\left|\beta_2\left(t\right)\right|^2\rightarrow \left\{ 
\renewcommand{\arraystretch}{1.5}
\begin{array}{l}
         \frac{1}{2}A^2\left[1+\cos\left(\tilde{\Omega} t\right)\right]
	\\
	\frac{1}{4}A'^2
	\\
	0
\end{array}\right.,
\label{HybridEP2}
\end{equation}
exhibiting single-frequency oscillation, constant steady-state excitation, and complete quenching, respectively. 
Thus, one obtains
three types of hybrid-type oscillating bound states cataloged in Table.~\ref{tablesep}: 
double-frequency oscillation in atom 1 coexisting with (i) single-frequency oscillation in atom 2 (H2-1 type),
(ii) static population in atom 2 (H2-0 type), and (iii) no excitation in atom 2. 
The atomic excitation probabilities of these hybrid-type oscillating bound states 
are displayed in Figs.~\ref{OscillatingBS_SepAmp}(c)-\ref{OscillatingBS_SepAmp}(e).

It should be noted that for the cases illustrated in Figs.~\ref{OscillatingBS_SepAmp}(c)-\ref{OscillatingBS_SepAmp}(e), the 
system requires an evolution time scale much longer than the dissipation time scale to reach the oscillating bound state as expected 
by Eqs.~\eqref{Probability3f} and \eqref{HybridEP2}. Prior to this, there exist additional quasi-dark modes in the system. This 
phenomenon will be discussed in detail in Sec.~\ref{PhenomenonQusiOBS}.
\subsubsection{\label{SepNSOBS-E}Atomic dynamics for asynchronous oscillating bound state \Rmnum{2}: exchange-type}
The formation of exchange-type oscillating bound states necessitates the coexistence of symmetric and antisymmetric dark modes 
with different frequencies. An example of the corresponding parameter points
is illustrated in the inset of Fig.~\ref{OscillatingBS_SepAmp}(f), satisfying $m\in\mathbb{E^+};~p=1,3,\cdots,m-1$ and 
$N\in\mathbb{E^+};~\tilde{q}=1,3,\cdots,N-1$. 
For the initial state $|+\rangle$, this point supports two symmetric dark modes 
[corresponding to the solid lines in the inset of Fig.~\ref{OscillatingBS_SepAmp}(f)] with frequencies 
$\omega_{n_1,n_2}=\bar\omega\mp\tilde\omega$ and amplitudes $A_{n_1} = A_{n_2} = A$.
In contrast, for the initial state $|-\rangle$, only a single antisymmetric mode [corresponding to the dashed line in the inset of 
Fig.~\ref{OscillatingBS_SepAmp}(f)] with frequency $\bar\omega=m\pi/\tau$ ($m\in\mathbb{E}^{+}$)  
and amplitude
\begin{equation}
\tilde{A}=\left[{1+\tfrac{1}{3}\left(2N^2+1\right)\tilde{\omega}\tau\cot\left(\tfrac{1}{2}\omega_{\tilde{q}}\tau\right)}\right]^{-1}
\end{equation}
[cf. Eqs.~\eqref{ASminus} and \eqref{SymmetricParameterPoints}], respectively.

Thus, when the system is initialized in the single-atom excitation state 
$|\text{eg}\rangle=(|+\rangle+|-\rangle)/\sqrt{2}$,
the corresponding atomic excitation probabilities take the form [cf. Eq.~\eqref{Beta12OSBGeneral}]
\begin{equation}
	\left|\beta_{1,2}\left(t\right)\right|^2\rightarrow B_{0}\pm B_1\cos\left(\tfrac{1}{2}\tilde{\Omega}t\right)+B_2\cos\left(\tilde\Omega t\right),
	\label{ProTwoFExchange}
\end{equation}
with $B_0=A^2/2+\tilde{A}^2/4$, $B_1=A\tilde{A}$, and $B_2=A^2/2$. 
It can be seen from the above results that for both the atoms, the excitation probabilities  
exhibit double-frequency (with $\tilde\Omega$ and $\tilde\Omega/2$) oscillation. Moreover, the components with frequency $\tilde\Omega$ oscillate in phase, while those with frequency $\tilde\Omega/2$ oscillate out of phase,
producing an exchange-type oscillating bound state (cataloged as E2-type in Table.~\ref{tablesep}).
The atomic excitation probabilities are displayed in Fig.~\ref{OscillatingBS_SepAmp}(f).
\begin{figure}[t]
	\centering
	\includegraphics[width=0.5\textwidth]{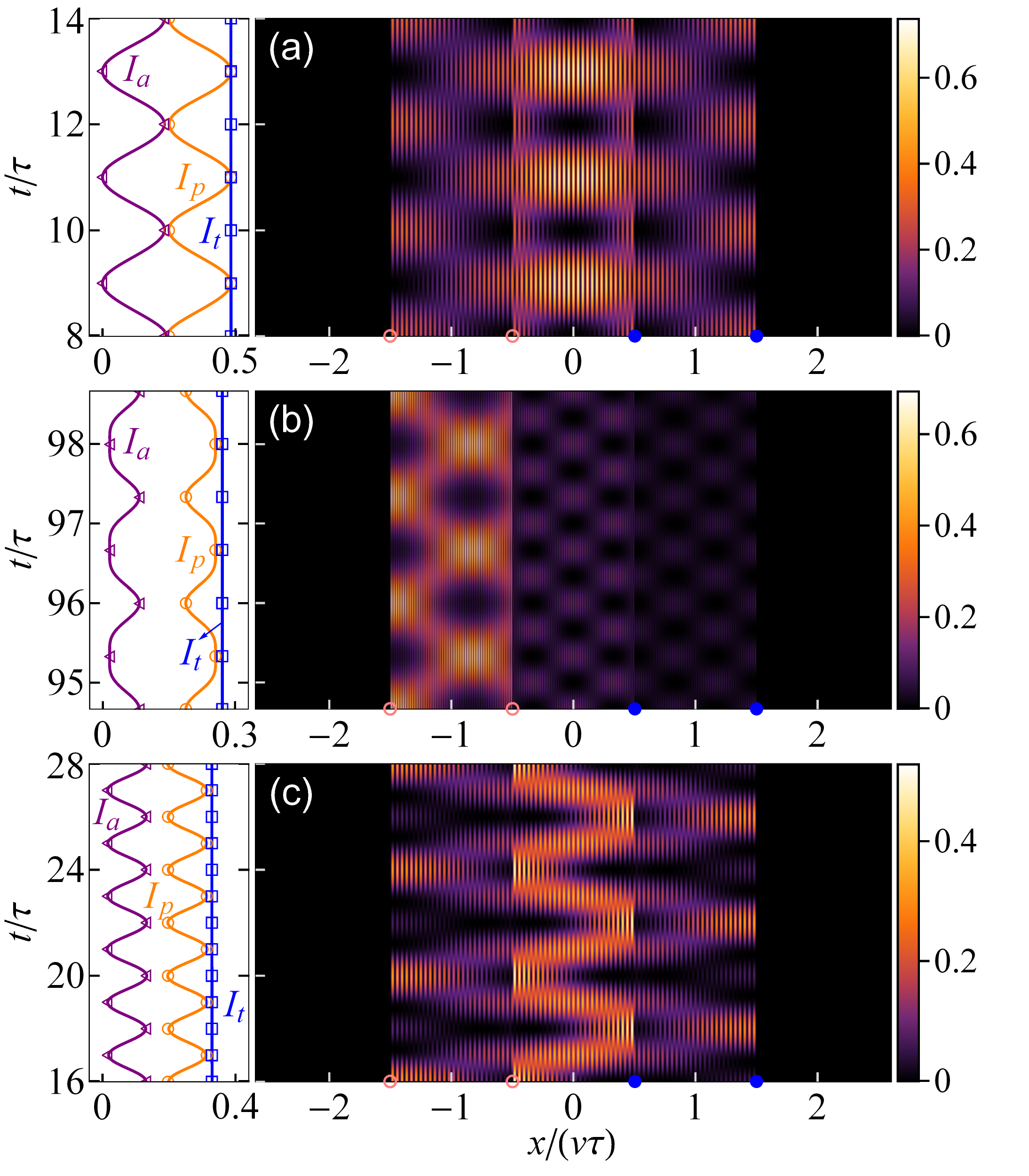}
	\caption{Long-time evolution of the field intensity $|\varphi(x,t)|^2$ [in units of 
	$(v\tau)^{-1}$] in the waveguide for two separate giant atoms is shown in the right panels, corresponding to 
	(a) synchronous oscillating bound state (S1-type), 
	(b) hybrid type oscillating bound state (H2-1 type), and 
	(c) exchange type oscillating bound state (E2-type).
	The pink circles (blue disks) in each right panel are used to label the 
	positions of the connection points of the atom 1 (2). 	
	The corresponding atomic, photonic, and total excitation probabilities are shown in the left panels: 
	solid lines represent numerical results, while triangles, circles, 
	and rectangles denote analytical results for comparison.	
	In (a)-(c), the parameters are the same as 
	those used in Figs.~\ref{OscillatingBS_SepAmp}(a), \ref{OscillatingBS_SepAmp}(c) and
	\ref{OscillatingBS_SepAmp}(f), respectively. 
	}	
	\label{OscillatingBS_SepPwf}
\end{figure}
\subsubsection{\label{FieldIntnsity}Long-time evolution of the field intensity}
The right panels of Figs.~\ref{OscillatingBS_SepPwf}(a)-\ref{OscillatingBS_SepPwf}(c) provide long-time evolution of the field 
intensity in the waveguide for two separate giant atoms.  The left panels show that the corresponding atomic and photonic excitations undergo periodic exchange, demonstrating the coherent coupling between atoms and photons in the effective cavity-QED structures. Since the atom-photon bound states do not decay, the total excitations of atoms and photons remain conserved. These results
are consistent with the analysis in Sec.~\ref{GeneralDescriptionOBS} [see Eqs.~\eqref{AtomExcitationOSBGeneral},
\eqref{PhotonExcitationOSBGeneral} and \eqref{TotalExcitationGeneral}].

Specifically, the right panel of Fig.~\ref{OscillatingBS_SepPwf}(a) shows the long-time evolution of the field intensity 
for synchronous oscillating bound state [corresponding to S1-type atomic dynamics in Fig.~\ref{OscillatingBS_SepAmp}(a)]. In this case, the photon distribution within the waveguide 
remains perfect reflection symmetry at all temporal stages due to parity conservation of system. 
This is a typical characteristic of the synchronous oscillating bound state.
Moreover, due to the coherent atom-photon interactions, the atomic and photonic excitations undergo 
periodic exchange following a cosine law, with
$I_{\mathrm{a}}(t)=2A^2[1+\cos(\tilde{\Omega}t)]$ and $I_{\mathrm{p}}(t)\simeq 2I-2A^2\cos(\tilde{\Omega}t)$, respectively
[see the left panel in Fig.~\ref{OscillatingBS_SepPwf}(a)]. Here we have defined $I\equiv A(1-A)\simeq I_{n_1}\simeq I_{n_2}$
under condition ${|\sin(\omega_{n_i}\tau)|}\ll\omega_{n_i}\tau$.
Particularly, when $\tilde{\Omega}t=(2n-1)\pi$, the atomic excitations are completely converted into field excitations, with the emitted photons predominantly localized in the interatomic region.  

The long-time evolution of the field intensity for hybrid-type oscillating bound state is shown 
in the right panel of Fig.~\ref{OscillatingBS_SepPwf}(b) [corresponding to the H2-1 type atomic dynamics in Fig.~\ref{OscillatingBS_SepAmp}(c)].
The field intensity between the coupling points of the left (right) atom varies with a period of 
$4\pi/\tilde{\Omega}$ ($2\pi/\tilde{\Omega}$), which is identical to that of the corresponding atomic excitation probabilities.
And the photons are mainly distributed in the region of the left atom. The atomic and photonic excitations for this case are $I_{\mathrm{a}}(t)=A^2+A'^2+2AA'\cos(\tilde{\Omega}t/2)+A^2\cos(\tilde{\Omega}t)$ and
$I_{\mathrm{p}}(t)=I+I'-2AA'\cos(\tilde{\Omega}t/2)-A^2\cos(\tilde{\Omega}t)$, respectively [see the left panel in Fig.~\ref{OscillatingBS_SepPwf}(b)]. 
Here we define $I'\equiv A'(1-A')\simeq I_{\bar{n}}$.

The long-time evolution of the field intensity for exchange-type oscillating bound state is shown in
the right panel of Fig.~\ref{OscillatingBS_SepPwf}(c) [corresponding to the E2-type atomic dynamics in Fig.~\ref{OscillatingBS_SepAmp}(f)]. According to Eq.~\eqref{ProTwoFExchange}, the components of 
$|\beta_{1}(t)|^2$ and $|\beta_{2}(t)|^2$ oscillating at frequency $\tilde{\Omega}/2$ exhibit out-of-phase behavior. 
This type of excitation exchange between atoms does not alter the total atomic or photonic 
excitation probabilities, but instead induces oscillations in the expectation 
value of the bound photon's position [see the right panel of Fig.~\ref{OscillatingBS_SepPwf}(c)]. 
In contrast, the components with frequency $\tilde{\Omega}$ 
oscillate in phase [see Eq.~\eqref{ProTwoFExchange}], leading to excitation exchange between atoms and photons. This is reflected in the expressions 
$I_{\mathrm{a}}(t)=A^2+\tilde{A}^2/2+A^2\cos(\tilde{\Omega}t)$ for total atomic excitations and 
$I_{\mathrm{p}}(t)=I+\tilde{I}/2-A^2\cos(\tilde{\Omega}t)$ for photonic excitations [see the left panels in 
Fig.~\ref{OscillatingBS_SepPwf}(c)], where we have defined $\tilde{I}\equiv\tilde{A}(1-\tilde{A})\simeq I_{\bar{n}}$.
The above results reveal that when the system is in such oscillating bound states, it simultaneously exhibits both photon-mediated interatomic interactions and coherent atom-photon interactions.
\subsection{\label{BraOBS}Oscillating bound state in the braided configuration}
\subsubsection{Exchange-type oscillating bound states} 
\begin{figure}[t]
	\centering
	\includegraphics[width=.5\textwidth]{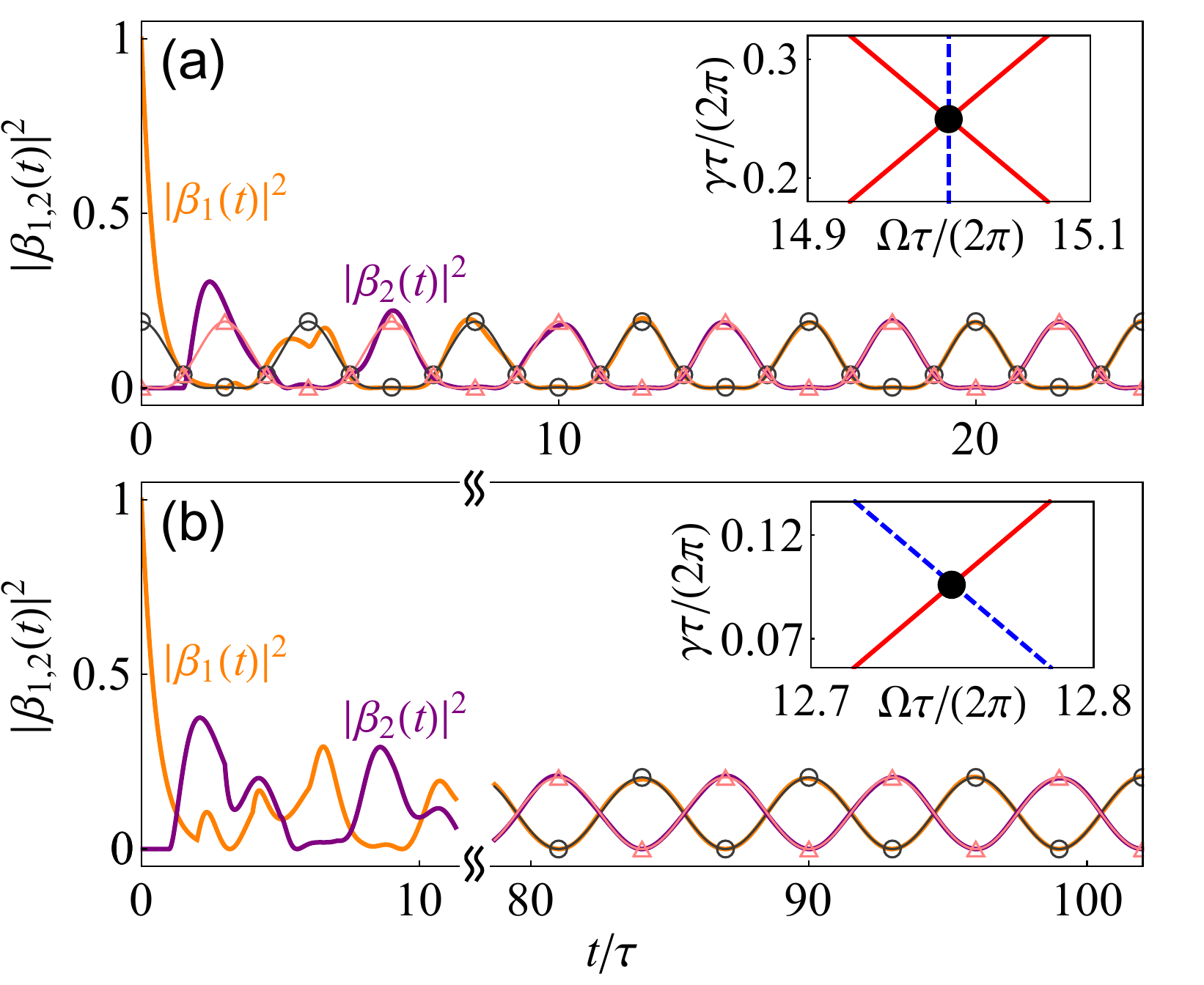}
	\caption{Time evolution of the atomic excitation probabilities of two braided giant atoms 
	under parameters for oscillating bound states. The initial states are $|\text{eg}\rangle$ in both panels. Parameters:
	(a) $N=2$, $m=30$, $p=1$ and $\tilde{q}=1$ (i.e., $n_1=59, n_2=61$), 
	corresponding to the parameter point $\Omega\tau/(2\pi)=15$, $\gamma\tau/(2\pi)=0.25$ [see the inset of panel (a)]. 
	(b) $N=3$, $m=51/2$, $p=1/2$ and $\tilde{q}=1$ 
	(i.e., $n_1=76, n_2=77$), corresponding to 
	$\Omega\tau/(2\pi)=12.75$, $\gamma\tau/(2\pi)=0.096$ [see the inset of panel (b)]. 
	The gray thin line marked with circles and the pink thin line marked with triangles represent the analytical 
	results under the long-time limit.}
	\label{OscillatingBS_BraAmp}
\end{figure}
\begin{figure}[t]
	\centering
	\includegraphics[width=.5\textwidth]{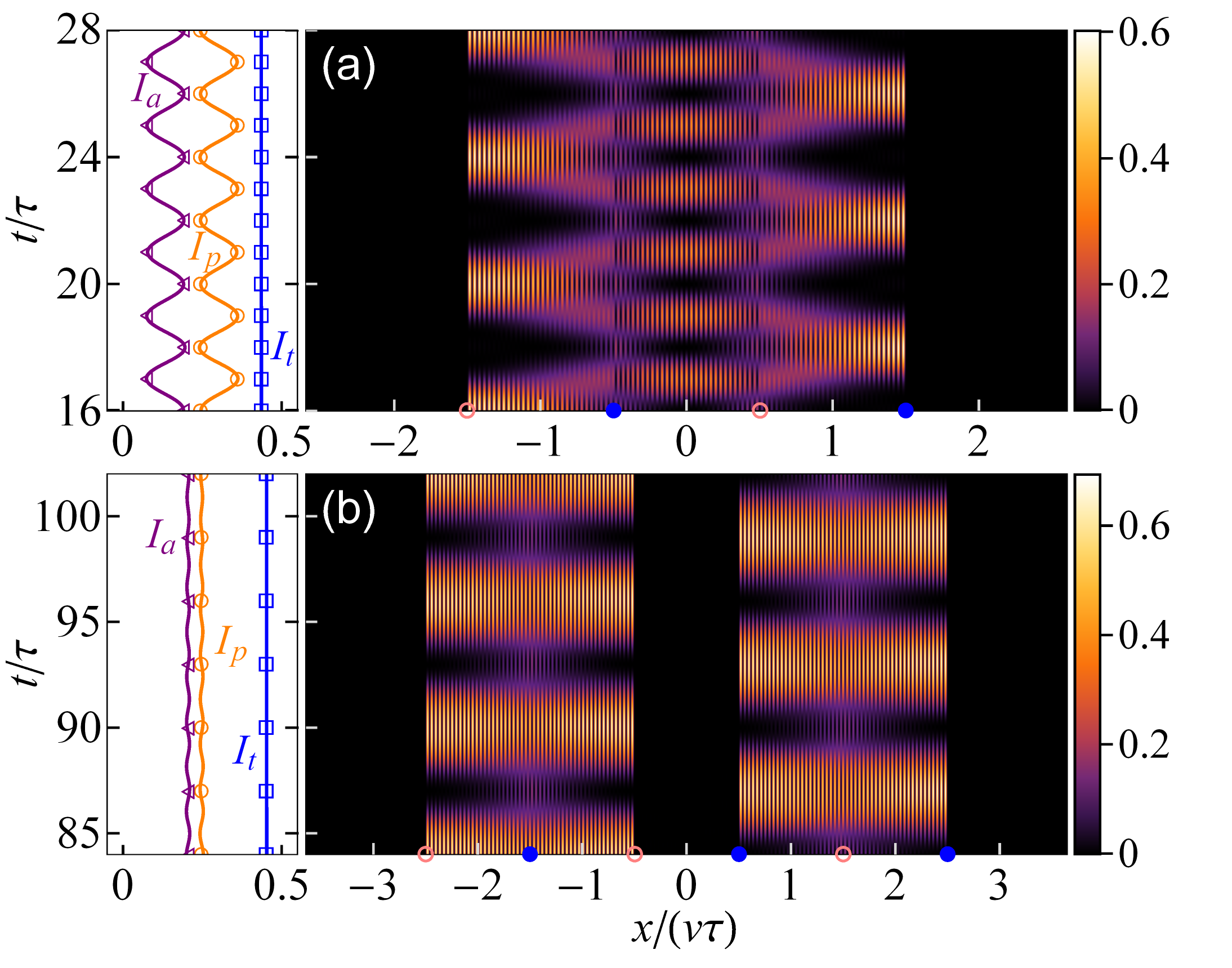}
	\caption{Long-time evolution of the field intensity $|\varphi(x,t)|^2$ [in units of 
	$(v\tau)^{-1}$] in the waveguide for two braided giant atoms is shown in the right panels, corresponding to 
	(a) E2-type oscillating bound state, and (b) E1-type oscillating bound state.
	The pink circles (blue disks) in each panel are used to label the positions of the connection points of the atom 1 (2). 	
	The corresponding atomic, photonic, and total excitation probabilities are shown in the left panels: 
	solid lines represent numerical results, while triangles, circles, and rectangles denote analytical results for comparison.
	In panels (a) and (b), the parameters are the same as 
	those used in Figs.~\ref{OscillatingBS_BraAmp}(a) and \ref{OscillatingBS_BraAmp}(b), respectively.
	}
	\label{OscillatingBS_BraPwf}
\end{figure}
Similar to the separate configuration, the synchronous oscillating bound states of two braided giant atoms are 
classified into two types: S1 and S2 (as summarized in Table~\ref{tablebra}). 
In the subsequent analysis, we focus on asynchronous oscillating bound states, particularly the novel E1-type oscillating bound states, which are unique to the braided configuration.

For the braided configuration, if a parameter point supports both symmetric and antisymmetric dark modes, these modes are 
always non-degenerate. Thus, at such a parameter point, initializing the system in a single-atom excitation state gives rise exclusively 
to exchange-type oscillating bound states.
For example, at the parameter point shown in the inset of Fig.~\ref{OscillatingBS_BraAmp}(a) [satisfying
$m\in\mathbb{Z^+};~p=1,3,\cdots~(p\le m)$], a similar analysis to that in Sec.~\ref{SepNSOBS-E} shows that the long-time atomic excitation dynamics remains described by Eq.~\eqref{ProTwoFExchange} 
upon replacing $\tilde{A}$ with $A'$, resulting in an E2-type oscillating bound state 
[see Figs.~\ref{OscillatingBS_BraAmp}(a)].

The parameter point selected in the inset of Fig.~\ref{OscillatingBS_BraAmp}(b) (satisfying
$m=\mathbb{O^+}/2;~p=1/2,3/2,\cdots,m$) gives rise to the E1-type oscillating bound state.
At this parameter point, the initial state $|+\rangle$ ($|-\rangle$) yields a symmetric (an antisymmetric) dark mode
with frequency $\omega_{n_1}=\bar\omega-\tilde\omega$ ($\omega_{n_2}=\bar\omega+\tilde\omega$)
and amplitude $A_{n_1}= A$ ($A_{n_2}= A$). Thus, for the initial state $|\text{eg}\rangle=(|+\rangle+|-\rangle)/\sqrt{2}$,
the long-time evolution of the atomic excitation probabilities are given by [cf. Eq.~\eqref{Beta12OSBGeneral}]:
\begin{equation}
	\left|\beta_{1,2}\left(t\right)\right|^2\rightarrow\frac{1}{2}A^2\left[1\pm\cos\left(\tilde{\Omega} t\right)\right].
	\label{beta12forE1}
\end{equation}
Clearly, the excitation probabilities exhibit sinusoidal excitation exchange between the two atoms, as shown in Fig.~\ref{OscillatingBS_BraAmp}(b). In other words, an E1-type oscillating bound state (as cataloged in Table~\ref{tablebra}) can be obtained.
 
The field intensities in the long-time limit are illustrated in Figs.~\ref{OscillatingBS_BraPwf}(a) and \ref{OscillatingBS_BraPwf}(b), 
corresponding to the atomic dynamics in Figs.~\ref{OscillatingBS_BraAmp}(a) and \ref{OscillatingBS_BraAmp}(b). 
Similar to its counterpart in the separate configuration, the field intensity of  
the E2-type bound state exhibits both synchronous and exchange-type oscillations, 
as shown in Fig.~\ref{OscillatingBS_BraPwf}(a).
Fig.~\ref{OscillatingBS_BraPwf}(b) shows the atomic and photonic excitations as well as the field intensity of the unique E1-type 
oscillating bound state for two braided giant atoms. 
Note that the cosine-modulated exchange oscillations of atomic excitations described by Eq.~\eqref{beta12forE1}
represents a \textit{pure exchange-type} oscillating bound state without any synchronous component.
Namely, this state exhibits only photon-mediated interatomic interactions and no coherent atom-photon interactions.
Thus, both the atomic and photonic excitation probabilities are constant, with $I_{\text{a}}=A^2$ and 
$I_{\text{p}}\simeq A(1-A)$ [see the left panel of Fig.~\ref{OscillatingBS_BraPwf}(b)]. 
The exchange dynamics of atomic excitations can be further understood from the temporal evolution of the 
field intensity distribution [see the right panel of Fig.~\ref{OscillatingBS_BraPwf}(b)]. Specifically, when atom 1 is in an excited state, the photons are confined 
between its two left coupling points, forming an optical cavity for atom 2. The excitation can then be transferred to atom 2 
through dipole-dipole interaction mediated by the effective cavity mode.  Meanwhile, atom 2 can pump the photon field, redistributing the optical confinement region to the space between its two right coupling points, which in turn act as an optical cavity for atom 1. This process cyclically repeats, forming an excitation exchange between the two atoms.
\subsubsection{\label{RelationDFI}Relation between exchange-type oscillating bound states and the decoherence free interactions}
The previously discussed various types of oscillating bound states are all located in the non-Markovian regime (with $\gamma\tau\gtrsim N^{-3}$, see the discussion in Sec.~\ref{SpecialParameterOBS}).
Here we will show that, for two braided atoms, the E1-type oscillating quasi-bound states 
can also occur in the Markovian regime $\gamma\tau\ll N^{-3}$. which is in perfect agreement with the predictions of decoherence-free interaction theory.

Under the RWA condition $\Omega\gg N^2\gamma$ and the Markovian condition \eqref{MarkvianCondition}, one can replace the complex frequency $s$ in $\Sigma_{\pm}(s)$ by $-\mathrm{i}\Omega$, and the solution to Eq.~\eqref{TranscendentalEQ} can be approximated as 
\begin{equation}
	s_{\pm}\simeq-\mathrm{i}(\Omega+\Delta_{\text{L}}\pm\mathcal{G})-\frac{1}{2}(\Gamma_{\text{eff}}\pm\Gamma_{\text{coll}}).
	\label{s-MarkovianRregime}
\end{equation}
For two braided atoms considered here, the Lamb shift and the effective decay of the individual atom are defined as~\cite{Kockum-PRL2018,Peng-PRA2023}:
\begin{subequations}
\begin{equation}
	\Delta_{\text{L}}\equiv\Im\left[\Sigma_{\mathrm{\Rmnum{1}}}(-\mathrm{i}\Omega)\right]
	=\frac{\gamma}{2}\frac{N\sin(2\Omega\tau)-\sin(2N\Omega\tau)}{1-\cos(2\Omega\tau)},
	\label{LambShift}
\end{equation}
\begin{equation}
	\Gamma_{\text{eff}}\equiv2\Re\left[\Sigma_{\mathrm{\Rmnum{1}}}(-\mathrm{i}\Omega)\right]
	=\gamma\frac{1-\cos(2N\Omega\tau)}{1-\cos(2\Omega\tau)}.
	\label{EffectiveDecay}
\end{equation}
\end{subequations}
The exchange interaction and the collective decay between the atoms are given by
\begin{subequations}
\begin{equation}
	\mathcal{G}\equiv\Im\left[\Sigma_{\mathrm{\Rmnum{2}}}(-\mathrm{i}\Omega)\right]
	=\frac{\gamma}{2}\frac{2N\sin(\Omega\tau)-\cos(\Omega\tau)\sin(2N\Omega\tau)}{1-\cos(2\Omega\tau)},
	\label{ExchangeInteraction}
\end{equation}
\begin{equation}
	\Gamma_{\text{coll}}\equiv2\Re
	\left[\Sigma_{\mathrm{\Rmnum{2}}}(-\mathrm{i}\Omega)\right]
	=\gamma\frac{\cos(\Omega\tau)[1-\cos(2N\Omega\tau)]}{1-\cos(2\Omega\tau)}.
	\label{CollectiveDecay}
\end{equation}
\end{subequations}

\begin{figure}[t]
	\centering
	\includegraphics[width=0.485\textwidth]{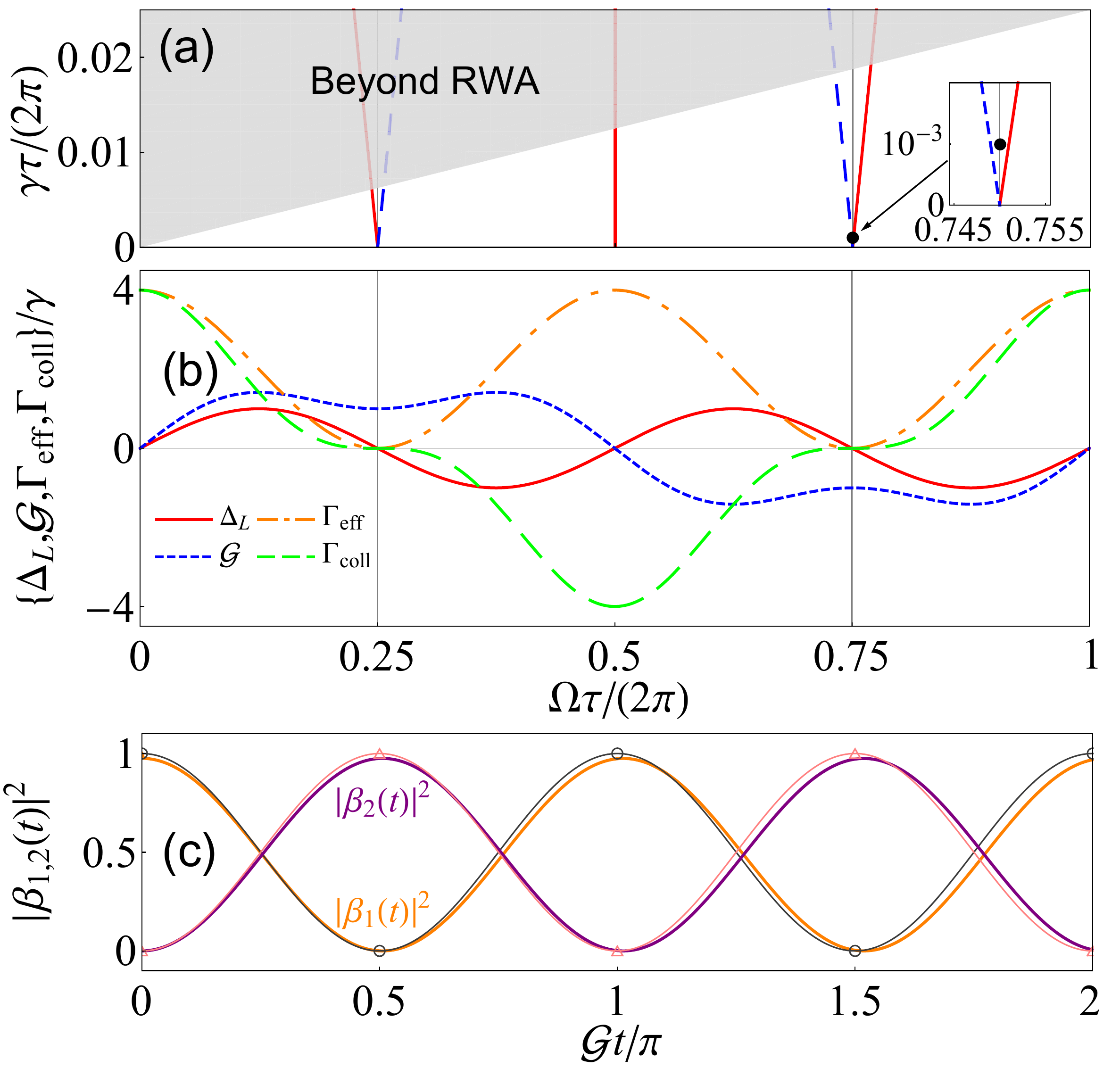}
	\caption{(a) The $\Omega$-$\gamma$ parameter lines that support bound states for two braided giant atoms with 
	$N=2$. The vertical gray lines indicate the $\Omega\tau$ values corresponding to decoherence-free interactions. 
	The inset highlights a specific parameter point [$\Omega\tau/\left(2\pi\right)=0.75$, $\gamma\tau/\left(2\pi\right)=0.001$] 
	that enables decoherence-free interactions.
	(b) Lamb shift $\Delta_{\text{L}}$, effective decay $\Gamma_{\text{eff}}$, coherent interaction $\mathcal{G}$, and collective
	decay $\Gamma_{\text{coll}}$ as functions of $\Omega\tau$ for the same configuration as (a). The vertical gray lines 
	correspond to those in (a), marking the $\Omega\tau$ values for decoherence-free interactions.
	(c) Time evolution of the atomic excitation probabilities at the parameter point shown in (a).
	The thick lines are the numerical results, and the thin lines with circular and triangular markers are analytical 
	approximate results according to Eq.~\eqref{DFIDynamics}.}
	\label{DecoherenceFI}
\end{figure}

Now we consider the parameter points satisfying 
\begin{equation}
\Omega\tau=\frac{n\pi}{N}
\label{DFICondition}
\end{equation}
($n\in\mathbb{Z}^{+}$, ${n}/{N}\notin\mathbb{Z}^{+}$) and $\gamma\tau\ll N^{-3}$, as depicted in the gray vertical lines in Fig.~\ref{DecoherenceFI}(a).
At these parameters, the quantities defined in Eqs.~\eqref{LambShift}-\eqref{CollectiveDecay} become 
$\Gamma_{\text{eff}}=\Gamma_{\text{coll}}=0$ and
\begin{equation}
	\Delta_{\text{L}}=\frac{1}{2}N\gamma\cot\left(\frac{n\pi}{N}\right),~~~
	\mathcal{G}=\frac{1}{2}N\gamma\csc\left(\frac{n\pi}{N}\right),
\end{equation}
as shown in Fig.~\ref{DecoherenceFI}(b). 
Thus, two quasi-dark modes with complex frequencies     
\begin{equation}
	s_{\pm}\simeq-\mathrm{i}(\Omega+\Delta_{\text{L}}\pm\mathcal{G})
	\label{quasi-darkMF}
\end{equation}
are permitted. In fact, such parameter points lie close to a pair of lines representing degenerate 
symmetric and antisymmetric dark modes with $s_{\pm}=-\mathrm{i}\Omega$ 
[see the red solid and blue dashed lines in Fig.~\ref{DecoherenceFI}(a)].
Consequently, a minor parameter deviation can enable the system to acquire a pair of quasi-dark modes featuring complex frequencies as described by Eq.~\eqref{quasi-darkMF}.
In addition, both modes attain nearly identical amplitudes of $A_{+}\simeq A_{-}\simeq{1}$
[cf. Eqs.~\eqref{AmplitudeDarkGeneral} and \eqref{MarkvianCondition}].
Thus, for an initial state $|\text{eg}\rangle$, the time-dependent excitation probabilities for the two atoms take the form
\begin{equation}
	\left|\beta_{1,2}\left(t\right)\right|^2\simeq\frac{1}{2}\left[1\pm\cos\left(2\mathcal{G} t\right)\right].
	\label{DFIDynamics}
\end{equation}
This result demonstrates E1-type oscillating quasi-bound state. 
Compared to oscillating bound state of the same type in the non-Markovian regime 
[see Fig.~\ref{OscillatingBS_BraAmp}(b)],
the atoms exhibit near-perfect excitation exchange at a frequency $2\mathcal{G}$ [see Fig.~\ref{DecoherenceFI}(c)], while maintaining an almost negligible photon population in the waveguide due to the extremely weak time-delay effect. On the other hand, in contrast to the ideal exchange
dynamics described by Eq.~\eqref{DFIDynamics} [see the thin lines with circular and triangular markers in Fig.~\ref{DecoherenceFI}(c)], this residual time-delay effect introduces subtle phase shifts and dissipative corrections in the exact atomic excitation dynamics [see the numerical results described by thick lines in Fig.~\ref{DecoherenceFI}(c)]. 

We have systematically examined the atomic excitation exchange dynamics under condition \eqref{DFICondition} 
within the framework of oscillating bound states. In fact, condition \eqref{DFICondition} precisely gives rise to 
both vanished individual and collective decays and a finite exchange interaction (the phenomenon of decoherence free interaction~\cite{Kockum-PRL2018}). 
As expected, when initialized in the single-atom excited state 
$|\text{eg}\rangle$, the corresponding effective interaction Hamiltonian 
$\mathcal{G}\left(\sigma_{1}^{+}\sigma_{2}^{-}+\sigma_{1}^{-}\sigma_{2}^{+}\right)$ produces Rabi-oscillation-type excitation exchange as described by Eq.~\eqref{DFIDynamics}.
These results conclusively demonstrate that the E1-type oscillating quasi-bound states in the Markovian regime fundamentally originate from decoherence-free interactions.
\subsection{\label{PhenomenonQusiOBS}Realizing multi-component oscillations
by using quasi-dark modes}
As elaborated in the preceding subsections,  at the strict parameter points for oscillating bound states (i.e., the precise intersections of 
different $\Omega\text{-}\gamma$ parameter lines), the allowed dark modes have at most three frequencies, which are equally 
spaced. This results in oscillations in atomic excitation probabilities at no more than two different frequencies. However, in the vicinity 
of some strict parameter points (particularly when $N$ is relatively large), there exist parameter lines that 
permit other dark modes. Owing to their influence, additional quasi-dark modes emerge at these parameter points, 
analogous to the scenario discussed in Sec.~\ref{RelationDFI}. Thus, multi-component oscillations for atomic excitation probabilities can be achieved.  

\begin{figure}[t]
	\centering
	\includegraphics[width=0.5\textwidth]{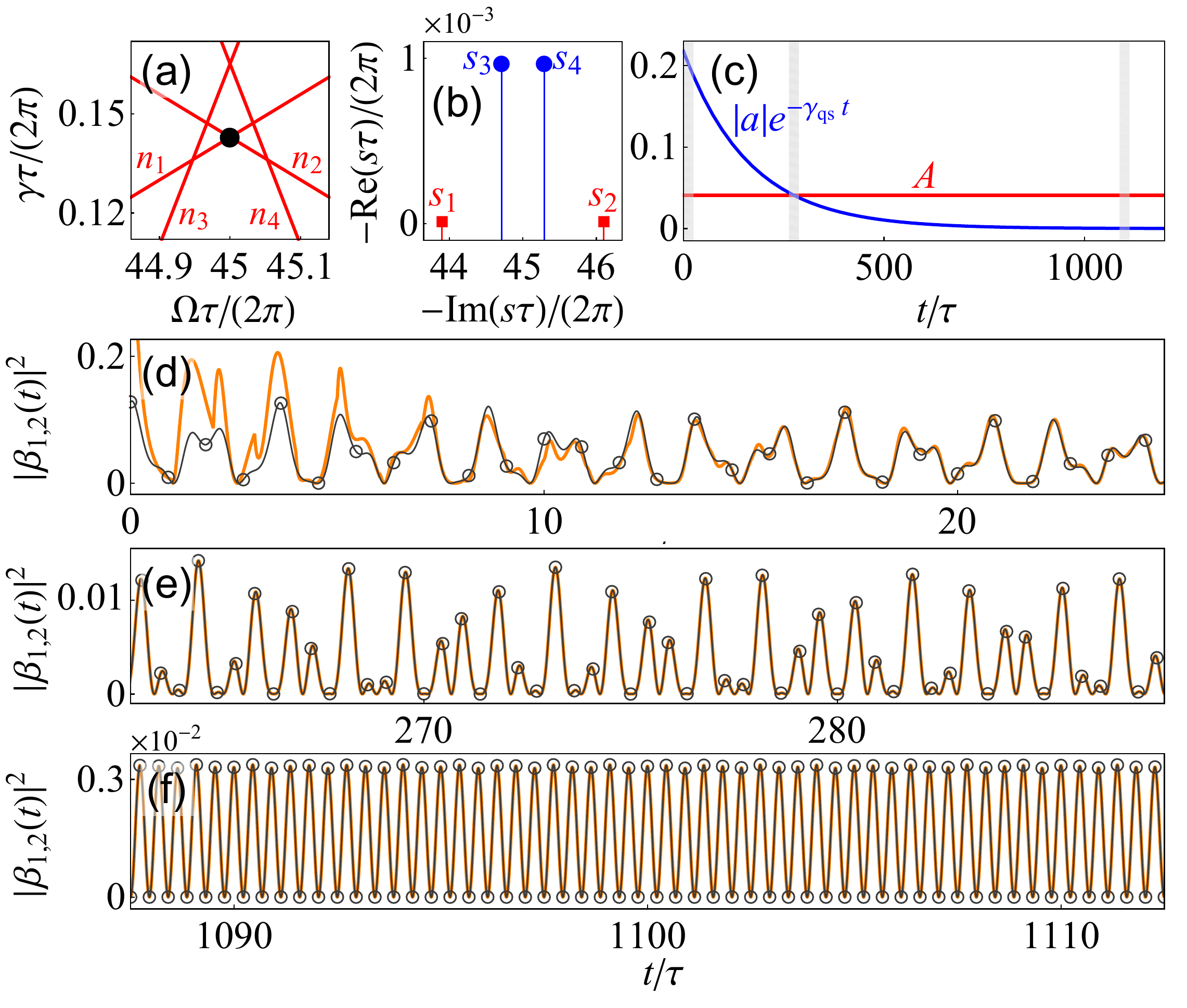}
	\caption{
	(a) The $\Omega$-$\gamma$ lines with $n_1=439, n_2=461$, $n_3=447$, and $n_4=453$ that supports
	symmetric dark modes for two separate giant atoms with $N=5$. Note that only the parameter lines for 
	the symmetric modes of interest are plotted, while those for antisymmetric modes are omitted.
	The parameter point marked by black disk [$\Omega\tau/(2\pi)=45$, $\gamma\tau/(2\pi)=0.143$] 
	supports both symmetric dark and symmetric quasi-dark modes.
	(b) The poles corresponding to dark and quasi-dark modes obtained by solving Eq.~\eqref{TranscendentalEQ} 
	for the symmetric branch $\Sigma_{+}(s)$.
	(c) Time evolution of the absolute value of the amplitude of the quasi-dark modes. 
	The constant amplitude of the dark mode is also included in the figure for comparison.
	(d)-(f) Time evolution of the atomic excitation probability in different stages with the parameter
	point shown in panel (a) and initial state $|+\rangle$. 
	The temporal ranges along the abscissae in panels (d)-(f) align with the gray-highlighted time intervals denoted in panel (c).
	The gray solid lines marked with circles in panels (d)-(f) represent the analytical results given by
	Eq.~\eqref{beta12forQOBS}.}
	\label{DynamicQBS}
\end{figure}

A specific example of such parameter points [$\Omega\tau/(2\pi)=45,~\gamma\tau/(2\pi)\simeq 0.143$] in 
separate configuration with $N=5$ is shown in Fig.~\ref{DynamicQBS}(a), where a pair of symmetric dark modes with 
$n_1=439$ and $n_2=461$ (i.e., $m=90$, $p=3$, and $\tilde{q}=4$) are allowed. Near this point, there are also lines 
representing another pair of symmetric dark modes with $n_3=447$ and $n_4=453$ (i.e., $m=90$, $p=1$, and $\tilde{q}=2$).
By solving the transcendental equation \eqref{TranscendentalEQ} for the symmetric branch $\Sigma_{+}(s)$, 
one obtains two purely imaginary poles with 
$-\Im(s_1\tau)/(2\pi)=\omega_{n_1}\tau/(2\pi)=43.9$ and $-\Im(s_2\tau)/(2\pi)=\omega_{n_2}\tau/(2\pi)=46.1$ 
exactly correspond to the two dark modes. The other two poles exhibit imaginary parts 
$-\Im(s_3\tau)/(2\pi)\equiv\omega^{\text{(qs)}}_{1}\tau/(2\pi)\simeq 44.709$ and 
$-\Im(s_4\tau)/(2\pi)\equiv\omega^{\text{(qs)}}_{2}\tau/(2\pi)\simeq 45.291$,
approximately matching $\omega_{n_3}\tau/(2\pi)=44.7$ and $\omega_{n_4}\tau/(2\pi)=45.3$, respectively. In addition, they share 
equally small negative real parts satisfying $-\Re(s_3\tau)/(2\pi)=-\Re(s_4\tau)/(2\pi)\equiv\gamma_{\text{qs}}\tau/(2\pi)\simeq 0.968\times10^{-3}$ [see Fig.~\ref{DynamicQBS}(b)], exhibiting the character of quasi-dark modes.
The atomic excitation amplitudes of the dark and quasi-dark modes are $A_{n_1}=A_{n_2}=A\simeq 0.041$ and
$A^{\text{(qs)}}_{1}=A^{\text{(qs)}*}_{2}=ae^{-\gamma_{\text{qs}}t}$ respectively, where $a\simeq 0.212-0.047\mathrm{i}$.  
Thus, for the initial state $|+\rangle$, the corresponding atomic excitation probabilities during the coexistence period of all four modes take the form:
\begin{equation}
	\left|\beta_{1,2}\left(t\right)\right|^2\rightarrow
	2\left[A\cos\left(\tfrac{\tilde{\Omega} t}{2}\right)+|a|e^{-\gamma_{\text{qs}}t}\cos\left(\tfrac{\tilde{\Omega}' t}{2}+\delta\right)
	\right]^2,
	\label{beta12forQOBS}
\end{equation}
with $\tilde{\Omega}=\omega_{n_2}-\omega_{n_1}$ and $\tilde{\Omega}’=\omega^{\text{(qs)}}_{2}-\omega^{\text{(qs)}}_{1}$.
Clearly, these two quasi-dark modes are characterized by dissipation time scales ($2\pi/\gamma_{\text{qs}}\sim10^3\tau$)
far exceeding the existence time of other rapid dissipation modes (not exceeding $2\pi/\gamma\sim10\tau$) as well as the oscillation period of the dark modes ($2\pi/\tilde{\Omega}\sim\tau$). Thus, over a long time range of $\gamma_{\text{qs}}^{-1}$, the amplitude of
the quasi-dark modes is comparable to that of the dark modes, as shown in Fig.~\ref{DynamicQBS}(c).
Consequently, during the existence of quasi-dark modes, the oscillations of atomic excitation probabilities will acquire additional harmonic components, thereby forming complex oscillations with multiple frequency components [with 
frequencies $\tilde{\Omega}$, $\tilde{\Omega}'$, and $(\tilde{\Omega}\pm\tilde{\Omega}')/2$, see Figs.~\ref{DynamicQBS}(d) and \ref{DynamicQBS}(e)].
Once these quasi-dark modes are completely dissipated, only the components 
of strict oscillating bound state remain in the end
[S1-type for this case, see Fig.~\ref{DynamicQBS}(f)].
\section{\label{Conclusion}Summary and conclusions}
In this work, we have systematically investigated the phenomenon of BIC in wQED system with two identical giant atoms.   
First, we have derived general dark-state conditions for the two-atom system, explicitly clarifying how coupling topology and atomic parameters (transition frequency, decay rate at coupling points, and number of coupling points) collectively regulate decay suppression, laying a foundation for understanding oscillating bound states in the systems.
Further, through analytical analysis of the long-time dynamical behaviors of atoms and bound photons under different topological configurations, we have carried out a detailed classification of bound states and explored the connections between these dynamical behaviors and the system's intrinsic light-matter interactions.
Our findings show that in a wQED system with giant-atoms, when the conditions for static bound states are satisfied, the system effectively behaves like an empty cavity, with bound photons forming stationary standing waves. In contrast, when the conditions allow for oscillating bound states, the system functions as an effective cavity QED architecture. This enables excitation exchange between atoms and bound photons, corresponding to coherent atom-photon interactions, or excitation exchange between distinct atoms, corresponding to photon-mediated coherent interatomic interactions.
Usually, the parameter points for such oscillating bound states lie within the non-Markovian regime, where time-delay feedback effects are non-negligible. However, for braided giant atoms, exchange-type oscillating quasi-bound states can also be achieved in the Markovian regime, consistent with the predictions of decoherence-free interactions. 
Furthermore, at certain appropriate parameter points, oscillating bound states can acquire additional harmonic components 
over a sufficiently long-time scale due to the presence of extra quasi-dark modes. This property holds promise for enhancing the capacity of quantum information storage and processing.
Overall, this study deepens the understanding of BIC in multi-giant-atom wQED systems and underscores the potential of such systems as versatile platforms for quantum control and quantum information applications.
%
%
\appendix
\section{\label{DerivationEOM}Hamiltonian and dynamical equation}
In this appendix, we first focus on the most general possible configurations containing $M$ atoms, with each atom possessing $N_i$ 
coupling points, situated at locations $x_{ij}$ ($i=1,2,\cdots,M$ and $j=1,2,\cdots,N_i$). At the point $x_{ij}$, the coupling strength 
between the $i$th atom and the photonic modes with wave vector $k$ is given by $g_{ij}(k)$.  
Under the RWA, the Hamiltonian of the entire system can be expressed as follows ($\hbar=1$)
\begin{eqnarray}
	\hat{H}&=&\sum_{i}{\Omega_i\sigma_{i}^{+}\!\:\sigma_{i}^{-}}+\int_{-\infty}^{+\infty}\mathrm{d}k\nu_{k}\hat{a}_{k}^{\dagger}\hat{a}_{k}
	\nonumber
	\\
	&&+\sum_{ij}\int_{-\infty}^{+\infty}\mathrm{d}k{g_{ij}(k)}\left(e^{\mathrm{i}k{x_{ij}}}\hat{a}_{k}\sigma_{i}^{+}+\mathrm{H.c.} \right),
	\label{HamiltonianGeneral}
\end{eqnarray}
where $\Omega_i$ is the transition frequency of the $i$th atom and $\nu_{k}$ are the frequencies of photonic fields. 

Assuming that the total system (atoms plus fields) is initially prepared in the single-excitation manifold, and noting that the Hamiltonian \eqref{HamiltonianGeneral} preserves the total number of excitations, the time-dependent state can be expressed as 
\begin{equation}
	|\Psi(t) \rangle =\sum_{i}\beta_i(t)\sigma_{i}^{+}|\emptyset\rangle+\int_{-\infty}^{+\infty}\mathrm{d}k\alpha_{k}(t)\hat{a}_{k}^{\dagger}|\emptyset\rangle,
	\label{GeneralTDS}
\end{equation}
where $|\emptyset\rangle=|\text{g}\rangle_1\otimes|\text{g}\rangle_2\otimes\cdots|\text{g}\rangle_M\otimes|\text{vac}\rangle$ is the ground state of the total system.
Substituting the state vector \eqref{GeneralTDS} into the Schr\"odinger equation
$\hat{H}|\Psi(t) \rangle=\mathrm{i}{\partial_t}|\Psi(t)\rangle$
yields the following dynamic equations of motion:
\begin{subequations}
\begin{equation}
	\dot\beta_i(t)=-\mathrm{i}\Omega_i\beta_i(t)-\mathrm{i}\sum_{j}\int_{-\infty}^{+\infty}\mathrm{d}k{g_{ij}(k)}\alpha_{k}(t)e^{\mathrm{i}k{x_{ij}}},
	\label{DynAtom}
\end{equation}
\begin{equation}
	\dot\alpha_k(t)=-\mathrm{i}\nu_{k}\alpha_k(t)-\mathrm{i}\sum_{ij}{g_{ij}(k)}\beta_i(t){e^{-\mathrm{i}k{x_{ij}}}}.	\label{DynPho}
\end{equation}
\end{subequations}
Equation \eqref{DynPho} can be formally integrated to yield
\begin{equation}
	\alpha_{k}(t)=\alpha_{k}(0)e^{-\mathrm{i}\nu_{k}{t}}
	-\mathrm{i}\sum_{ij}{g_{ij}(k)}\int_{0}^{t}\mathrm{d}t'\beta_{i}(t')e^{\mathrm{i}[\nu_{k}(t'-t)-kx_{ij}]}.
	\label{PhoSol}
\end{equation}
By substituting this formal solution of $\alpha_k(t)$ into Eq.~\eqref{DynAtom}, we can eliminate the photonic degrees of freedom and obtain
%
%
\begin{eqnarray}
	\dot\beta_{i}(t)&=&-\mathrm{i}\Omega_{i}\beta_{i}(t)-\mathrm{i}\sum_{j}\int_{-\infty}^{\infty}\mathrm{d}k{\tilde{g}_{ij}\left(\nu_{k}\right)}\alpha_{k}(0)e^{\mathrm{i}({k}{x_{ij}}-{\nu_{k}}{t})}
	\nonumber
	\\
	&&-\frac{1}{v}\sum_{i'jj'}\int_{0}^{\infty}\mathrm{d}\nu_{k}\left\{{\tilde{g}_{ij}\left(\nu_{k}\right)}{\tilde{g}_{i'j'}\left(\nu_{k}\right)}
	\int_{0}^{t}\mathrm{d}{t'}\beta_{i'}(t')\right.
	\nonumber
	\\
	&&\left.\times\left[e^{\mathrm{i}\nu_{k}(t'-t+\tau_{ij,i'j'})}+e^{\mathrm{i}\nu_{k}(t'-t-\tau_{ij,i'j'})}\right]\right\},
	\label{EOMAtomNS}
\end{eqnarray}
where $\tau_{ij,i'j'}=(x_{ij}-x_{i'j'})/v$ is the time delay between two coupling points. To derive the above equation, we have assumed $g_{ij}(k)=g_{ij}(-k)$ and used the dispersion relation $\nu_{k}=v\left|k\right|$. 
The coupling strength is redefined as $\tilde{g}_{ij}\left(\nu_{k}\right)\equiv g_{ij}(\nu_{k}/v)$.
To further simplify Eq.~\eqref{EOMAtomNS}, the well-known Weisskopf-Wigner approximation is employed, under the assumption that the emitted radiation is concentrated within a narrow range around atomic transition frequency and that the spectral density of the field modes in this area varies little. We can therefore replace ${\tilde{g}_{ij}(\nu_{k})}$ by ${\tilde{g}_{ij}(\Omega_{i})}$ and the lower limit in the $\nu_{k}$ integration by $-\infty$. The integral $\int_{-\infty}^{\infty}\mathrm{d}\nu_{k}e^{\mathrm{i}\nu_{k}t}=2\pi\delta(t)$ and
\begin{eqnarray}
	&&\int_{0}^{t}\mathrm{d}t'\beta_{i'}(t')[\delta(t'-t+\tau_{ij,i'j'})+\delta(t'-t-\tau_{ij,i'j'})]
	\nonumber
	\\
	&&=\beta_{i'}(t-\left|\tau_{ij,i'j'}\right|)\Theta(t-\left|
	\tau_{ij,i'j'}\right|)
\end{eqnarray}
yield the following equation for the atomic excitation amplitudes: 
\begin{eqnarray}
\dot\beta_{i}&=&-\mathrm{i}\sum_{j}\sqrt{\frac{\gamma_{ij}{v}}{2}}\varphi_{\text{in}}(x_{ij},t)-\mathrm{i}\Omega_{i}\beta_{i}(t)\nonumber
\\
&&-\frac{1}{2}\sum_{i'jj'}\sqrt{\gamma_{ij}\gamma_{i'j'}}\beta_{i'}\left(t-\left|\tau_{ij,i'j'}\right|\right)
\Theta\left(t-\left|\tau_{ij,i'j'}\right|\right),
\nonumber
\\
\label{EOMAtomWWA}
\end{eqnarray}
where $\Theta(\bullet)$ is the Heaviside step function, 
\begin{equation}
	\gamma_{ij}\equiv\frac{4\pi{\tilde{g}_{ij}^2(\Omega_{i})}}{v}
\end{equation}
is the bare relaxation rate at the $j$th connection point of the $i$th atom, and
\begin{equation}
	\varphi_{\text{in}}(x,t)\equiv\frac{1}{\sqrt{2\pi}}\int_{-\infty}^{\infty}\mathrm{d}k\alpha_{k}(0)e^{\mathrm{i}(kx-\nu_{k}{t})}
\end{equation}
is defined as the wave function of the input field.
	
Finally, using Eq.~\eqref{PhoSol} and the Weisskopf-Wigner approximation again, the total time-dependent field function in the waveguide can be obtained:
\begin{eqnarray}
	\varphi(x,t)&\equiv&\langle x|\Psi(t)\rangle=\frac{1}{\sqrt{2\pi}}\int_{-\infty}^{\infty}\mathrm{d}k\alpha_{k}(t)e^{\mathrm{i}kx}
	\nonumber
	\\
	&=&\varphi_{\text{in}}(x,t)-\mathrm{i}\sum_{ij}\sqrt{\frac{\gamma_{ij}}{2v}}\beta_{i}\left(t-\frac{\left|x-x_{ij}\right|}{v}\right)
	\nonumber
	\\
	&&\times\Theta\left(t-\frac{\left|x-x_{ij}\right|}{v}\right).
	\label{WFPhoton}
\end{eqnarray}
Here $|x\rangle\equiv\hat{a}^{\dagger}(x)|\emptyset\rangle$ is the base vector in coordinate space, where $\hat{a}^{\dagger}(x)\equiv\int_{-\infty}^{\infty}\mathrm{d}k\hat{a}^{\dag}_{k}e^{-\mathrm{i}kx}/{\sqrt{2\pi}}$ is the field operator creating a photon at $x$.
Thus according to Eq.~\eqref{GeneralTDS}, the time-dependent state of the total system can be re-expressed as
\begin{equation}
	|\Psi(t) \rangle =\sum_{i}\beta_i(t)\sigma_{i}^{+}|\emptyset\rangle+\int_{-\infty}^{+\infty}\mathrm{d}x\varphi(x,t)\hat{a}^{\dagger}(x)|\emptyset\rangle.
	\label{GeneralTDSRealSpace2}
\end{equation}

For the case of two identical giant atoms considered in the main text, we have $\Omega_{1}=\Omega_{2}\equiv\Omega$, $\gamma_{1j}=\gamma_{2j'}\equiv\gamma$, and $N_1=N_2\equiv N$. Equations \eqref{EOMAtomWWA} and \eqref{WFPhoton} reduce to Eqs.~\eqref{EOMAtom2A} and \eqref{WFPhoton2A}. 
%
%
\section{\label{DerivationTM}Derivation of atomic excitation amplitudes}
Next, we apply the Laplace transform $E_{i}(s)=\int_{0}^{\infty}\beta_{i}(t)e^{-st}{\mathrm{d}t}$ to solve Eq.~\eqref{EOMAtomWWA}. 
Straightforwardly, one can find the Laplace transformed coefficients $\mathbf{E}=[E_{1}(s),E_{2}(s),\cdots,E_{M}]^\top$ satisfy the following
linear equations
\begin{equation}
	\mathbf{T}\mathbf{E}=\mathbf{V}.
	\label{linearEQforE}
\end{equation}
The elements of $\mathbf{T}$ and $\mathbf{V}$ take the forms
\begin{subequations}
\begin{equation}
	T_{ii'}=\left(s+\mathrm{i}\Omega_{i}\right)\delta_{ii'}+\frac{1}{2}\sum_{jj'}\sqrt{\gamma_{ij}\gamma_{i'j'}}e^{-s\left|\tau_{ij,i'j'}\right|},
	\label{Telement}
\end{equation}
\begin{equation}
	V_{i}=\beta_{i}(0)-\mathrm{i}\sum_{j}\sqrt{\frac{\gamma_{ij}v}{4\pi}}\int_{-\infty}^{\infty}{\mathrm{d}k}\frac{e^{\mathrm{i}kx_{ij}}\alpha_{k}(0)}{s+\mathrm{i}\nu_{k}}.	
	\label{Velement}
\end{equation}
\end{subequations}
One can further solve $E_i(s)$ from Eq.~\eqref{linearEQforE} and take an inverse Laplace transform of $E_i(s)$ to obtain the time dependent atomic excitation amplitudes $\beta_i(s)$:
\begin{eqnarray}
	\beta_{i}\left(t\right)&=&\frac{1}{2\pi\mathrm{i}}\int_{\sigma-\mathrm{i}\infty}^{\sigma+\mathrm{i}\infty}\mathrm{d}sE_{i}(s)e^{st}
	=\sum\operatorname{Res}\left[e^{st}E_i(s)\right]
	\nonumber
	\\
	&=&\sum e^{s_\mathrm{pole~}t}\lim_{\epsilon\to0}E_i(s_\mathrm{pole}+\epsilon)\epsilon.
	\label{InverseLaplace}
\end{eqnarray}
Here, we focus on the case of two identical giant atoms with all connection points equally spaced, as shown in Fig.~\ref{Schematics}. 
The total system is initially prepared in the state
$|\Psi(0)\rangle=\left(c_1\sigma_1^{+}+c_2\sigma_2^{+}\right)|\emptyset\rangle$.
Namely, the atoms are prepared in a superposition state $c_1|\text{e}\text{g}\rangle+c_2|\text{g}\text{e}\rangle$ (with $|c_1|^2+|c_2|^2=1$), and the field in the waveguide is in the vacuum state $|\text{vac}\rangle$. Thus we have $\beta_{i}(0)=c_{i}$ ($i=1,2$) and $\alpha_k(0)=0$, and consequently $V_i=c_i$ [see Eq.~\eqref{Velement}]. According to Eq.~\eqref{linearEQforE},
 the Laplace coefficients for this case takes the form
\begin{equation}
	E_{1,2}(s)=\frac{1}{\sqrt{2}}\left[c_{+}E_{+}(s)\pm c_{-}E_{-}(s)\right],
	\label{LaplaceCoefficients12}
\end{equation}
with $c_{\pm}=(c_1\pm c_2)/\sqrt{2}$ and 
\begin{equation}
	E_{\pm}(s)=\left[{s+\mathrm{i}\Omega+\Sigma_{\pm}(s)}\right]^{-1}.
	\label{LaplaceCoefficientsPM}
\end{equation}
From taking an inverse Laplace transform of Eq.~\eqref{LaplaceCoefficients12}, we obtain the time-dependent atomic excitation amplitudes  
given in Sec.~\ref{Model}.
\section{\label{ExpressionSE}Analytical expressions and symmetries of the self-energy functions}
For separate configuration [see Fig.~\ref{Schematics}(a)], the possible time delays between the connection points of the same 
atom are $|\tau_{j,j'}|=l\tau$ ($l=0,1,2,\cdots,N-1$). The combination numbers of these time delays 
are $N$ for $l = 0$ and $2(N-l)$ for 
$l>0$. Thus the function $\Sigma_{\mathrm{\Rmnum{1}}}(s)$ [defined in Eq.~\eqref{Sigmaiip}] can be written as  
\begin{equation}
	\Sigma_{\mathrm{\Rmnum{1}}}(s)=\gamma\sum_{l=1}^{N-1}\left(N-l\right)e^{-ls\tau}
	+\frac{1}{2}N\gamma.
	\label{SigmaSep1EqualSpace}
\end{equation}
The time delays between the coupling points of atoms 1 and 2 take the form $|\tau_{1j,2j'}|=l\tau$ ($l=1,2,\cdots,2N-1$). The combination number of the time delays are $l$ for $l\leq N $ and $2N-l$ for $l>N$.
Thus the function $\Sigma_{\mathrm{\Rmnum{2}}}(s)$ [defined in Eq.~\eqref{Sigmaiip}] related to these time delays can be written as 
\begin{equation}
	\Sigma_{\mathrm{\Rmnum{2}}}(s)
	=\frac{\gamma}{2}\sum_{l=1}^Nle^{-ls\tau}+\frac{\gamma}{2}\sum_{l=N+1}^{2N-1}\left(2N-l\right)e^{-ls\tau}
	\label{SigmaSep2EqualSpace}
\end{equation}
For braided configuration [see Fig.~\ref{Schematics}(b)], 
the corresponding energy correlation functions $\Sigma_{\mathrm{\Rmnum{1}}}(s)$
and $\Sigma_{\mathrm{\Rmnum{2}}}(s)$ can be calculated similarly:   
\begin{eqnarray}
	\Sigma_{\mathrm{\Rmnum{1}}}(s)&=&\gamma\sum_{l=1}^{N-1}\left(N-l\right)e^{-2ls\tau}
	+\frac{1}{2}N\gamma,
	\label{SigmaBra1EqualSpace}
	\\
         \Sigma_{\mathrm{\Rmnum{2}}}(s)
	&=&\frac{\gamma}{2}\sum_{l=1}^N\left(2N-2l+1\right)e^{-\left(2l-1\right)s\tau}.
	\label{SigmaBra2EqualSpace}
\end{eqnarray}

Substituting Eqs.~\eqref{SigmaSep1EqualSpace} and \eqref{SigmaSep2EqualSpace} [or Eqs.~\eqref{SigmaBra1EqualSpace} and \eqref{SigmaBra2EqualSpace}] into Eq.~\eqref{SigmaPM} and subsequently summing the series allows for the determination of the self-energy functions
$\Sigma_{\pm}(s)$ for each configuration. Specifically, $\Sigma_{+}(s)$ takes the same form for both the separate and the braided configurations:
\begin{equation}
	\Sigma_{+}(s)=\frac{\left[N\left(e^{2s\tau}-1\right)-e^{s\tau}\left(1-e^{-2Ns\tau}\right)\right]\gamma}{2\left(e^{s\tau}-1\right)^2}.
	\label{SigmaPMSepAndBra}
\end{equation}

The function $\Sigma_{-}(s)$ for the separate and the braided configurations can be expressed as 
\begin{equation}
         \Sigma_{-}(s)=\frac{\left[N\left(e^{2s\tau}-1\right)-e^{s\tau}\left(3-4e^{-Ns\tau}\right)+e^{-2Ns\tau}\right]\gamma}{2\left(e^{s\tau}-1\right)^2}
	\label{SigmaMSep}
\end{equation}
and
\begin{equation}
	\Sigma_{-}(s)=\frac{\left[N\left(e^{2s\tau}-1\right)+e^{s\tau}\left(1-e^{-2Ns\tau}\right)\right]\gamma}{2\left(e^{s\tau}+1\right)^2},
	\label{SigmaMBra}
\end{equation}
respectively.

Using the explicit forms of 
$\Sigma_{\pm}(s)$ [see Eqs.~\eqref{SigmaPMSepAndBra}-\eqref{SigmaMBra}], one can verify that for both the separate and the braided configurations, the 
functions $\Sigma_{\pm}(s)$ and $\Sigma'_{\pm}(s)$ satisfy
\begin{equation}
f_{\pm}(s)=f_{\pm}\left(s+\frac{2\mathrm{i}n\pi}{\tau}\right).
\label{SymmetrySB}
\end{equation}
Here $f=\Sigma,~\Sigma'$ and $n\in\mathbb{Z}$. These symmetries imply that for a pair of parameter points $(\Omega,\gamma)$ and $(\Omega',\gamma)$ satisfying
\begin{equation}
\Omega'\tau\mp\Omega\tau=2n\pi,
\label{DynIdenSB} 
\end{equation}
the identity $\beta'_{\pm}(t)=e^{-2\mathrm{i}n\pi t/\tau}\beta_{\pm}(t)$ or $\beta'_{\pm}(t)=e^{-2\mathrm{i}n\pi t/\tau}\beta^{*}_{\pm}(t)$ 
holds [see Eqs.~\eqref{BetaPMtGeneral}-\eqref{AmplitudeFun}], yielding  $|\beta'_{\pm}(t)|^2=|\beta_{\pm}(t)|^2$.

Furthermore, the braided configuration exhibits symmetry properties:
\begin{equation}
f_{\pm}\left(s\right)=f_{\mp}\left[s+\frac{\mathrm{i}(2n-1)\pi}{\tau}\right],
\label{SymmetryB}
\end{equation}
Thus for a pair of parameter points $(\Omega,\gamma)$ and $(\Omega',\gamma)$ satisfying
\begin{equation}
\Omega'\tau\mp\Omega\tau=\left(2n-1\right)\pi,
\label{DynIdenBraid} 
\end{equation}
the relation $\beta'_{\pm}(t)=e^{-\mathrm{i}(2n-1)\pi t/\tau}\beta_{\mp}(t)$ or 
$\beta'_{\pm}(t)=e^{-\mathrm{i}(2n-1)\pi t/\tau}\beta^{*}_{\mp}(t)$ holds [see Eqs.~\eqref{BetaPMtGeneral}-\eqref{AmplitudeFun}], resulting in $|\beta'_{\pm}(t)|^2=|\beta_{\mp}(t)|^2$. 
\section{\label{DerivationDSCondition}Derivation of dark-state conditions}
Our goal is to identify the dark states that correspond to purely imaginary solutions to Eq.~\eqref{TranscendentalEQ}. 
To this end, we plug the expression of $\Sigma_{+}(s)$ [$\Sigma_{-}(s)$]
into the transcendental equation \eqref{TranscendentalEQ} and replace $s$ by $-\mathrm{i}\omega_{+}$ ($-\mathrm{i}\omega_{-}$). After dividing the resulting equations into real and imaginary parts, we obtain the following equations.

(i) For both the separate and braided configurations, the equations satisfied by $\omega_{+}$ are of the form 
\begin{subequations}
	\begin{equation}
	\frac{\sin^2\left(N\omega_{+}\tau\right)}{\sin^2\left(\frac{1}{2}\omega_{+}\tau\right)}=0,
	\label{SepBraSupRe}
	\end{equation}
	\begin{equation}
	\frac{N}{2}\cot\left(\frac{\omega_{+}\tau}{2}\right)-\frac{\sin\left(2N\omega_{+}\tau\right)}{8\sin^2\left(\frac{1}{2}\omega_{+}\tau\right)}=\frac{\omega_{+}-\Omega}{\gamma}.
	\label{SepBraSupIm}
	\end{equation}
\end{subequations}

(ii) For the separate configuration, the equations satisfied by $\omega_{-}$ take the form    
\begin{subequations}	
	\begin{equation}
	\frac{\sin^4\left(\frac{1}{2}N\omega_{-}\tau\right)}{\sin^2\left(\frac{1}{2}\omega_{-}\tau\right)}=0,
	\label{SepSubRe}
	\end{equation}
	\begin{eqnarray}
        \frac{N}{2}\cot\left(\frac{\omega_{-}\tau}{2}\right)+\frac{\frac{1}{2}\sin\left(2N\omega_{-}\tau\right)-2\sin\left(N\omega_{-}\tau\right)}{4\sin^2\left(\frac{1}{2}\omega_{-}\tau\right)}
        =\frac{\omega_{-}-\Omega}{\gamma}.
        \nonumber
        \\
	\label{SepSubIm}
	\end{eqnarray}
\end{subequations}

(iii) For the braided configuration, the equations satisfied by $\omega_{-}$ take the form     
\begin{subequations}
	\begin{equation}
	\frac{\sin^2\left(N\omega_{-}\tau\right)}{\cos^2\left(\frac{1}{2}\omega_{-}\tau\right)}=0,
	\label{BraSubRe}
	\end{equation}
	\begin{equation}
	-\frac{N}{2}\tan\left(\frac{\omega_{-}\tau}{2}\right)-\frac{\sin\left(2N\omega_{-}\tau\right)}{8\cos^2\left(\frac{1}{2}\omega_{-}\tau\right)}=\frac{\omega_{-}-\Omega}{\gamma}.
	\label{BraSubIm}
	\end{equation}
\end{subequations}

The solutions $\omega_{+,n}$ ($\omega_{-,n}$) to Eq.~\eqref{SepBraSupRe} [\eqref{SepSubRe} or \eqref{BraSubRe}] represent the 
frequencies of the symmetric (antisymmetric) dark states. Subsequently, by substituting these solutions into Eq.~\eqref{SepBraSupIm} [\eqref{SepSubIm} or  
\eqref{BraSubIm}], the symmetric (antisymmetric) dark-state conditions satisfied by the parameters $\Omega$ and $\gamma$ can be obtained. Additionally, substituting the values 
$s_{+,n}=-\mathrm{i}\omega_{+,n}$ ($s_{-,n}=-\mathrm{i}\omega_{-,n}$) into Eq.~\eqref{AmplitudeFun}, the corresponding atomic excitation amplitude can be obtained. The results are summarized as follows.

(i) For both the separate and braided configurations, the frequencies of the symmetric dark modes take the form
\begin{equation}
	\omega_{+,n}=\frac{n\pi}{N\tau}\equiv\omega_{n},~~~n\in\mathbb{Z}^{+},~{n}/{N}\notin\mathbb{E}.
	\label{DarkFrequencySepBraP}
\end{equation}
Corresponding to the mode with frequenccy $\omega_{n}$, the  dark-state condition satisfied by the parameters $\Omega$ and $\gamma$ is  
\begin{equation}
	\Omega+\frac{1}{2}N\gamma\mathrm{cot}\left(\frac{1}{2}\omega_{n}\tau\right)=\omega_{n},
	\label{DarkConditionSepBraP}
\end{equation}
and the atomic excitation amplitude is
\begin{equation}
	A_{n}=\left[{1+\tfrac{1}{2}N\gamma\tau\csc^2\left(\tfrac{1}{2}\omega_{n}\tau\right)}\right]^{-1}.
	\label{DarkAmpSepBraP}
\end{equation}

(ii) For the separate configuration, the frequencies of the antisymmetric dark modes are
\begin{equation}
	\omega_{-,n}=\omega_{n},~~~n\in\mathbb{E}^{+}.
	\label{DarkFrequencySepM}
\end{equation}
When ${n}/{N}\notin\mathbb{E}$, the dark-state condition 
and the atomic amplitude $A_{n}$ for the mode $\omega_{n}$  take the same form as in 
Eqs.~\eqref{DarkConditionSepBraP} and \eqref{DarkAmpSepBraP}, respectively.
And when ${n}/{N}=2m$ ($m\in\mathbb{Z}^{+}$), the dark-state condition is
\begin{equation}
	\Omega\tau=2m\pi,
\end{equation}
with corresponding atomic excitation amplitude 
\begin{equation}
	A_n=\left[{1+\tfrac{1}{6}N\left(2N^2+1\right)\gamma\tau}\right]^{-1}.
\end{equation}

(iii) For the braided configuration, the frequencies of  the antisymmetric dark modes are	
\begin{equation}
      \omega_{-,n}=\omega_{n},~~n\in\mathbb{Z}^{+},{n}/{N}\notin\mathbb{O}.
	\label{DarkFrequencyBraM}
\end{equation}
The corresponding dark-state condition is
\begin{equation}
	\Omega-\frac{1}{2}N\gamma\tan\left(\frac{1}{2}\omega_{n}\tau\right)=\omega_{n},
	\label{DarkConditionBraM}
\end{equation}
and the atomic excitation amplitude is
\begin{equation}
	A_{n}=\left[{1+\tfrac{1}{2}N\gamma\tau\sec^2\left(\tfrac{1}{2}\omega_{n}\tau\right)}\right]^{-1}.
\end{equation}

The preceding results can be unified through the periodicity of trigonometric functions. By setting $n=2lN+q$ ($l=0,1,2,\cdots$) to rewrite the terms $\cot(\omega_{n}\tau/2)$ and $\csc(\omega_{n}\tau/2)$ in the cases (i) and (ii), and by setting $n=(2l-1)N+q$ to rewrite the terms $\tan(\omega_{n}\tau/2)$ and $\sec(\omega_{n}\tau/2)$ in the case (iii), one can finally obtain the results in Sec.~\ref{DSCondition} in the main text.
\section{\label{ParameterOBSGeneral}Conditions for oscillating bound states}
The intersection 
point of (at least) two $\Omega\text{-}\gamma$ lines $\omega_{n_1}$ and $\omega_{n_2}$ ($n_1<n_2$) gives rise 
to the parameter point of the oscillating bound state. According to Eq.~\eqref{DarkConditionGeneral}, 
the required parameters $\Omega$ and $\gamma$ are
\begin{equation}
	\Omega=\bar{\omega}-\frac{\sin(\omega_{\bar{q}}\tau)}{\sin(\omega_{\tilde{q}}\tau)}\tilde{\omega},
	~~\gamma=\frac{2\tilde{\omega}}{N}\frac{\cos(\omega_{\tilde{q}}\tau)-\cos(\omega_{\bar{q}}\tau)}{\sin(\omega_{\tilde{q}}\tau)},
\end{equation}
with 
\begin{subequations}
\begin{equation}
\omega_{\bar{q}}=\frac{\omega_{q_1}+\omega_{q_2}}{2}=\frac{\bar{q}\pi}{N\tau},
\end{equation}
\begin{equation}
\omega_{\tilde{q}}=\frac{\omega_{q_1}-\omega_{q_2}}{2}=\frac{\tilde{q}\pi}{N\tau},
\end{equation}
\begin{equation}
\bar{\omega}=\frac{\omega_{n_1}+\omega_{n_2}}{2}=\frac{r\pi}{\tau}+\omega_{\bar{q}},
\end{equation}
\begin{equation}
\tilde{\omega}=\frac{\omega_{n_2}-\omega_{n_1}}{2}=\frac{p\pi}{\tau}-\omega_{\tilde{q}}.
\end{equation}
\end{subequations}
Here $\bar{q}=(q_1+q_2)/{2}$ and $\tilde{q}=(q_1-q_2)/{2}$ are satisfied, where $q_1$ and $q_2$ are determined by $n_1$ and $n_2$ via Eq.~\eqref{nANDq}. $q_1>q_2$ is required to ensure that the lines labeled by $q_1$ and $q_2$ intersect in the region 
$\gamma>0$. By defining
\begin{equation}
L_{i}\equiv\left\lfloor\frac{n_i}{2N}\right\rfloor,~~~\tilde{L}_{i}\equiv\left\lfloor\frac{n_i+N}{2N}\right\rfloor,
\label{Lparameter}
\end{equation}
the rules to fix the values for $r$ and $p$ can be summarized as follows:  

(i) If both  $n_1$ and $n_2$ ($n_1<n_2$) represent modes belonging to the same category among $\text{S}_{+}$, $\text{B}_{+}$, and $\text{S}_{-}$, we have  
\begin{equation}
r=L_1+L_2,~~~p=L_2-L_1. 
\label{rp1}
\end{equation}
Note that for the separate configuration, the set of intersection points of the $\Omega\text{-}\gamma$ lines associated with symmetric dark modes $\text{S}_{+}$ contains all possible parameter points for oscillating bound states, since any $\text{S}_{-}$ line either coincides with an $\text{S}_{+}$ line or passes through the intersection
point of two $\text{S}_{+}$ lines [see Figs.~\ref{DarkStateCondition}(a) and \ref{DarkStateCondition}(b) for the cases of $N=2$ and $N=3$].

(ii) For the braided configuration, if both $n_1$ and $n_2$ ($n_1<n_2$) represent antisymmetric dark modes  ($\text{B}_{-}$ modes), we have
\begin{equation}
r=\tilde{L}_1+\tilde{L}_2-1,~·~p=\tilde{L}_2-\tilde{L}_1.
\label{rp2}
\end{equation}
This rule can be employed to determine the intersection points of the dashed lines in Figs.~\ref{DarkStateCondition}(c) and \ref{DarkStateCondition}(d) (for the cases of $N=2$ and $N=3$).

(iii) For the braided configuration, if $n_1$ and $n_2$ ($n_1<n_2$) correspond to a symmetric and an antisymmetric dark modes ($\text{B}_{+}$ and $\text{B}_{-}$ modes), respectively, we have
\begin{equation}
r=L_1+\tilde{L}_2-\frac{1}{2},~~~p=\tilde{L}_2-L_1-\frac{1}{2}.
\label{rp3}
\end{equation}
In contrast, if $n_1$ represents an antisymmetric dark mode and $n_2$ a symmetric one, we obtain
\begin{equation}
r=\tilde{L}_1+L_2-\frac{1}{2},~~~p=L_2-\tilde{L}_1+\frac{1}{2}.
\label{rp4}
\end{equation}
The rules given by Eqs.~\eqref{rp3} and \eqref{rp4} can be employed to determine the intersection points between the solid and dashed lines in Figs.~\ref{DarkStateCondition}(c) and \ref{DarkStateCondition}(d) (for the cases of $N=2$ and $N=3$).

Here we are interested in the case of $\bar{q}=N$, i.e., $q_1=N+\tilde{q}$, $q_2=N-\tilde{q}$. The parameters $\Omega$ and $\gamma$ under 
these conditions should be
\begin{equation}
	\Omega=\bar{\omega}=\frac{m\pi}{\tau},
	~~~\gamma=\frac{2\tilde{\omega}}{N}\cot\left(\frac{1}{2}\omega_{\tilde{q}}\tau\right),
\end{equation}
with $m=r+1$. According to Eq.~\eqref{Slope}, the slopes of the pair of lines are negatives of each other, with $K_{N-\tilde{q}}=-K_{N+\tilde{q}}$. Thus they are symmetric with respect to the vertical line $\Omega\tau=m\pi$, representing
dark modes with frequencies $\omega_{n_{1,2}}=\bar{\omega}\mp\tilde{\omega}=(m\mp p)\pi/\tau\pm\omega_{\tilde{q}}$ [i.e., with
indices $n_{1,2}=(m\mp p)N\pm\tilde{q}$].
According to Eqs.~\eqref{Lparameter}-\eqref{rp4}, one can derive both the allowable values of 
$m$ and, for each specified $m$, the corresponding values of 
$p$. Additionally, the value of $\tilde{q}$ can be determined by Eqs.~\eqref{nANDq} and \eqref{tilde_q}. 
The corresponding results are summarized in Eqs.~\eqref{mpq1}-\eqref{mpq3} in the main text. 

\bibliography{MS-LFJ-06-01-2025}

\providecommand{\noopsort}[1]{}\providecommand{\singleletter}[1]{#1}%
\begin{thebibliography}{52}%
\makeatletter
\providecommand \@ifxundefined [1]{%
 \@ifx{#1\undefined}
}%
\providecommand \@ifnum [1]{%
 \ifnum #1\expandafter \@firstoftwo
 \else \expandafter \@secondoftwo
 \fi
}%
\providecommand \@ifx [1]{%
 \ifx #1\expandafter \@firstoftwo
 \else \expandafter \@secondoftwo
 \fi
}%
\providecommand \natexlab [1]{#1}%
\providecommand \enquote  [1]{``#1''}%
\providecommand \bibnamefont  [1]{#1}%
\providecommand \bibfnamefont [1]{#1}%
\providecommand \citenamefont [1]{#1}%
\providecommand \href@noop [0]{\@secondoftwo}%
\providecommand \href [0]{\begingroup \@sanitize@url \@href}%
\providecommand \@href[1]{\@@startlink{#1}\@@href}%
\providecommand \@@href[1]{\endgroup#1\@@endlink}%
\providecommand \@sanitize@url [0]{\catcode `\\12\catcode `\$12\catcode
  `\&12\catcode `\#12\catcode `\^12\catcode `\_12\catcode `\%12\relax}%
\providecommand \@@startlink[1]{}%
\providecommand \@@endlink[0]{}%
\providecommand \url  [0]{\begingroup\@sanitize@url \@url }%
\providecommand \@url [1]{\endgroup\@href {#1}{\urlprefix }}%
\providecommand \urlprefix  [0]{URL }%
\providecommand \Eprint [0]{\href }%
\providecommand \doibase [0]{https://doi.org/}%
\providecommand \selectlanguage [0]{\@gobble}%
\providecommand \bibinfo  [0]{\@secondoftwo}%
\providecommand \bibfield  [0]{\@secondoftwo}%
\providecommand \translation [1]{[#1]}%
\providecommand \BibitemOpen [0]{}%
\providecommand \bibitemStop [0]{}%
\providecommand \bibitemNoStop [0]{.\EOS\space}%
\providecommand \EOS [0]{\spacefactor3000\relax}%
\providecommand \BibitemShut  [1]{\csname bibitem#1\endcsname}%
\let\auto@bib@innerbib\@empty
\bibitem [{\citenamefont {Kockum}(2021)}]{Kockum-MI2021}%
  \BibitemOpen
  \bibfield  {author} {\bibinfo {author} {\bibfnamefont {A.~F.}\ \bibnamefont
  {Kockum}},\ }\bibfield  {title} {\bibinfo {title} {Quantum optics with giant
  atoms—the first five years},\ }in\ \href
  {https://doi.org/10.1007/978-981-15-5191-8_12} {\emph {\bibinfo {booktitle}
  {International Symposium on Mathematics, Quantum Theory, and
  Cryptography}}},\ \bibinfo {series and number} {Mathematics for Industry},\
  \bibinfo {editor} {edited by\ \bibinfo {editor} {\bibfnamefont
  {T.}~\bibnamefont {Takagi}}, \bibinfo {editor} {\bibfnamefont
  {M.}~\bibnamefont {Wakayama}}, \bibinfo {editor} {\bibfnamefont
  {K.}~\bibnamefont {Tanaka}}, \bibinfo {editor} {\bibfnamefont
  {N.}~\bibnamefont {Kunihiro}}, \bibinfo {editor} {\bibfnamefont
  {K.}~\bibnamefont {Kimoto}},\ and\ \bibinfo {editor} {\bibfnamefont
  {Y.}~\bibnamefont {Ikematsu}}}\ (\bibinfo  {publisher} {Springer Singapore},\
  \bibinfo {address} {Singapore},\ \bibinfo {year} {2021})\ pp.\ \bibinfo
  {pages} {125--146}\BibitemShut {NoStop}%
\bibitem [{\citenamefont {Roy}\ \emph {et~al.}(2017)\citenamefont {Roy},
  \citenamefont {Wilson},\ and\ \citenamefont {Firstenberg}}]{Roy-RMP2017}%
  \BibitemOpen
  \bibfield  {author} {\bibinfo {author} {\bibfnamefont {D.}~\bibnamefont
  {Roy}}, \bibinfo {author} {\bibfnamefont {C.~M.}\ \bibnamefont {Wilson}},\
  and\ \bibinfo {author} {\bibfnamefont {O.}~\bibnamefont {Firstenberg}},\
  }\bibfield  {title} {\bibinfo {title} {Colloquium: Strongly interacting
  photons in one-dimensional continuum},\ }\href
  {https://doi.org/10.1103/RevModPhys.89.021001} {\bibfield  {journal}
  {\bibinfo  {journal} {Rev. Mod. Phys.}\ }\textbf {\bibinfo {volume} {89}},\
  \bibinfo {pages} {021001} (\bibinfo {year} {2017})}\BibitemShut {NoStop}%
\bibitem [{\citenamefont {Gu}\ \emph {et~al.}(2017)\citenamefont {Gu},
  \citenamefont {Kockum}, \citenamefont {Miranowicz}, \citenamefont {Liu},\
  and\ \citenamefont {Nori}}]{Gu-PhysReports2017}%
  \BibitemOpen
  \bibfield  {author} {\bibinfo {author} {\bibfnamefont {X.}~\bibnamefont
  {Gu}}, \bibinfo {author} {\bibfnamefont {A.~F.}\ \bibnamefont {Kockum}},
  \bibinfo {author} {\bibfnamefont {A.}~\bibnamefont {Miranowicz}}, \bibinfo
  {author} {\bibfnamefont {Y.-x.}\ \bibnamefont {Liu}},\ and\ \bibinfo {author}
  {\bibfnamefont {F.}~\bibnamefont {Nori}},\ }\bibfield  {title} {\bibinfo
  {title} {Microwave photonics with superconducting quantum circuits},\ }\href
  {https://doi.org/https://doi.org/10.1016/j.physrep.2017.10.002} {\bibfield
  {journal} {\bibinfo  {journal} {Phys. Rep.}\ }\textbf {\bibinfo {volume}
  {718-719}},\ \bibinfo {pages} {1 } (\bibinfo {year} {2017})}\BibitemShut
  {NoStop}%
\bibitem [{\citenamefont {Sheremet}\ \emph {et~al.}(2023)\citenamefont
  {Sheremet}, \citenamefont {Petrov}, \citenamefont {Iorsh}, \citenamefont
  {Poshakinskiy},\ and\ \citenamefont {Poddubny}}]{Sheremet-RMP2023}%
  \BibitemOpen
  \bibfield  {author} {\bibinfo {author} {\bibfnamefont {A.~S.}\ \bibnamefont
  {Sheremet}}, \bibinfo {author} {\bibfnamefont {M.~I.}\ \bibnamefont
  {Petrov}}, \bibinfo {author} {\bibfnamefont {I.~V.}\ \bibnamefont {Iorsh}},
  \bibinfo {author} {\bibfnamefont {A.~V.}\ \bibnamefont {Poshakinskiy}},\ and\
  \bibinfo {author} {\bibfnamefont {A.~N.}\ \bibnamefont {Poddubny}},\
  }\bibfield  {title} {\bibinfo {title} {Waveguide quantum electrodynamics:
  Collective radiance and photon-photon correlations},\ }\href
  {https://doi.org/10.1103/RevModPhys.95.015002} {\bibfield  {journal}
  {\bibinfo  {journal} {Rev. Mod. Phys.}\ }\textbf {\bibinfo {volume} {95}},\
  \bibinfo {pages} {015002} (\bibinfo {year} {2023})}\BibitemShut {NoStop}%
\bibitem [{\citenamefont {Andersson}\ \emph {et~al.}(2019)\citenamefont
  {Andersson}, \citenamefont {Suri}, \citenamefont {Guo}, \citenamefont
  {Aref},\ and\ \citenamefont {Delsing}}]{Andersson-NatPhys2019}%
  \BibitemOpen
  \bibfield  {author} {\bibinfo {author} {\bibfnamefont {G.}~\bibnamefont
  {Andersson}}, \bibinfo {author} {\bibfnamefont {B.}~\bibnamefont {Suri}},
  \bibinfo {author} {\bibfnamefont {L.}~\bibnamefont {Guo}}, \bibinfo {author}
  {\bibfnamefont {T.}~\bibnamefont {Aref}},\ and\ \bibinfo {author}
  {\bibfnamefont {P.}~\bibnamefont {Delsing}},\ }\bibfield  {title} {\bibinfo
  {title} {Non-exponential decay of a giant artificial atom},\ }\href
  {https://doi.org/10.1038/s41567-019-0605-6} {\bibfield  {journal} {\bibinfo
  {journal} {Nat. Phys.}\ }\textbf {\bibinfo {volume} {15}},\ \bibinfo {pages}
  {1123} (\bibinfo {year} {2019})}\BibitemShut {NoStop}%
\bibitem [{\citenamefont {Kannan}\ \emph {et~al.}(2020)\citenamefont {Kannan},
  \citenamefont {Ruckriegel}, \citenamefont {Campbell}, \citenamefont {Kockum},
  \citenamefont {Braum\"uller}, \citenamefont {Kim}, \citenamefont
  {Kjaergaard}, \citenamefont {Krantz}, \citenamefont {Melville}, \citenamefont
  {Niedzielski}, \citenamefont {Veps\"al\"ainen}, \citenamefont {Winik},
  \citenamefont {Yoder}, \citenamefont {Nori}, \citenamefont {Orlando},
  \citenamefont {Gustavsson},\ and\ \citenamefont
  {Oliver}}]{Kannan-Nature2020}%
  \BibitemOpen
  \bibfield  {author} {\bibinfo {author} {\bibfnamefont {B.}~\bibnamefont
  {Kannan}}, \bibinfo {author} {\bibfnamefont {M.~J.}\ \bibnamefont
  {Ruckriegel}}, \bibinfo {author} {\bibfnamefont {D.~L.}\ \bibnamefont
  {Campbell}}, \bibinfo {author} {\bibfnamefont {A.~F.}\ \bibnamefont
  {Kockum}}, \bibinfo {author} {\bibfnamefont {J.}~\bibnamefont
  {Braum\"uller}}, \bibinfo {author} {\bibfnamefont {D.~K.}\ \bibnamefont
  {Kim}}, \bibinfo {author} {\bibfnamefont {M.}~\bibnamefont {Kjaergaard}},
  \bibinfo {author} {\bibfnamefont {P.}~\bibnamefont {Krantz}}, \bibinfo
  {author} {\bibfnamefont {A.}~\bibnamefont {Melville}}, \bibinfo {author}
  {\bibfnamefont {B.~M.}\ \bibnamefont {Niedzielski}}, \bibinfo {author}
  {\bibfnamefont {A.}~\bibnamefont {Veps\"al\"ainen}}, \bibinfo {author}
  {\bibfnamefont {R.}~\bibnamefont {Winik}}, \bibinfo {author} {\bibfnamefont
  {J.~L.}\ \bibnamefont {Yoder}}, \bibinfo {author} {\bibfnamefont
  {F.}~\bibnamefont {Nori}}, \bibinfo {author} {\bibfnamefont {T.~P.}\
  \bibnamefont {Orlando}}, \bibinfo {author} {\bibfnamefont {S.}~\bibnamefont
  {Gustavsson}},\ and\ \bibinfo {author} {\bibfnamefont {W.~D.}\ \bibnamefont
  {Oliver}},\ }\bibfield  {title} {\bibinfo {title} {Waveguide quantum
  electrodynamics with superconducting artificial giant atoms},\ }\href
  {https://doi.org/10.1038/s41586-020-2529-9} {\bibfield  {journal} {\bibinfo
  {journal} {Nature (London)}\ }\textbf {\bibinfo {volume} {583}},\ \bibinfo
  {pages} {775} (\bibinfo {year} {2020})}\BibitemShut {NoStop}%
\bibitem [{\citenamefont {Wang}\ \emph {et~al.}(2022)\citenamefont {Wang},
  \citenamefont {Wang}, \citenamefont {Yao}, \citenamefont {Shen},
  \citenamefont {Wu}, \citenamefont {Qian}, \citenamefont {Li}, \citenamefont
  {Zhu},\ and\ \citenamefont {You}}]{Wang-Natcom2022}%
  \BibitemOpen
  \bibfield  {author} {\bibinfo {author} {\bibfnamefont {Z.-Q.}\ \bibnamefont
  {Wang}}, \bibinfo {author} {\bibfnamefont {Y.-P.}\ \bibnamefont {Wang}},
  \bibinfo {author} {\bibfnamefont {J.}~\bibnamefont {Yao}}, \bibinfo {author}
  {\bibfnamefont {R.-C.}\ \bibnamefont {Shen}}, \bibinfo {author}
  {\bibfnamefont {W.-J.}\ \bibnamefont {Wu}}, \bibinfo {author} {\bibfnamefont
  {J.}~\bibnamefont {Qian}}, \bibinfo {author} {\bibfnamefont {J.}~\bibnamefont
  {Li}}, \bibinfo {author} {\bibfnamefont {S.-Y.}\ \bibnamefont {Zhu}},\ and\
  \bibinfo {author} {\bibfnamefont {J.~Q.}\ \bibnamefont {You}},\ }\bibfield
  {title} {\bibinfo {title} {Giant spin ensembles in waveguide magnonics},\
  }\href {https://doi.org/10.1038/s41467-022-35174-9} {\bibfield  {journal}
  {\bibinfo  {journal} {Nat. Commun.}\ }\textbf {\bibinfo {volume} {13}},\
  \bibinfo {pages} {7580} (\bibinfo {year} {2022})}\BibitemShut {NoStop}%
\bibitem [{\citenamefont {Kockum}\ \emph {et~al.}(2014)\citenamefont {Kockum},
  \citenamefont {Delsing},\ and\ \citenamefont {Johansson}}]{Kockum-PRA2014}%
  \BibitemOpen
  \bibfield  {author} {\bibinfo {author} {\bibfnamefont {A.~F.}\ \bibnamefont
  {Kockum}}, \bibinfo {author} {\bibfnamefont {P.}~\bibnamefont {Delsing}},\
  and\ \bibinfo {author} {\bibfnamefont {G.}~\bibnamefont {Johansson}},\
  }\bibfield  {title} {\bibinfo {title} {Designing frequency-dependent
  relaxation rates and lamb shifts for a giant artificial atom},\ }\href
  {https://doi.org/10.1103/PhysRevA.90.013837} {\bibfield  {journal} {\bibinfo
  {journal} {Phys. Rev. A}\ }\textbf {\bibinfo {volume} {90}},\ \bibinfo
  {pages} {013837} (\bibinfo {year} {2014})}\BibitemShut {NoStop}%
\bibitem [{\citenamefont {Kockum}\ \emph {et~al.}(2018)\citenamefont {Kockum},
  \citenamefont {Johansson},\ and\ \citenamefont {Nori}}]{Kockum-PRL2018}%
  \BibitemOpen
  \bibfield  {author} {\bibinfo {author} {\bibfnamefont {A.~F.}\ \bibnamefont
  {Kockum}}, \bibinfo {author} {\bibfnamefont {G.}~\bibnamefont {Johansson}},\
  and\ \bibinfo {author} {\bibfnamefont {F.}~\bibnamefont {Nori}},\ }\bibfield
  {title} {\bibinfo {title} {Decoherence-free interaction between giant atoms
  in waveguide quantum electrodynamics},\ }\href
  {https://doi.org/10.1103/PhysRevLett.120.140404} {\bibfield  {journal}
  {\bibinfo  {journal} {Phys. Rev. Lett.}\ }\textbf {\bibinfo {volume} {120}},\
  \bibinfo {pages} {140404} (\bibinfo {year} {2018})}\BibitemShut {NoStop}%
\bibitem [{\citenamefont {Carollo}\ \emph {et~al.}(2020)\citenamefont
  {Carollo}, \citenamefont {Cilluffo},\ and\ \citenamefont
  {Ciccarello}}]{Carollo-PRR2020}%
  \BibitemOpen
  \bibfield  {author} {\bibinfo {author} {\bibfnamefont {A.}~\bibnamefont
  {Carollo}}, \bibinfo {author} {\bibfnamefont {D.}~\bibnamefont {Cilluffo}},\
  and\ \bibinfo {author} {\bibfnamefont {F.}~\bibnamefont {Ciccarello}},\
  }\bibfield  {title} {\bibinfo {title} {Mechanism of decoherence-free coupling
  between giant atoms},\ }\href
  {https://doi.org/10.1103/PhysRevResearch.2.043184} {\bibfield  {journal}
  {\bibinfo  {journal} {Phys. Rev. Res.}\ }\textbf {\bibinfo {volume} {2}},\
  \bibinfo {pages} {043184} (\bibinfo {year} {2020})}\BibitemShut {NoStop}%
\bibitem [{\citenamefont {Du}\ \emph {et~al.}(2023)\citenamefont {Du},
  \citenamefont {Guo},\ and\ \citenamefont {Li}}]{Du-PRA2023}%
  \BibitemOpen
  \bibfield  {author} {\bibinfo {author} {\bibfnamefont {L.}~\bibnamefont
  {Du}}, \bibinfo {author} {\bibfnamefont {L.}~\bibnamefont {Guo}},\ and\
  \bibinfo {author} {\bibfnamefont {Y.}~\bibnamefont {Li}},\ }\bibfield
  {title} {\bibinfo {title} {Complex decoherence-free interactions between
  giant atoms},\ }\href {https://doi.org/10.1103/PhysRevA.107.023705}
  {\bibfield  {journal} {\bibinfo  {journal} {Phys. Rev. A}\ }\textbf {\bibinfo
  {volume} {107}},\ \bibinfo {pages} {023705} (\bibinfo {year}
  {2023})}\BibitemShut {NoStop}%
\bibitem [{\citenamefont {Ingelsten}\ \emph {et~al.}(2024)\citenamefont
  {Ingelsten}, \citenamefont {Kockum},\ and\ \citenamefont
  {Soro}}]{Ingelsten-PRR2024}%
  \BibitemOpen
  \bibfield  {author} {\bibinfo {author} {\bibfnamefont {R.~E.}\ \bibnamefont
  {Ingelsten}}, \bibinfo {author} {\bibfnamefont {A.~F.}\ \bibnamefont
  {Kockum}},\ and\ \bibinfo {author} {\bibfnamefont {A.}~\bibnamefont {Soro}},\
  }\bibfield  {title} {\bibinfo {title} {Avoiding decoherence with giant atoms
  in a two-dimensional structured environment},\ }\href
  {https://doi.org/10.1103/PhysRevResearch.6.043222} {\bibfield  {journal}
  {\bibinfo  {journal} {Phys. Rev. Res.}\ }\textbf {\bibinfo {volume} {6}},\
  \bibinfo {pages} {043222} (\bibinfo {year} {2024})}\BibitemShut {NoStop}%
\bibitem [{\citenamefont {Guo}\ \emph {et~al.}(2017)\citenamefont {Guo},
  \citenamefont {Grimsmo}, \citenamefont {Kockum}, \citenamefont {Pletyukhov},\
  and\ \citenamefont {Johansson}}]{Guo-PRA2017}%
  \BibitemOpen
  \bibfield  {author} {\bibinfo {author} {\bibfnamefont {L.}~\bibnamefont
  {Guo}}, \bibinfo {author} {\bibfnamefont {A.}~\bibnamefont {Grimsmo}},
  \bibinfo {author} {\bibfnamefont {A.~F.}\ \bibnamefont {Kockum}}, \bibinfo
  {author} {\bibfnamefont {M.}~\bibnamefont {Pletyukhov}},\ and\ \bibinfo
  {author} {\bibfnamefont {G.}~\bibnamefont {Johansson}},\ }\bibfield  {title}
  {\bibinfo {title} {Giant acoustic atom: A single quantum system with a
  deterministic time delay},\ }\href
  {https://doi.org/10.1103/PhysRevA.95.053821} {\bibfield  {journal} {\bibinfo
  {journal} {Phys. Rev. A}\ }\textbf {\bibinfo {volume} {95}},\ \bibinfo
  {pages} {053821} (\bibinfo {year} {2017})}\BibitemShut {NoStop}%
\bibitem [{\citenamefont {Guo}\ \emph {et~al.}(2020{\natexlab{a}})\citenamefont
  {Guo}, \citenamefont {Kockum}, \citenamefont {Marquardt},\ and\ \citenamefont
  {Johansson}}]{Guo-PRR2020}%
  \BibitemOpen
  \bibfield  {author} {\bibinfo {author} {\bibfnamefont {L.}~\bibnamefont
  {Guo}}, \bibinfo {author} {\bibfnamefont {A.~F.}\ \bibnamefont {Kockum}},
  \bibinfo {author} {\bibfnamefont {F.}~\bibnamefont {Marquardt}},\ and\
  \bibinfo {author} {\bibfnamefont {G.}~\bibnamefont {Johansson}},\ }\bibfield
  {title} {\bibinfo {title} {Oscillating bound states for a giant atom},\
  }\href {https://doi.org/10.1103/PhysRevResearch.2.043014} {\bibfield
  {journal} {\bibinfo  {journal} {Phys. Rev. Res.}\ }\textbf {\bibinfo {volume}
  {2}},\ \bibinfo {pages} {043014} (\bibinfo {year}
  {2020}{\natexlab{a}})}\BibitemShut {NoStop}%
\bibitem [{\citenamefont {Xu}\ and\ \citenamefont {Guo}(2024)}]{Xu-NJP2024}%
  \BibitemOpen
  \bibfield  {author} {\bibinfo {author} {\bibfnamefont {L.}~\bibnamefont
  {Xu}}\ and\ \bibinfo {author} {\bibfnamefont {L.}~\bibnamefont {Guo}},\
  }\bibfield  {title} {\bibinfo {title} {Catch and release of propagating
  bosonic field with non-markovian giant atom},\ }\href
  {https://doi.org/10.1088/1367-2630/ad18ed} {\bibfield  {journal} {\bibinfo
  {journal} {New J. Phys.}\ }\textbf {\bibinfo {volume} {26}},\ \bibinfo
  {pages} {013025} (\bibinfo {year} {2024})}\BibitemShut {NoStop}%
\bibitem [{\citenamefont {Roccati}\ and\ \citenamefont
  {Cilluffo}(2024)}]{Roccati-PRL2024}%
  \BibitemOpen
  \bibfield  {author} {\bibinfo {author} {\bibfnamefont {F.}~\bibnamefont
  {Roccati}}\ and\ \bibinfo {author} {\bibfnamefont {D.}~\bibnamefont
  {Cilluffo}},\ }\bibfield  {title} {\bibinfo {title} {Controlling markovianity
  with chiral giant atoms},\ }\href
  {https://doi.org/10.1103/PhysRevLett.133.063603} {\bibfield  {journal}
  {\bibinfo  {journal} {Phys. Rev. Lett.}\ }\textbf {\bibinfo {volume} {133}},\
  \bibinfo {pages} {063603} (\bibinfo {year} {2024})}\BibitemShut {NoStop}%
\bibitem [{\citenamefont {Qiu}\ and\ \citenamefont {L\"u}(2024)}]{Qiu-PRR2024}%
  \BibitemOpen
  \bibfield  {author} {\bibinfo {author} {\bibfnamefont {Q.-Y.}\ \bibnamefont
  {Qiu}}\ and\ \bibinfo {author} {\bibfnamefont {X.-Y.}\ \bibnamefont {L\"u}},\
  }\bibfield  {title} {\bibinfo {title} {Non-markovian collective emission of
  giant emitters in the zeno regime},\ }\href
  {https://doi.org/10.1103/PhysRevResearch.6.033243} {\bibfield  {journal}
  {\bibinfo  {journal} {Phys. Rev. Res.}\ }\textbf {\bibinfo {volume} {6}},\
  \bibinfo {pages} {033243} (\bibinfo {year} {2024})}\BibitemShut {NoStop}%
\bibitem [{\citenamefont {Yannopapas}(2025)}]{Yannopapas-Photonics2025}%
  \BibitemOpen
  \bibfield  {author} {\bibinfo {author} {\bibfnamefont {V.}~\bibnamefont
  {Yannopapas}},\ }\bibfield  {title} {\bibinfo {title} {Entanglement dynamics
  of two giant atoms embedded in a one-dimensional photonic lattice with a
  synthetic gauge field},\ }\href {https://doi.org/10.3390/photonics12060612}
  {\bibfield  {journal} {\bibinfo  {journal} {Photonics}\ }\textbf {\bibinfo
  {volume} {12}},\ \bibinfo {pages} {612} (\bibinfo {year} {2025})}\BibitemShut
  {NoStop}%
\bibitem [{\citenamefont {Sivakumar}\ \emph {et~al.}(2025)\citenamefont
  {Sivakumar}, \citenamefont {Liu},\ and\ \citenamefont
  {Zubairy}}]{Sivakumar-PRA2025}%
  \BibitemOpen
  \bibfield  {author} {\bibinfo {author} {\bibfnamefont {A.~V.}\ \bibnamefont
  {Sivakumar}}, \bibinfo {author} {\bibfnamefont {J.}~\bibnamefont {Liu}},\
  and\ \bibinfo {author} {\bibfnamefont {M.~S.}\ \bibnamefont {Zubairy}},\
  }\bibfield  {title} {\bibinfo {title} {Single-photon trapping in a giant-atom
  cavity coupled to a one-dimensional waveguide},\ }\href
  {https://doi.org/10.1103/PhysRevA.111.033714} {\bibfield  {journal} {\bibinfo
   {journal} {Phys. Rev. A}\ }\textbf {\bibinfo {volume} {111}},\ \bibinfo
  {pages} {033714} (\bibinfo {year} {2025})}\BibitemShut {NoStop}%
\bibitem [{\citenamefont {John}\ and\ \citenamefont
  {Wang}(1990)}]{John-PRL1990}%
  \BibitemOpen
  \bibfield  {author} {\bibinfo {author} {\bibfnamefont {S.}~\bibnamefont
  {John}}\ and\ \bibinfo {author} {\bibfnamefont {J.}~\bibnamefont {Wang}},\
  }\bibfield  {title} {\bibinfo {title} {Quantum electrodynamics near a
  photonic band gap: Photon bound states and dressed atoms},\ }\href
  {https://doi.org/10.1103/PhysRevLett.64.2418} {\bibfield  {journal} {\bibinfo
   {journal} {Phys. Rev. Lett.}\ }\textbf {\bibinfo {volume} {64}},\ \bibinfo
  {pages} {2418} (\bibinfo {year} {1990})}\BibitemShut {NoStop}%
\bibitem [{\citenamefont {Tong}\ \emph {et~al.}(2010)\citenamefont {Tong},
  \citenamefont {An}, \citenamefont {Luo},\ and\ \citenamefont
  {Oh}}]{Tong-JPhyB2010}%
  \BibitemOpen
  \bibfield  {author} {\bibinfo {author} {\bibfnamefont {Q.-J.}\ \bibnamefont
  {Tong}}, \bibinfo {author} {\bibfnamefont {J.-H.}\ \bibnamefont {An}},
  \bibinfo {author} {\bibfnamefont {H.-G.}\ \bibnamefont {Luo}},\ and\ \bibinfo
  {author} {\bibfnamefont {C.~H.}\ \bibnamefont {Oh}},\ }\bibfield  {title}
  {\bibinfo {title} {Decoherence suppression of a dissipative qubit by the
  non-markovian effect},\ }\href
  {https://doi.org/10.1088/0953-4075/43/15/155501} {\bibfield  {journal}
  {\bibinfo  {journal} {J. Phys. B}\ }\textbf {\bibinfo {volume} {43}},\
  \bibinfo {pages} {155501} (\bibinfo {year} {2010})}\BibitemShut {NoStop}%
\bibitem [{\citenamefont {Calaj\'o}\ \emph {et~al.}(2016)\citenamefont
  {Calaj\'o}, \citenamefont {Ciccarello}, \citenamefont {Chang},\ and\
  \citenamefont {Rabl}}]{Calajo-PRA2016}%
  \BibitemOpen
  \bibfield  {author} {\bibinfo {author} {\bibfnamefont {G.}~\bibnamefont
  {Calaj\'o}}, \bibinfo {author} {\bibfnamefont {F.}~\bibnamefont
  {Ciccarello}}, \bibinfo {author} {\bibfnamefont {D.}~\bibnamefont {Chang}},\
  and\ \bibinfo {author} {\bibfnamefont {P.}~\bibnamefont {Rabl}},\ }\bibfield
  {title} {\bibinfo {title} {Atom-field dressed states in slow-light waveguide
  qed},\ }\href {https://doi.org/10.1103/PhysRevA.93.033833} {\bibfield
  {journal} {\bibinfo  {journal} {Phys. Rev. A}\ }\textbf {\bibinfo {volume}
  {93}},\ \bibinfo {pages} {033833} (\bibinfo {year} {2016})}\BibitemShut
  {NoStop}%
\bibitem [{\citenamefont {Shi}\ \emph {et~al.}(2016)\citenamefont {Shi},
  \citenamefont {Wu}, \citenamefont {Gonz\'alez-Tudela},\ and\ \citenamefont
  {Cirac}}]{Shi-PRX2016}%
  \BibitemOpen
  \bibfield  {author} {\bibinfo {author} {\bibfnamefont {T.}~\bibnamefont
  {Shi}}, \bibinfo {author} {\bibfnamefont {Y.-H.}\ \bibnamefont {Wu}},
  \bibinfo {author} {\bibfnamefont {A.}~\bibnamefont {Gonz\'alez-Tudela}},\
  and\ \bibinfo {author} {\bibfnamefont {J.~I.}\ \bibnamefont {Cirac}},\
  }\bibfield  {title} {\bibinfo {title} {Bound states in boson impurity
  models},\ }\href {https://doi.org/10.1103/PhysRevX.6.021027} {\bibfield
  {journal} {\bibinfo  {journal} {Phys. Rev. X}\ }\textbf {\bibinfo {volume}
  {6}},\ \bibinfo {pages} {021027} (\bibinfo {year} {2016})}\BibitemShut
  {NoStop}%
\bibitem [{\citenamefont {Hood}\ \emph {et~al.}(2016)\citenamefont {Hood},
  \citenamefont {Goban}, \citenamefont {Asenjo-Garcia}, \citenamefont {Lu},
  \citenamefont {Yu}, \citenamefont {Chang},\ and\ \citenamefont
  {Kimble}}]{Hood-PNAS2016}%
  \BibitemOpen
  \bibfield  {author} {\bibinfo {author} {\bibfnamefont {J.~D.}\ \bibnamefont
  {Hood}}, \bibinfo {author} {\bibfnamefont {A.}~\bibnamefont {Goban}},
  \bibinfo {author} {\bibfnamefont {A.}~\bibnamefont {Asenjo-Garcia}}, \bibinfo
  {author} {\bibfnamefont {M.}~\bibnamefont {Lu}}, \bibinfo {author}
  {\bibfnamefont {S.-P.}\ \bibnamefont {Yu}}, \bibinfo {author} {\bibfnamefont
  {D.~E.}\ \bibnamefont {Chang}},\ and\ \bibinfo {author} {\bibfnamefont
  {H.~J.}\ \bibnamefont {Kimble}},\ }\bibfield  {title} {\bibinfo {title}
  {Atom–atom interactions around the band edge of a photonic crystal
  waveguide},\ }\href {https://doi.org/10.1073/pnas.1603788113} {\bibfield
  {journal} {\bibinfo  {journal} {Proc. Natl. Acad. Sci. USA}\ }\textbf
  {\bibinfo {volume} {113}},\ \bibinfo {pages} {10507} (\bibinfo {year}
  {2016})}\BibitemShut {NoStop}%
\bibitem [{\citenamefont {Liu}\ and\ \citenamefont
  {Houck}(2017)}]{Liu-NatPhy2017}%
  \BibitemOpen
  \bibfield  {author} {\bibinfo {author} {\bibfnamefont {Y.}~\bibnamefont
  {Liu}}\ and\ \bibinfo {author} {\bibfnamefont {A.~A.}\ \bibnamefont
  {Houck}},\ }\bibfield  {title} {\bibinfo {title} {Quantum electrodynamics
  near a photonic bandgap},\ }\href {https://doi.org/10.1038/nphys3834}
  {\bibfield  {journal} {\bibinfo  {journal} {Nat. Phys}\ }\textbf {\bibinfo
  {volume} {13}},\ \bibinfo {pages} {48} (\bibinfo {year} {2017})}\BibitemShut
  {NoStop}%
\bibitem [{\citenamefont {Sundaresan}\ \emph {et~al.}(2019)\citenamefont
  {Sundaresan}, \citenamefont {Lundgren}, \citenamefont {Zhu}, \citenamefont
  {Gorshkov},\ and\ \citenamefont {Houck}}]{Sundaresan-PRX2019}%
  \BibitemOpen
  \bibfield  {author} {\bibinfo {author} {\bibfnamefont {N.~M.}\ \bibnamefont
  {Sundaresan}}, \bibinfo {author} {\bibfnamefont {R.}~\bibnamefont
  {Lundgren}}, \bibinfo {author} {\bibfnamefont {G.}~\bibnamefont {Zhu}},
  \bibinfo {author} {\bibfnamefont {A.~V.}\ \bibnamefont {Gorshkov}},\ and\
  \bibinfo {author} {\bibfnamefont {A.~A.}\ \bibnamefont {Houck}},\ }\bibfield
  {title} {\bibinfo {title} {Interacting qubit-photon bound states with
  superconducting circuits},\ }\href
  {https://doi.org/10.1103/PhysRevX.9.011021} {\bibfield  {journal} {\bibinfo
  {journal} {Phys. Rev. X}\ }\textbf {\bibinfo {volume} {9}},\ \bibinfo {pages}
  {011021} (\bibinfo {year} {2019})}\BibitemShut {NoStop}%
\bibitem [{\citenamefont {Zhao}\ and\ \citenamefont
  {Wang}(2020)}]{Zhao-PRA2020}%
  \BibitemOpen
  \bibfield  {author} {\bibinfo {author} {\bibfnamefont {W.}~\bibnamefont
  {Zhao}}\ and\ \bibinfo {author} {\bibfnamefont {Z.}~\bibnamefont {Wang}},\
  }\bibfield  {title} {\bibinfo {title} {Single-photon scattering and bound
  states in an atom-waveguide system with two or multiple coupling points},\
  }\href {https://doi.org/10.1103/PhysRevA.101.053855} {\bibfield  {journal}
  {\bibinfo  {journal} {Phys. Rev. A}\ }\textbf {\bibinfo {volume} {101}},\
  \bibinfo {pages} {053855} (\bibinfo {year} {2020})}\BibitemShut {NoStop}%
\bibitem [{\citenamefont {Ferreira}\ \emph {et~al.}(2021)\citenamefont
  {Ferreira}, \citenamefont {Banker}, \citenamefont {Sipahigil}, \citenamefont
  {Matheny}, \citenamefont {Keller}, \citenamefont {Kim}, \citenamefont
  {Mirhosseini},\ and\ \citenamefont {Painter}}]{Ferreira-PRX2021}%
  \BibitemOpen
  \bibfield  {author} {\bibinfo {author} {\bibfnamefont {V.~S.}\ \bibnamefont
  {Ferreira}}, \bibinfo {author} {\bibfnamefont {J.}~\bibnamefont {Banker}},
  \bibinfo {author} {\bibfnamefont {A.}~\bibnamefont {Sipahigil}}, \bibinfo
  {author} {\bibfnamefont {M.~H.}\ \bibnamefont {Matheny}}, \bibinfo {author}
  {\bibfnamefont {A.~J.}\ \bibnamefont {Keller}}, \bibinfo {author}
  {\bibfnamefont {E.}~\bibnamefont {Kim}}, \bibinfo {author} {\bibfnamefont
  {M.}~\bibnamefont {Mirhosseini}},\ and\ \bibinfo {author} {\bibfnamefont
  {O.}~\bibnamefont {Painter}},\ }\bibfield  {title} {\bibinfo {title}
  {Collapse and revival of an artificial atom coupled to a structured photonic
  reservoir},\ }\href {https://doi.org/10.1103/PhysRevX.11.041043} {\bibfield
  {journal} {\bibinfo  {journal} {Phys. Rev. X}\ }\textbf {\bibinfo {volume}
  {11}},\ \bibinfo {pages} {041043} (\bibinfo {year} {2021})}\BibitemShut
  {NoStop}%
\bibitem [{\citenamefont {Wang}\ \emph {et~al.}(2021)\citenamefont {Wang},
  \citenamefont {Liu}, \citenamefont {Kockum}, \citenamefont {Li},\ and\
  \citenamefont {Nori}}]{Wang-PRL2021}%
  \BibitemOpen
  \bibfield  {author} {\bibinfo {author} {\bibfnamefont {X.}~\bibnamefont
  {Wang}}, \bibinfo {author} {\bibfnamefont {T.}~\bibnamefont {Liu}}, \bibinfo
  {author} {\bibfnamefont {A.~F.}\ \bibnamefont {Kockum}}, \bibinfo {author}
  {\bibfnamefont {H.-R.}\ \bibnamefont {Li}},\ and\ \bibinfo {author}
  {\bibfnamefont {F.}~\bibnamefont {Nori}},\ }\bibfield  {title} {\bibinfo
  {title} {Tunable chiral bound states with giant atoms},\ }\href
  {https://doi.org/10.1103/PhysRevLett.126.043602} {\bibfield  {journal}
  {\bibinfo  {journal} {Phys. Rev. Lett.}\ }\textbf {\bibinfo {volume} {126}},\
  \bibinfo {pages} {043602} (\bibinfo {year} {2021})}\BibitemShut {NoStop}%
\bibitem [{\citenamefont {Scigliuzzo}\ \emph {et~al.}(2022)\citenamefont
  {Scigliuzzo}, \citenamefont {Calaj\`o}, \citenamefont {Ciccarello},
  \citenamefont {Perez~Lozano}, \citenamefont {Bengtsson}, \citenamefont
  {Scarlino}, \citenamefont {Wallraff}, \citenamefont {Chang}, \citenamefont
  {Delsing},\ and\ \citenamefont {Gasparinetti}}]{Scigliuzzo-PRX2022}%
  \BibitemOpen
  \bibfield  {author} {\bibinfo {author} {\bibfnamefont {M.}~\bibnamefont
  {Scigliuzzo}}, \bibinfo {author} {\bibfnamefont {G.}~\bibnamefont
  {Calaj\`o}}, \bibinfo {author} {\bibfnamefont {F.}~\bibnamefont
  {Ciccarello}}, \bibinfo {author} {\bibfnamefont {D.}~\bibnamefont
  {Perez~Lozano}}, \bibinfo {author} {\bibfnamefont {A.}~\bibnamefont
  {Bengtsson}}, \bibinfo {author} {\bibfnamefont {P.}~\bibnamefont {Scarlino}},
  \bibinfo {author} {\bibfnamefont {A.}~\bibnamefont {Wallraff}}, \bibinfo
  {author} {\bibfnamefont {D.}~\bibnamefont {Chang}}, \bibinfo {author}
  {\bibfnamefont {P.}~\bibnamefont {Delsing}},\ and\ \bibinfo {author}
  {\bibfnamefont {S.}~\bibnamefont {Gasparinetti}},\ }\bibfield  {title}
  {\bibinfo {title} {Controlling atom-photon bound states in an array of
  josephson-junction resonators},\ }\href
  {https://doi.org/10.1103/PhysRevX.12.031036} {\bibfield  {journal} {\bibinfo
  {journal} {Phys. Rev. X}\ }\textbf {\bibinfo {volume} {12}},\ \bibinfo
  {pages} {031036} (\bibinfo {year} {2022})}\BibitemShut {NoStop}%
\bibitem [{\citenamefont {Wang}\ and\ \citenamefont {Li}(2022)}]{Wang-QST2022}%
  \BibitemOpen
  \bibfield  {author} {\bibinfo {author} {\bibfnamefont {X.}~\bibnamefont
  {Wang}}\ and\ \bibinfo {author} {\bibfnamefont {H.-R.}\ \bibnamefont {Li}},\
  }\bibfield  {title} {\bibinfo {title} {Chiral quantum network with giant
  atoms},\ }\href {https://doi.org/10.1088/2058-9565/ac6a04} {\bibfield
  {journal} {\bibinfo  {journal} {Quantum Sci. Technol}\ }\textbf {\bibinfo
  {volume} {7}},\ \bibinfo {pages} {035007} (\bibinfo {year}
  {2022})}\BibitemShut {NoStop}%
\bibitem [{\citenamefont {Soro}\ \emph {et~al.}(2023)\citenamefont {Soro},
  \citenamefont {Mu\~noz},\ and\ \citenamefont {Kockum}}]{Soro-PRA2023}%
  \BibitemOpen
  \bibfield  {author} {\bibinfo {author} {\bibfnamefont {A.}~\bibnamefont
  {Soro}}, \bibinfo {author} {\bibfnamefont {C.~S.}\ \bibnamefont {Mu\~noz}},\
  and\ \bibinfo {author} {\bibfnamefont {A.~F.}\ \bibnamefont {Kockum}},\
  }\bibfield  {title} {\bibinfo {title} {Interaction between giant atoms in a
  one-dimensional structured environment},\ }\href
  {https://doi.org/10.1103/PhysRevA.107.013710} {\bibfield  {journal} {\bibinfo
   {journal} {Phys. Rev. A}\ }\textbf {\bibinfo {volume} {107}},\ \bibinfo
  {pages} {013710} (\bibinfo {year} {2023})}\BibitemShut {NoStop}%
\bibitem [{\citenamefont {Jia}\ and\ \citenamefont {Yu}(2024)}]{Jia-OE2024}%
  \BibitemOpen
  \bibfield  {author} {\bibinfo {author} {\bibfnamefont {W.~Z.}\ \bibnamefont
  {Jia}}\ and\ \bibinfo {author} {\bibfnamefont {M.~T.}\ \bibnamefont {Yu}},\
  }\bibfield  {title} {\bibinfo {title} {Atom-photon dressed states in a
  waveguide-qed system with multiple giant atoms},\ }\href
  {https://doi.org/10.1364/OE.518325} {\bibfield  {journal} {\bibinfo
  {journal} {Opt. Express}\ }\textbf {\bibinfo {volume} {32}},\ \bibinfo
  {pages} {9495} (\bibinfo {year} {2024})}\BibitemShut {NoStop}%
\bibitem [{\citenamefont {Tufarelli}\ \emph {et~al.}(2013)\citenamefont
  {Tufarelli}, \citenamefont {Ciccarello},\ and\ \citenamefont
  {Kim}}]{Tufarelli-PRA2013}%
  \BibitemOpen
  \bibfield  {author} {\bibinfo {author} {\bibfnamefont {T.}~\bibnamefont
  {Tufarelli}}, \bibinfo {author} {\bibfnamefont {F.}~\bibnamefont
  {Ciccarello}},\ and\ \bibinfo {author} {\bibfnamefont {M.~S.}\ \bibnamefont
  {Kim}},\ }\bibfield  {title} {\bibinfo {title} {Dynamics of spontaneous
  emission in a single-end photonic waveguide},\ }\href
  {https://doi.org/10.1103/PhysRevA.87.013820} {\bibfield  {journal} {\bibinfo
  {journal} {Phys. Rev. A}\ }\textbf {\bibinfo {volume} {87}},\ \bibinfo
  {pages} {013820} (\bibinfo {year} {2013})}\BibitemShut {NoStop}%
\bibitem [{\citenamefont {Hsu}\ \emph {et~al.}(2016)\citenamefont {Hsu},
  \citenamefont {Zhen}, \citenamefont {Stone}, \citenamefont {Joannopoulos},\
  and\ \citenamefont {Solja\v{c}i\'c}}]{Hsu-NatRevMats2016}%
  \BibitemOpen
  \bibfield  {author} {\bibinfo {author} {\bibfnamefont {C.~W.}\ \bibnamefont
  {Hsu}}, \bibinfo {author} {\bibfnamefont {B.}~\bibnamefont {Zhen}}, \bibinfo
  {author} {\bibfnamefont {A.~D.}\ \bibnamefont {Stone}}, \bibinfo {author}
  {\bibfnamefont {J.~D.}\ \bibnamefont {Joannopoulos}},\ and\ \bibinfo {author}
  {\bibfnamefont {M.}~\bibnamefont {Solja\v{c}i\'c}},\ }\bibfield  {title}
  {\bibinfo {title} {Bound states in the continuum},\ }\href
  {https://doi.org/10.1038/natrevmats.2016.48} {\bibfield  {journal} {\bibinfo
  {journal} {Nat. Rev. Mater.}\ }\textbf {\bibinfo {volume} {1}},\ \bibinfo
  {pages} {16048} (\bibinfo {year} {2016})}\BibitemShut {NoStop}%
\bibitem [{\citenamefont {Fong}\ and\ \citenamefont
  {Law}(2017)}]{Fong-PRA2017}%
  \BibitemOpen
  \bibfield  {author} {\bibinfo {author} {\bibfnamefont {P.~T.}\ \bibnamefont
  {Fong}}\ and\ \bibinfo {author} {\bibfnamefont {C.~K.}\ \bibnamefont {Law}},\
  }\bibfield  {title} {\bibinfo {title} {Bound state in the continuum by
  spatially separated ensembles of atoms in a coupled-cavity array},\ }\href
  {https://doi.org/10.1103/PhysRevA.96.023842} {\bibfield  {journal} {\bibinfo
  {journal} {Phys. Rev. A}\ }\textbf {\bibinfo {volume} {96}},\ \bibinfo
  {pages} {023842} (\bibinfo {year} {2017})}\BibitemShut {NoStop}%
\bibitem [{\citenamefont {Facchi}\ \emph {et~al.}(2019)\citenamefont {Facchi},
  \citenamefont {Lonigro}, \citenamefont {Pascazio}, \citenamefont {Pepe},\
  and\ \citenamefont {Pomarico}}]{Facchi-PRA2019}%
  \BibitemOpen
  \bibfield  {author} {\bibinfo {author} {\bibfnamefont {P.}~\bibnamefont
  {Facchi}}, \bibinfo {author} {\bibfnamefont {D.}~\bibnamefont {Lonigro}},
  \bibinfo {author} {\bibfnamefont {S.}~\bibnamefont {Pascazio}}, \bibinfo
  {author} {\bibfnamefont {F.~V.}\ \bibnamefont {Pepe}},\ and\ \bibinfo
  {author} {\bibfnamefont {D.}~\bibnamefont {Pomarico}},\ }\bibfield  {title}
  {\bibinfo {title} {Bound states in the continuum for an array of quantum
  emitters},\ }\href {https://doi.org/10.1103/PhysRevA.100.023834} {\bibfield
  {journal} {\bibinfo  {journal} {Phys. Rev. A}\ }\textbf {\bibinfo {volume}
  {100}},\ \bibinfo {pages} {023834} (\bibinfo {year} {2019})}\BibitemShut
  {NoStop}%
\bibitem [{\citenamefont {Calajó}\ \emph {et~al.}(2019)\citenamefont
  {Calajó}, \citenamefont {Fang}, \citenamefont {Baranger},\ and\
  \citenamefont {Ciccarello}}]{Calajo-PRL2019}%
  \BibitemOpen
  \bibfield  {author} {\bibinfo {author} {\bibfnamefont {G.}~\bibnamefont
  {Calajó}}, \bibinfo {author} {\bibfnamefont {Y.-L.~L.}\ \bibnamefont
  {Fang}}, \bibinfo {author} {\bibfnamefont {H.~U.}\ \bibnamefont {Baranger}},\
  and\ \bibinfo {author} {\bibfnamefont {F.}~\bibnamefont {Ciccarello}},\
  }\bibfield  {title} {\bibinfo {title} {Exciting a bound state in the
  continuum through multiphoton scattering plus delayed quantum feedback},\
  }\href {https://doi.org/10.1103/PhysRevLett.122.073601} {\bibfield  {journal}
  {\bibinfo  {journal} {Phys. Rev. Lett.}\ }\textbf {\bibinfo {volume} {122}},\
  \bibinfo {pages} {073601} (\bibinfo {year} {2019})}\BibitemShut {NoStop}%
\bibitem [{\citenamefont {Sinha}\ \emph {et~al.}(2020)\citenamefont {Sinha},
  \citenamefont {Meystre}, \citenamefont {Goldschmidt}, \citenamefont {Fatemi},
  \citenamefont {Rolston},\ and\ \citenamefont {Solano}}]{Sinha-PRL2020}%
  \BibitemOpen
  \bibfield  {author} {\bibinfo {author} {\bibfnamefont {K.}~\bibnamefont
  {Sinha}}, \bibinfo {author} {\bibfnamefont {P.}~\bibnamefont {Meystre}},
  \bibinfo {author} {\bibfnamefont {E.~A.}\ \bibnamefont {Goldschmidt}},
  \bibinfo {author} {\bibfnamefont {F.~K.}\ \bibnamefont {Fatemi}}, \bibinfo
  {author} {\bibfnamefont {S.~L.}\ \bibnamefont {Rolston}},\ and\ \bibinfo
  {author} {\bibfnamefont {P.}~\bibnamefont {Solano}},\ }\bibfield  {title}
  {\bibinfo {title} {Non-markovian collective emission from macroscopically
  separated emitters},\ }\href {https://doi.org/10.1103/PhysRevLett.124.043603}
  {\bibfield  {journal} {\bibinfo  {journal} {Phys. Rev. Lett.}\ }\textbf
  {\bibinfo {volume} {124}},\ \bibinfo {pages} {043603} (\bibinfo {year}
  {2020})}\BibitemShut {NoStop}%
\bibitem [{\citenamefont {Qiu}\ \emph {et~al.}(2023)\citenamefont {Qiu},
  \citenamefont {Wu},\ and\ \citenamefont {L\"u}}]{Qiu-SciChina2023}%
  \BibitemOpen
  \bibfield  {author} {\bibinfo {author} {\bibfnamefont {Q.-Y.}\ \bibnamefont
  {Qiu}}, \bibinfo {author} {\bibfnamefont {Y.}~\bibnamefont {Wu}},\ and\
  \bibinfo {author} {\bibfnamefont {X.-Y.}\ \bibnamefont {L\"u}},\ }\bibfield
  {title} {\bibinfo {title} {Collective radiance of giant atoms in
  non-markovian regime},\ }\href {https://doi.org/10.1007/s11433-022-1990-x}
  {\bibfield  {journal} {\bibinfo  {journal} {Sci. China-Phys. Mech. Astron.}\
  }\textbf {\bibinfo {volume} {66}},\ \bibinfo {pages} {224212} (\bibinfo
  {year} {2023})}\BibitemShut {NoStop}%
\bibitem [{\citenamefont {Yu}\ \emph {et~al.}(2025)\citenamefont {Yu},
  \citenamefont {Zhang}, \citenamefont {Wang},\ and\ \citenamefont
  {Wang}}]{Yu-PRA2025}%
  \BibitemOpen
  \bibfield  {author} {\bibinfo {author} {\bibfnamefont {H.}~\bibnamefont
  {Yu}}, \bibinfo {author} {\bibfnamefont {X.}~\bibnamefont {Zhang}}, \bibinfo
  {author} {\bibfnamefont {Z.}~\bibnamefont {Wang}},\ and\ \bibinfo {author}
  {\bibfnamefont {J.}~\bibnamefont {Wang}},\ }\bibfield  {title} {\bibinfo
  {title} {Rabi oscillation and fractional population via the bound states in
  the continuum in a giant-atom waveguide qed setup},\ }\href
  {https://doi.org/10.1103/PhysRevA.111.053710} {\bibfield  {journal} {\bibinfo
   {journal} {Phys. Rev. A}\ }\textbf {\bibinfo {volume} {111}},\ \bibinfo
  {pages} {053710} (\bibinfo {year} {2025})}\BibitemShut {NoStop}%
\bibitem [{\citenamefont {Lombardo}\ \emph {et~al.}(2014)\citenamefont
  {Lombardo}, \citenamefont {Ciccarello},\ and\ \citenamefont
  {Palma}}]{Lombardo-PRA2014}%
  \BibitemOpen
  \bibfield  {author} {\bibinfo {author} {\bibfnamefont {F.}~\bibnamefont
  {Lombardo}}, \bibinfo {author} {\bibfnamefont {F.}~\bibnamefont
  {Ciccarello}},\ and\ \bibinfo {author} {\bibfnamefont {G.~M.}\ \bibnamefont
  {Palma}},\ }\bibfield  {title} {\bibinfo {title} {Photon localization versus
  population trapping in a coupled-cavity array},\ }\href
  {https://doi.org/10.1103/PhysRevA.89.053826} {\bibfield  {journal} {\bibinfo
  {journal} {Phys. Rev. A}\ }\textbf {\bibinfo {volume} {89}},\ \bibinfo
  {pages} {053826} (\bibinfo {year} {2014})}\BibitemShut {NoStop}%
\bibitem [{\citenamefont {He}\ \emph {et~al.}(2025)\citenamefont {He},
  \citenamefont {Sun}, \citenamefont {Li}, \citenamefont {Yang}, \citenamefont
  {Lu},\ and\ \citenamefont {Zhou}}]{He-QT2025}%
  \BibitemOpen
  \bibfield  {author} {\bibinfo {author} {\bibfnamefont {Y.}~\bibnamefont
  {He}}, \bibinfo {author} {\bibfnamefont {G.}~\bibnamefont {Sun}}, \bibinfo
  {author} {\bibfnamefont {J.}~\bibnamefont {Li}}, \bibinfo {author}
  {\bibfnamefont {Y.}~\bibnamefont {Yang}}, \bibinfo {author} {\bibfnamefont
  {J.}~\bibnamefont {Lu}},\ and\ \bibinfo {author} {\bibfnamefont
  {L.}~\bibnamefont {Zhou}},\ }\bibfield  {title} {\bibinfo {title} {Emergent
  oscillating bound states in a semi-infinite linear waveguide with a
  point-like $\lambda$-type quantum emitter driven by a classical field},\
  }\href {https://doi.org/https://doi.org/10.1002/qute.202400535} {\bibfield
  {journal} {\bibinfo  {journal} {Adv. Quantum Technol.}\ }\textbf {\bibinfo
  {volume} {8}},\ \bibinfo {pages} {2400535} (\bibinfo {year}
  {2025})}\BibitemShut {NoStop}%
\bibitem [{\citenamefont {Guo}\ \emph {et~al.}(2020{\natexlab{b}})\citenamefont
  {Guo}, \citenamefont {Wang}, \citenamefont {Purdy},\ and\ \citenamefont
  {Taylor}}]{Guo-PRA2020}%
  \BibitemOpen
  \bibfield  {author} {\bibinfo {author} {\bibfnamefont {S.}~\bibnamefont
  {Guo}}, \bibinfo {author} {\bibfnamefont {Y.}~\bibnamefont {Wang}}, \bibinfo
  {author} {\bibfnamefont {T.}~\bibnamefont {Purdy}},\ and\ \bibinfo {author}
  {\bibfnamefont {J.}~\bibnamefont {Taylor}},\ }\bibfield  {title} {\bibinfo
  {title} {Beyond spontaneous emission: Giant atom bounded in the continuum},\
  }\href {https://doi.org/10.1103/PhysRevA.102.033706} {\bibfield  {journal}
  {\bibinfo  {journal} {Phys. Rev. A}\ }\textbf {\bibinfo {volume} {102}},\
  \bibinfo {pages} {033706} (\bibinfo {year} {2020}{\natexlab{b}})}\BibitemShut
  {NoStop}%
\bibitem [{\citenamefont {Lim}\ \emph {et~al.}(2023)\citenamefont {Lim},
  \citenamefont {Mok},\ and\ \citenamefont {Kwek}}]{Lim-PRA2023}%
  \BibitemOpen
  \bibfield  {author} {\bibinfo {author} {\bibfnamefont {K.~H.}\ \bibnamefont
  {Lim}}, \bibinfo {author} {\bibfnamefont {W.-K.}\ \bibnamefont {Mok}},\ and\
  \bibinfo {author} {\bibfnamefont {L.-C.}\ \bibnamefont {Kwek}},\ }\bibfield
  {title} {\bibinfo {title} {Oscillating bound states in non-markovian photonic
  lattices},\ }\href {https://doi.org/10.1103/PhysRevA.107.023716} {\bibfield
  {journal} {\bibinfo  {journal} {Phys. Rev. A}\ }\textbf {\bibinfo {volume}
  {107}},\ \bibinfo {pages} {023716} (\bibinfo {year} {2023})}\BibitemShut
  {NoStop}%
\bibitem [{\citenamefont {Zhang}\ \emph {et~al.}(2023)\citenamefont {Zhang},
  \citenamefont {Liu}, \citenamefont {Gong},\ and\ \citenamefont
  {Wang}}]{Zhang-PRA2023}%
  \BibitemOpen
  \bibfield  {author} {\bibinfo {author} {\bibfnamefont {X.}~\bibnamefont
  {Zhang}}, \bibinfo {author} {\bibfnamefont {C.}~\bibnamefont {Liu}}, \bibinfo
  {author} {\bibfnamefont {Z.}~\bibnamefont {Gong}},\ and\ \bibinfo {author}
  {\bibfnamefont {Z.}~\bibnamefont {Wang}},\ }\bibfield  {title} {\bibinfo
  {title} {Quantum interference and controllable magic cavity qed via a giant
  atom in a coupled resonator waveguide},\ }\href
  {https://doi.org/10.1103/PhysRevA.108.013704} {\bibfield  {journal} {\bibinfo
   {journal} {Phys. Rev. A}\ }\textbf {\bibinfo {volume} {108}},\ \bibinfo
  {pages} {013704} (\bibinfo {year} {2023})}\BibitemShut {NoStop}%
\bibitem [{\citenamefont {Li}\ and\ \citenamefont {Shen}(2024)}]{Li-PRA2024}%
  \BibitemOpen
  \bibfield  {author} {\bibinfo {author} {\bibfnamefont {Z.~Y.}\ \bibnamefont
  {Li}}\ and\ \bibinfo {author} {\bibfnamefont {H.~Z.}\ \bibnamefont {Shen}},\
  }\bibfield  {title} {\bibinfo {title} {Non-markovian dynamics with a giant
  atom coupled to a semi-infinite photonic waveguide},\ }\href
  {https://doi.org/10.1103/PhysRevA.109.023712} {\bibfield  {journal} {\bibinfo
   {journal} {Phys. Rev. A}\ }\textbf {\bibinfo {volume} {109}},\ \bibinfo
  {pages} {023712} (\bibinfo {year} {2024})}\BibitemShut {NoStop}%
\bibitem [{\citenamefont {Yang}\ \emph {et~al.}(2025)\citenamefont {Yang},
  \citenamefont {Sun}, \citenamefont {Li}, \citenamefont {Lu},\ and\
  \citenamefont {Zhou}}]{Yang-PRA2025}%
  \BibitemOpen
  \bibfield  {author} {\bibinfo {author} {\bibfnamefont {Y.}~\bibnamefont
  {Yang}}, \bibinfo {author} {\bibfnamefont {G.}~\bibnamefont {Sun}}, \bibinfo
  {author} {\bibfnamefont {J.}~\bibnamefont {Li}}, \bibinfo {author}
  {\bibfnamefont {J.}~\bibnamefont {Lu}},\ and\ \bibinfo {author}
  {\bibfnamefont {L.}~\bibnamefont {Zhou}},\ }\bibfield  {title} {\bibinfo
  {title} {Coherent control of spontaneous emission for a giant driven
  $\mathrm{\ensuremath{\Lambda}}$-type three-level atom},\ }\href
  {https://doi.org/10.1103/PhysRevA.111.013707} {\bibfield  {journal} {\bibinfo
   {journal} {Phys. Rev. A}\ }\textbf {\bibinfo {volume} {111}},\ \bibinfo
  {pages} {013707} (\bibinfo {year} {2025})}\BibitemShut {NoStop}%
\bibitem [{\citenamefont {Sun}\ \emph {et~al.}(2025)\citenamefont {Sun},
  \citenamefont {Yang}, \citenamefont {Li}, \citenamefont {Lu},\ and\
  \citenamefont {Zhou}}]{Sun-PRA2025}%
  \BibitemOpen
  \bibfield  {author} {\bibinfo {author} {\bibfnamefont {G.}~\bibnamefont
  {Sun}}, \bibinfo {author} {\bibfnamefont {Y.}~\bibnamefont {Yang}}, \bibinfo
  {author} {\bibfnamefont {J.}~\bibnamefont {Li}}, \bibinfo {author}
  {\bibfnamefont {J.}~\bibnamefont {Lu}},\ and\ \bibinfo {author}
  {\bibfnamefont {L.}~\bibnamefont {Zhou}},\ }\bibfield  {title} {\bibinfo
  {title} {Cavity-modified oscillating bound states with a
  $\mathrm{\ensuremath{\Lambda}}$-type giant emitter in a linear waveguide},\
  }\href {https://doi.org/10.1103/PhysRevA.111.033701} {\bibfield  {journal}
  {\bibinfo  {journal} {Phys. Rev. A}\ }\textbf {\bibinfo {volume} {111}},\
  \bibinfo {pages} {033701} (\bibinfo {year} {2025})}\BibitemShut {NoStop}%
\bibitem [{\citenamefont {Noachtar}\ \emph {et~al.}(2022)\citenamefont
  {Noachtar}, \citenamefont {Kn\"orzer},\ and\ \citenamefont
  {Jonsson}}]{Noachtar-PRA2022}%
  \BibitemOpen
  \bibfield  {author} {\bibinfo {author} {\bibfnamefont {D.~D.}\ \bibnamefont
  {Noachtar}}, \bibinfo {author} {\bibfnamefont {J.}~\bibnamefont
  {Kn\"orzer}},\ and\ \bibinfo {author} {\bibfnamefont {R.~H.}\ \bibnamefont
  {Jonsson}},\ }\bibfield  {title} {\bibinfo {title} {Nonperturbative treatment
  of giant atoms using chain transformations},\ }\href
  {https://doi.org/10.1103/PhysRevA.106.013702} {\bibfield  {journal} {\bibinfo
   {journal} {Phys. Rev. A}\ }\textbf {\bibinfo {volume} {106}},\ \bibinfo
  {pages} {013702} (\bibinfo {year} {2022})}\BibitemShut {NoStop}%
\bibitem [{\citenamefont {Terradas-Briansó}\ \emph {et~al.}(2022)\citenamefont
  {Terradas-Briansó}, \citenamefont {Gonz\'alez-Guti\'errez}, \citenamefont
  {Nori}, \citenamefont {Mart\'{\i}n-Moreno},\ and\ \citenamefont
  {Zueco}}]{Terradas-Brianso-PRA2022}%
  \BibitemOpen
  \bibfield  {author} {\bibinfo {author} {\bibfnamefont {S.}~\bibnamefont
  {Terradas-Briansó}}, \bibinfo {author} {\bibfnamefont {C.~A.}\ \bibnamefont
  {Gonz\'alez-Guti\'errez}}, \bibinfo {author} {\bibfnamefont {F.}~\bibnamefont
  {Nori}}, \bibinfo {author} {\bibfnamefont {L.}~\bibnamefont
  {Mart\'{\i}n-Moreno}},\ and\ \bibinfo {author} {\bibfnamefont
  {D.}~\bibnamefont {Zueco}},\ }\bibfield  {title} {\bibinfo {title}
  {Ultrastrong waveguide qed with giant atoms},\ }\href
  {https://doi.org/10.1103/PhysRevA.106.063717} {\bibfield  {journal} {\bibinfo
   {journal} {Phys. Rev. A}\ }\textbf {\bibinfo {volume} {106}},\ \bibinfo
  {pages} {063717} (\bibinfo {year} {2022})}\BibitemShut {NoStop}%
\bibitem [{\citenamefont {Peng}\ and\ \citenamefont
  {Jia}(2023)}]{Peng-PRA2023}%
  \BibitemOpen
  \bibfield  {author} {\bibinfo {author} {\bibfnamefont {Y.~P.}\ \bibnamefont
  {Peng}}\ and\ \bibinfo {author} {\bibfnamefont {W.~Z.}\ \bibnamefont {Jia}},\
  }\bibfield  {title} {\bibinfo {title} {Single-photon scattering from a chain
  of giant atoms coupled to a one-dimensional waveguide},\ }\href
  {https://doi.org/10.1103/PhysRevA.108.043709} {\bibfield  {journal} {\bibinfo
   {journal} {Phys. Rev. A}\ }\textbf {\bibinfo {volume} {108}},\ \bibinfo
  {pages} {043709} (\bibinfo {year} {2023})}\BibitemShut {NoStop}%
\end{thebibliography}%

\end{document}